%% file: scalar_v2.tex
\documentclass[11pt]{article}
\usepackage{latexsym}
\usepackage{chapterbib}
\usepackage{graphics,graphicx}
\usepackage{amssymb,amsmath,amsfonts,amstext}
\usepackage{color,xcolor}
\usepackage{hyperref}
\usepackage{geometry}  
\usepackage{braket}
\usepackage{cancel}
\usepackage[utf8]{inputenc}
\usepackage{slashed}
\usepackage{tikz}
\usepackage{accents}
\usepackage{cite}
\usepackage{framed}
\usepackage{bm}

\allowdisplaybreaks 

\include{macros}

\begin{document}

\begin{titlepage}

\pagestyle{empty}

\begin{flushright}
 {\small SISSA 13/2017/FISI}
\end{flushright}
\vskip1.5in

\begin{center}
\textbf{\LARGE Anomaly-corrected supersymmetry algebra \vskip0.2cm and supersymmetric holographic renormalization}
\end{center}
\vskip0.2in

\begin{center}
{\large Ok Song An}
\end{center}
\vskip0.2in

\begin{abstract}
We present a systematic approach to supersymmetric holographic renormalization for a generic 5D $\mathcal{N}=2$ gauged supergravity theory with matter multiplets, including its fermionic sector, with all gauge fields consistently set to zero. We determine  the complete set of supersymmetric local boundary counterterms, including the finite counterterms that parameterize the choice of supersymmetric renormalization scheme. This allows us to derive holographically the superconformal Ward identities of a 4D superconformal field theory on a generic background, including the Weyl and super-Weyl anomalies. Moreover, we show that these anomalies satisfy the Wess-Zumino consistency condition. The super-Weyl anomaly implies that the fermionic operators of the dual field theory, such as the supercurrent, do not transform as tensors under rigid supersymmetry on backgrounds that admit a conformal Killing spinor, and their anticommutator with the conserved supercharge contains anomalous terms. This property is explicitly checked for a toy model. Finally, using the anomalous transformation of the supercurrent, we obtain the anomaly-corrected supersymmetry algebra on curved backgrounds admitting a conformal Killing spinor.
\end{abstract}

\end{titlepage}

\tableofcontents
\addtocontents{toc}{\protect\setcounter{tocdepth}{3}}
\renewcommand{\theequation}{\arabic{section}.\arabic{equation}}

\section{Introduction}
\setcounter{equation}{0}

Supersymmetric (SUSY) field theories in curved backgrounds \cite{Festuccia:2011ws,Dumitrescu:2012ha,Klare:2012gn} (see also \cite{Dumitrescu:2016ltq} for a recent review) have received much attention in recent years, since they provide a playground where physically interesting, non-perturbative, results can often be obtained through localization techniques \cite{Nekrasov:2002qd,Pestun:2007rz}.

Formulating consistent SUSY field theories in curved space usually consists of two steps \cite{Festuccia:2011ws}; the first one is to find the classical supergravity theory (SUGRA) by coupling a flat-space SUSY field theory to the gravity multiplet, and the second one is to take a rigid limit of SUGRA such that the gravity multiplet becomes non-dynamical, but maintains a non-trivial background value. Consistency requires that there exists at least one SUSY transformation of the SUGRA under which this background gravity multiplet should be invariant, namely
\be \label{SUSY-background}
\d_\h e^a_{(0)i}=0,\quad \d_\h\J_{(0)+i}=0,\quad \cdots, 
\ee 
where $e^a_{(0)i}$ refers to the vielbein and $\J_{(0)+i}$ is the gravitino field and $\h$ refers to the spinor parameter of the preserved SUSY. We refer to appendix \ref{gamma-matrices} and \ref{ADM-SBC} for notations and conventions. The requirement that variation of the bosonic fields vanish is trivially satisfied on bosonic backgrounds.

One then derives the SUSY transformation of the local operators and the SUSY algebra in curved space from the corresponding ones of SUGRA. However, they are \emph{classical} in the sense that the SUSY transformation laws and algebra derived in this way do not reflect any quantum effects.

To clarify this point, let us schematically discuss these quantum effects for a theory with an $\cn=1$ 4D superconformal field theory (SCFT) as a UV fixed point. For this aim, we derive the Ward identities which contain UV data of quantum field theories. These Ward identities can be obtained in a local renormalization group language \cite{Osborn:1991gm} without relying on a classical Lagrangian description, see e.g. section 2.3 in \cite{Papadimitriou:2016yit} for a recent review. In $\cn=1$ SCFT, we have two local fermionic transformations, supersymmetry and super-Weyl, respectively
\bsub 
\bal 
& \d_{\e_+}e^{a}_{(0)i}=-\frac12\lbar\J_{(0)+i}\G^a\e_+,\quad \d_{\e_+}\J_{(0)+i}=\bb D_i\e_++\cdots,\quad \cdots \\
& \d_{\e_-}e^{a}_{(0)i}=0,\quad \d_{\e_-}\J_{(0)+i}=-\Hat\G_i\e_-+\cdots,\quad  \cdots 
\eal
\esub 
where the ellipses indicate possible contributions from other fields in the gravity multiplet and higer-order terms in fermions. Requiring the generating functional of connected correlation functions, $W[g_{(0)ij},\J_{(0)+i},\cdots]$, to be invariant under these local transformations up to a possible anomaly, we obtain two local operator equations, namely
\bsub\label{intro-e+e-ward}
\bal 
& \frac12\ct_a^i\lbar\J_{(0)+i}\G^a-\lbar\cs^i\labD_i+\cdots =\lbar \ca_s,\\
& -\lbar\cs^i\Hat\G_{(0)i}+\cdots=\lbar \ca_{\rm sW},
\eal 
\esub 
where $\ct_a^i$ and $\cs^i$ refer to the energy-momentum tensor and supercurrent operator, respectively. Note that the Ward identities hold for generic background such as ones where the fermionic sources are turned on. Combining these two Ward identities with the parameters $\h_+$ and $\h_-$ which satisfy conformal Killing spinor (CKS) condition
\be \label{intro-CKS}
\d_\h\J_{(0)+i}\equiv \d_{\h_+}\J_{(0)+i}+\d_{\h_-}\J_{(0)+i}=\bb D_i\h_+-\Hat\G_i\h_-=0,
\ee 
to the lowest order in fermions, we obtain the SUSY-$\h$ Ward identity
\be \label{intro-SUSY-ward}
-\frac12\ct^i_a\lbar\J_{(0)+i}\G^a\h_+ + D_i(\lbar\cs^i\h_+)+\cdots=-(\lbar\ca_s\h_++\lbar\ca_{\rm sW}\h_-)\equiv \ca_\h,
\ee 
where the fermionic sources are still turned on, because the CKS equation \eqref{intro-CKS} to the lowest order in fermions does not require the background to be bosonic. One can see from the operator equation \eqref{intro-SUSY-ward} that the SUSY-$\h$ anomaly $\ca_\h$ should be dependent on fermionic background sources such as the gravitini field $\J_{+i}$. Therefore, one may not notice the existence of $\ca_\h$ on the bosonic background.

The Ward identities such as \eqref{intro-SUSY-ward} turn out to be rather useful.\footnote{One should keep in mind that the conservation law which allows to construct the conserved supercharge with non-covariantly-constant rigid parameter $\h_+$ is $D_i(\lbar\cs^i\h_+)=0$, not $\lbar\cs^i\labD_i\h_+=0$.} For instance, they determine the variation of quantum operators under the corresponding symmetry transformations, see e.g. (2.3.7) in \cite{Polchinski:1998rq}. It then follows from \eqref{intro-SUSY-ward} that on the (bosonic) supersymmetric vacua the supercurrent operator $\cs^i$ transforms under the SUSY-$\h$ transformation as
\be \label{intro-SUSY-var-sCur}
\d_\h\cs^i\Big|_{\rm susy-vacua}=\Big(-\frac12\ct^i_a\G^a\h_+-\frac{\d}{\d\lbar\J_{(0)+i}}\ca_\h+\cdots\Big)_{\rm susy-vacua}.
\ee 
We emphasize that the anomalous term $\frac{\d}{\d\lbar\J_{(0)+i}}\ca_\h$ does not appear in the `classical' SUSY variation of the supercurrent operator $\cs^i$, and it is non-zero in generic curved backgrounds admitting a conformal Killing spinor. Moreover, by integrating \eqref{intro-SUSY-var-sCur} over a Cauchy surface, one can obtain the commutator of two supercharges (see e.g. (2.6.14) and (2.6.15) in \cite{Polchinski:1998rq}) and find that it is also corrected by the anomalous term.

The upshot is that once the Ward identities \eqref{intro-e+e-ward} are found, one can see immediately all these quantum corrections. The main obstacle in obtaining \eqref{intro-e+e-ward} is to find out the anomalies $\ca_s$ and $\ca_{\rm sW}$. Fortunately, we have a nice tool for computing the anomalies, i.e. AdS/CFT correspondence \cite{Maldacena:1997re,Gubser:1998bc,Witten:1998qj}. The holographic computation of the quantum anomalies, such as the computation of the Weyl anomaly in \cite{Henningson:1998gx}, results in specific values for the anomaly coefficients. For instance, for the Weyl anomaly one gets $a=c$ from a holographic calculation, which is valid for $\cn=4$ super Yang-Mills theory in the large $N$ limit. This is due to the fact that we use  two-derivative action on the gravity side. To obtain the whole class of anomalies one should consider the higher-derivative action. We emphasize that since the anomalies belonging to the same multiplet are related by SUSY transformations, the super-Weyl anomaly $\ca_{\rm sW}$ obtained by a holographic computation also has specific values for the anomaly coefficients.

Henceforth, in order to obtain the Ward identities of 4D $\cn=1$ SCFT by AdS/CFT, we consider generic $\cn=2$ 5D gauged SUGRA including its fermionic sector in asymptotically locally AdS (AlAdS) spaces, particular examples of which were studied in \cite{Bena:2010pr,Cassani:2010na,Liu:2011dw,Halmagyi:2011yd,Argurio:2014uca,Bertolini:2015hua}.\footnote{Even though the solution considered in \cite{Bertolini:2015hua} is not AlAdS due to existence of the massless scalars, the general form of the action given there is the same with the one here.} More specifically, the theory we consider has a superpotential $\cw$ and its field content consists of a vielbein, two gravitini, as well as an equal number of spin-$1/2$ and scalar fields with negative mass-squared in order for the space to be asymptotically AdS. All gauge fields are consistently set to zero for simplicity. We study this theory up to quadratic order in the fermions. Having a stable AlAdS solution requires that $\cw$ has an isolated local extremum. We also demand that $\cw$ is a local function around that point.

As indicated in \cite{Ceresole:2001wi,Halmagyi:2011yd}, the $\cn=2$ 5D gauged SUGRA can have a superpotential $\cw$ in several cases. A typical case is when there are only vector multiplets and $U(1)_R$ (subgroup of $SU(2)_R$ R-symmetry group) is gauged \cite{Gunaydin:1999zx}. When there are also hypermultiplets, the gauged SUGRA can have a superpotential under a certain constraint related to the `very special geometry' on the scalar manifold of the vector multiplets, which we do not discuss here in detail.

As in field theory, renormalization is indispensable also in the bulk holographic computation. Although it has been studied since the early period of AdS/CFT, many  works on holographic renormalization (HR) \cite{Henningson:1998gx,Balasubramanian:1999re,deBoer:1999tgo,Kraus:1999di,deHaro:2000vlm,Bianchi:2001de,Bianchi:2001kw,Martelli:2002sp,Skenderis:2002wp,Papadimitriou:2004ap} have focused on the bosonic sector. \cite{Henningson:1998cd,Henneaux:1998ch,Arutyunov:1998ve,Volovich:1998tj,Kalkkinen:2000uk,Amsel:2009rr} obtained some boundary counterterms for the fermionic sector, but typically these were limited to either lower dimensional spacetime (mainly 3 or 4 dimensions) or to homogeneous solutions which do not depend on the transverse directions. We note that in a context different from this paper, 4D $\cn=1$ SUGRA including the fermionic sector was treated in \cite{Freedman:2016yue} by a somehow \emph{ad hoc} method.

We perform HR along the lines of \cite{deBoer:1999tgo,Papadimitriou:2004ap,Papadimitriou:2011qb}. By formulating the theory in radial Hamiltonian language, we obtain the radial Hamiltonian, which gives the first class constraints. From the Hamiltonian constraint we obtain the Hamilton-Jacobi (HJ) equation, enabling us to determine the divergent counterterms in a covariant way without relying on a specific solution of the classical SUGRA. We emphasize that the counterterms, as the solution of HJ equation, should satisfy other first class constraints. General covariance of the counterterms is a necessary and sufficient condition to satisfy diffeomorphism constraint, which is one of the first class constraints.

Once the counterterms are obtained, one can renormalize the canonical momenta of the radial Hamiltonian and thus obtain the \emph{renormalized canonical momenta}. According to AdS/CFT dictionary, the renormalized canonical momenta correspond to local operators of the field theory in the local renormalization group language \cite{Osborn:1991gm}. The first class constraints turn out to be relations between local sources and operators, from which we obtain the Ward identities (see \eqref{ward-identities}) that in fact reflect the symmetries of the dual field theory and do not rely on a Lagrangian description of the quantum field theory. Since the bulk theory is 5D $\cn=2$ SUGRA, the dual field theory is supposed to have 4D $\cn=1$ superconformal symmetry and we obtain the corresponding Ward identities. Note that here we cannot see the $U(1)_R$ symmetry because we truncate all gauge fields. In a related work \cite{Papadimitriou:2017kzw} the $U(1)_R$ gauge field is included in the model.

It turns out that the $\cn=1$ superconformal symmetry is broken by anomalies. From the bulk point of view, these anomalies are due to the fact that the first class constraints are non-linear functions of canonical momenta, implying that corresponding symmetries are broken by the radial cut-off. From the dual field theory point of view, of course, the global anomalies are a quantum effect. We obtain not only the SUSY-completion of the trace-anomaly but also the holographic super-Weyl anomaly,\footnote{Notice that the existence of super-Weyl anomaly is natural, due to the existence of Weyl anomaly that is related to the super-Weyl anomaly by SUSY transformation.} which are rather interesting by themselves, since they can provide another tool for testing AdS/CFT.\footnote{As we will see in the main text, our result of the super-Weyl anomaly is different from \cite{Abbott:1977xk} where they obtained it on the field theory by using Feynman diagrams. In \cite{Chaichian:2003kr}, they tried to obtain the holographic super-Weyl anomaly. In any case, we show that our result satisfies Wess-Zumino (WZ) consistency conditions. One can check that the result of \cite{Abbott:1977xk} does not satisfy the consistency condition. See \cite{Bilal:2008qx} for a review of WZ consistency condition.\label{fn:sW-anomaly}}
As discussed before, we find that due to the anomaly operators do not transform as tensors under super-Weyl transformation and the variation of operators gets an anomalous contribution, see \eqref{var-e-pm-remomenta}. Hence, the $Q$-transformation of the operators also becomes anomalous, since it is obtained by putting together supersymmetry and super-Weyl transformations. Here $Q$ refers to the preserved supercharge. This is rather remarkable, since it implies that the `classical' SUSY variation cannot become a total derivative in the path integral of SUSY field theories in curved space, unless the anomaly effects disappear. In this regard, it is shown in \cite{Papadimitriou:2017kzw} that the `new' non-covariant finite counterterms suggested in \cite{Genolini:2016sxe,Genolini:2016ecx} should be discarded since they were introduced in order to match with field theory without taking into account the anomaly-effect. From the anomalous transformation of the supercurrent operator, we find that the supersymmetry algebra in curved space is corrected by anomalous terms, see \eqref{curved-susy-algebra}.

We finally note that the boundary conditions consistent with SUSY should be specified before the main computation of HR. HR has a direct relationship with having a well-defined variational problem \cite{Papadimitriou:2010as}, which requires the boundary condition \emph{a priori}. In this work we always keep Dirichlet boundary conditions for the metric and gravitino field. As we will see, consistency with SUSY requires that either Dirichlet or Neumann boundary conditions should be imposed for scalars and their SUSY-partner spin $1/2$ fields, together at the same time.

The rest of this paper is organized as follows. In section \ref{sugra-5d-action} we review the generic $\cn=2$ 5D gauged SUGRA action and SUSY variation of the fields. In section \ref{radial-Hamiltonian-Holography}, we first present the radial Hamiltonian and other first class constraints for Dirichlet boundary conditions. We then explain a systematic way of holographic renormalization and obtain the flow equations. In section \ref{generic-HJ} we determine the divergent counterterms and possible finite counterterms. In particular, the complete set of counterterms are shown for a toy model. By means of these counterterms, in section \ref{holographic-dic-ward} we obtain the holographic Ward identities and anomalies and show that the anomalies satisfy the Wess-Zumino consistency condition. We then define constraint functions of the phase space by using Ward identities and see that symmetry transformation of the sources and operators are simply described in terms of the Poisson bracket and constraint functions. Finally, we provide important subsequent results which hold on supersymmetric backgrounds and present anomaly-corrected supersymmetry algebra. In section \ref{Neumann-BC} we observe from consistency with SUSY that scalars and their SUSY-partner fields should have the same boundary condition. In appendix \ref{gamma-matrices}, we describe our notations and present some useful identities, and in appendix \ref{ADM-SBC} we develop the preliminary steps to obtain radial Hamiltonian, which contains definition of ADM decomposition, strong Fefferman-Graham (FG) gauge and generalized Penrose-Brown-Henneaux (gPBH) transformations. In appendix \ref{ADM-decomposition}, we show ADM decomposition of the radial Lagrangian part by part and in appendix \ref{momenta-var-derivation} we prove that gPBH transformation of the operators can be obtained from the holographic Ward identities. In appendix \ref{susyalgebra-without-Poisson} we derive the anomaly-corrected SUSY algebra in an alternative way.

\section{$\cn=2$ gauged SUGRA action in 5D}
\label{sugra-5d-action}
\setcounter{equation}{0}

The action of gauged on-shell SUGRA possessing a single superpotential with all gauge fields consistently truncated ($D=d+1=5$) is given by \cite{Bertolini:2015hua}
\be \label{SUGRA-action}
S=S_b+S_f,
\ee
where
\bal
S_b=\;&\frac{1}{2\k^2}\int_{\cm}d^{d+1}x\sqrt{-g}\,\left(R[g]-\cg_{IJ}
(\vf)\pa_\m\vf^I\pa^\m\vf^J-\cv(\vf)\right),\\
S_f=\;&-\frac{1}{2\k^2}\int_{\cm}d^{d+1}x\sqrt{-g}\,\bigg\{\left(\lbar{\J}_\m\G^
{\m\n\r}\nabla_\n\J_\r-\lbar \J_\m \overleftarrow \nabla_\n
\G^{\m\n\r}\J_\r-\cw\lbar\J_\m\G^{\m\n}\J_\n \right)\NO\\
& \hskip0.5in
+\left(i\cg_{IJ}\lbar\z^I\G^\m\left(\slashed\pa\vf^J-\cg^{JK}
\pa_K\cw\right)\J_\m-i \cg_{IJ}\lbar \J_\m (\slashed \pa \vf^I+\cg^{IK}\pa_K
\cw)\G^\m \z^J  \right)\NO\\
&\hskip0.5in
+\left(\cg_{IJ}\lbar\z^I\left(\d^J_K\slashed\nabla+\G^{J}_{KL}[\cg]
\slashed\pa\vf^L\right)\z^K-\cg_{IJ}\[ \lbar \z^I\slash \overleftarrow
\nabla\z^J +\lbar \z^K(\slashed \pa \vf^L)\G^J_{\;KL}\z^I  \]  \right)\NO\\
& \hskip0.5in +2\cm_{IJ}(\vf)\lbar\z^I\z^J+\text{quartic terms}\bigg\},
\eal
and the scalar potential and the mass matrix $\cm_{IJ}$ are expressed in terms
of the superpotential as
\bal\label{sup-eq}
\cv(\vf)=&\, \cg^{IJ}\pa_I{\cal W}(\vf)\pa_J{\cal W}(\vf)-\frac{d}{d-1}{\cal
W}(\vf)^2,\\
\cm_{IJ}(\vf)=&\, \pa_I\pa_J\cw-\G^{K}_{IJ}[\cg]\pa_K\cw-\frac12\cg_{IJ}\cw.
\eal
Here $\k^2$ is related to the gravitational constant by $\k^2=8\p G_{(d+1)}$. Note that in AlAdS spaces (with radius 1) which we are interested in the scalar potential and the superpotential are given by
\be 
\cv(\vf)=-d(d-1)+\co\(\vf^2\),\quad \cw(\vf)=-(d-1)+\co\(\vf^2\).
\ee
The action \eqref{SUGRA-action} is invariant under the supersymmetry transformation\footnote{In \cite{Bertolini:2015hua} the transformation rule of the gravitino field is given by $\d_\e\J_\m=(\nabla_\m+\frac16\cw\G_\m)\e$, which is obtained by setting $D=5$ explicitly in \eqref{susy-var-fermion}.}
\bsub
\label{susy-var-boson}
\bal
\d_\e\vf^I=&\,\frac i2 \bar\e\z^I+\text{h.c.}=\frac i2 \(\lbar
\e\z^I-\lbar\z^I\e\),\\
\d_\e E^\a_\m=&\,\frac12\bar \e\G^\a\J_\m+\text{h.c.}=\frac 12\(
\lbar\e\G^\a\J_\m-\lbar\J_\m\G^\a\e\),
\eal
\esub
where $\rm h.c.$ refers to hermitian conjugation, and
\bsub 
\label{susy-var-fermion}
\bal
& \d_\e\z^I=\, -\frac i2\left(\slashed\pa\vf^I-\cg^{IJ}\pa_J\cw\right)\e,\\
& \d_\e\J_\m=\,\left(\nabla_\m+\frac{1}{2(d-1)}\cw\G_\m\right)\e.
\eal
\esub
for any value of $d$.

Two comments are in order about the action \eqref{SUGRA-action}. Firstly, all the fermions here including the supersymmetry transformation parameter $\e$ are Dirac fermions. In fact, in $\cn=2$ 5 dimensional SUGRA, the gravitino field is expressed in terms of a symplectic Majorana spinor \cite{Freedman:2012zz}, which can also be described in terms of Dirac fermion \cite{Liu:2011dw}. Other fermions in the theory can also be expressed in the same way. Secondly, we would like to be as general as possible and thus, we keep $d$ generic in most of the following computations.

\section{Radial Hamiltonian dynamics}
\label{radial-Hamiltonian-Holography}
\setcounter{equation}{0}

According to the holographic dictionary \cite{Witten:1998qj} the on-shell action of the supergravity theory is the generating functional of the dual field theory. Therefore, the first step of the holographic computation is usually to consider the on-shell action on the bulk side. As is well-known, this on-shell action always suffers from the long-distance divergence which corresponds to the UV divergence of the dual field theory, and thus we need to renormalize the on-shell action of the supergravity theory, which is called as holographic renormalization \cite{Henningson:1998gx}. 

The Hamiltonian formulation is one of the powerful approach in holographic renormalization \cite{deBoer:1999tgo,Papadimitriou:2004ap}. The Hamiltonian constraint, one of the first class constraints obtained from the radial Hamiltonian, gives the Hamilton-Jacobi (HJ) equation by which we can obtain all the infinite counterterms for generic sources and curved background. The holographic renormalization is basically done once we find all the divergent counterterms and subtract them from the on-shell action for generic background and sources. Depending on the problem under consideration one can add some extra finite counterterms which actually correspond to the choice of scheme in the boundary field theory.

In this section we obtain the radial Hamiltonian, from which we extract the first class constraints.
Afterwards, we explain a general algorithm for obtaining the full counterterms from the HJ equation. We then present the flow equations which are needed to form a complete set of equations of motion.

\subsection{Radial Hamiltonian}

The Gibbons-Hawking term \cite{Gibbons:1976ue}
\be \label{GH}
\frac{1}{\k^2}\int_{\pa \cm} d^dx\sqrt{-\g}\;K,
\ee 
where $K$ is the extrinsic curvature on the boundary $\pa\cm$,
was introduced to have a well-defined variational problem for the Einstein-Hilbert action
\be 
S_{EH}=\frac{1}{2\k^2}\int_\cm  d^{d+1}x\sqrt{-g}\;R.
\ee 
As indicated in \cite{Henneaux:1998ch,Henningson:1998cd,Kalkkinen:2000uk,Arutyunov:1998ve}, by the same reason some additional boundary terms are needed when the theory involves the fermionic fields. It turns out that regarding the action \eqref{SUGRA-action} we have to add the boundary terms (for details, see appendix \ref{hyperino-kinetic-decomp} and \ref{gravitino-kinetic-decomp})
\bsub\label{GH-terms} 
\bal 
& \pm \frac{1}{2\k^2}\int_{\pa \cm} d^dx\sqrt{-\g}\;\lbar\J_i\Hat\G^{ij}\J_j,\label{GH-J}\\
& \pm \frac{1}{2\k^2}\int_{\pa\cm} d^dx\sqrt{-\g}\;\cg_{IJ}\lbar\z^I\z^J,\label{GH-Z}
\eal
\esub 
where the signs in front of the terms bilinear in fermionic fields fixes which radiality (see \eqref{radiality}) of the fermion should be used as a generalized coordinate. Note that, however, the sign depends on mass of the fields and choice of the boundary condition \cite{Arutyunov:1998ve}. Since mass of the gravitino $\J_\m$ in our case is $(d-1)/2>0$, sign of \eqref{GH-J} should be positive (see also appendix \ref{asymptotics-fermion} and \ref{gPBH}). Sign of the mass of $\z^I$ changes according to the model, and thus we can not choose sign of \eqref{GH-Z} \emph{a priori}. 

For the time being, however, let us pick the $+$ sign. As we will discuss the opposite case in section \ref{Neumann-BC}, picking minus sign corresponds to imposing Neumann boundary condition on spin-1/2 field $\z^I$. We emphasize that this choice of the sign will not affect our claim later about determination of the scalar fields' leading asymptotics.  The whole action including the terms \eqref{GH} and \eqref{GH-terms} is then given by
\be
S_{\rm full}=S+\frac{1}{2\k^2}\int_{\pa\cm} d^d{\rm x}\sqrt{-\g}\;\(2K+\lbar\J_i\Hat\G^{ij}\J_j+\cg_{IJ}\lbar\z^I\z^J\).
\ee
The full action $S_{\rm full}$ can be written as $S_{\rm full}=\int dr\;L$, where the radial Lagrangian $L$ is
\bala
L=\;& \frac{1}{2\k^2}\int_{\S_r} d^dx\;N\sqrt{-\g}\Bigg\{R[\g]-\cg_{IJ}\pa_i\vf^I\pa^i
\vf^J-\cv(\vf)+(\g^{ij}\g^{kl}-\g^{ik}\g^{jl})K_{ij}K_{kl}\\
&-\frac{\cg_{IJ}}{N^2}
(\dot \vf^I-N^i\pa_i\vf^I)(\dot \vf^J-N^j\pa_j\vf^J)
+\frac
2N\(\dot{\lbar\J}_{+i}\Hat\G^{ij}  \J_{-j}+\lbar\J_{-i} \Hat \G^{ij} \dot\J_{+j}
\)\\
&+\frac1N\dot e_a^i e_b^j\(\lbar\J_i\G^{ab}\J_j+\lbar\J_j\G^{ba}\J_i\) + \(K+\frac 1N D_k N^k \)\lbar \J_i\Hat\G^{ij}\J_j+\frac{1}{4N}e_{ak}\dot
e_b^{\;k}\lbar \J_i \G\{ \Hat \G^{ij},\G^{ab}\}\J_j\\
&
+\frac{1}{2N}K_{kl}\[\(\lbar\J_r-N^i\lbar\J_i\)[\Hat\G^{kj},\Hat\G^l]
\J_j-\lbar\J_j[\Hat\G^{kj},\Hat\G^l]\(\J_r-N^i\J_i\)  \] \\
&  +\frac{1}{4N}\lbar\J_i \(2\pa_k N[\Hat \G^{ij},\Hat \G^k]-(D_k N_l)\G\{\Hat
\G^{ij},\Hat\G^{kl} \} \)\J_j\\
& -\frac{N^i}{N}\(\lbar\J_j\G\Hat\G^{jk}\bb D_i\J_k-\lbar\J_j\overleftarrow{\bb
D}_i\G\Hat\G^{jk}\J_k \)-\lbar\J_i\Hat\G^{ijk}\bb
D_j\J_k+\lbar\J_i\overleftarrow{\bb D}_j\Hat\G^{ijk}\J_k \\
&-\frac 1N\lbar\J_k\overleftarrow{\bb D}_j\G\Hat\G^{jk}\(\J_r-N^i\J_i\)-\frac
1N\(\lbar\J_r-N^i\lbar\J_i\)\G\Hat\G^{jk}\bb D_j\J_k \\
&+\frac 1N\lbar\J_k\G\Hat\G^{jk}\(\bb D_j\J_r-N^i\bb
D_j\J_i\)+\frac1N\(\lbar\J_r\overleftarrow{\bb
D}_j-N^i\lbar\J_i\overleftarrow{\bb D}_j \)\G\Hat\G^{jk}\J_k \\
& +\frac
1N\cw\[
\(\lbar\J_r-N^i\lbar\J_i\)\G\Hat\G^j\J_j+\lbar\J_j\Hat\G^j\G\(\J_r-N^i\J_i\)
\]+\cw\lbar\J_i\Hat\G^{ij}\J_j\\
&+\frac{2
}{N}\cg_{IJ}\left(\lbar\z^I_+\dot\z^J_-+\dot{\lbar\z}^I_-\z^J_+\right)+\left(K+
\frac{1}{N}D_k N^k\right)\cg_{IJ}\lbar\z^I\z^J -\frac{1}{2N}\cg_{IJ}e_{ai}\dot e_b^{i}\lbar\z^I\G^{ab}\G\z^J\\
& +\frac{1}{N}\(
\dot\vf^K -N^i\pa_i \vf^K+N^i\pa_i\vf^K \)\pa_K\cg_{IJ}\lbar\z^I\z^J - \cg_{IJ}\(\lbar\z^I\Hat \G^i \bb D_i \z^J-\lbar\z^I \overleftarrow{\bb
D_i}\Hat \G^i \z^J \)\\
& -\frac{1}{N}\cg_{IJ} \[-\frac12 D_i N_j \(\lbar \z^I\Hat\G^{ij}\G\z^J
\)-N^i\lbar \z^I\G \bb D_i \z^J +N^i (\lbar\z^I\overleftarrow{\bb D}_i)\G\z^J
\]\\
& -\frac{i}{N}\cg_{IJ}\Bigg[ \frac 1N \(\dot \vf^J-N^j\pa_j \vf^J\)\Big[
\lbar\z^I \(\J_r-N^i\J_i+ N\Hat\G^i\G\J_i \)-\(\lbar \J_r-N^i \lbar\J_i
+N\lbar\J_i \G \Hat\G^i \)\z^I\Big]\\
& \hskip0.2in+\pa_i \vf^J\[\lbar\z^I\G\Hat\G^i\(\J_r-N^j \J_j
\)-\(\lbar\J_r-N^j\lbar\J_j \)\Hat\G^i\G\z^I \]+N\pa_i\vf^J
\(\lbar\z^I\Hat\G^j\Hat\G^i\J_j-\lbar\J_j\Hat\G^i\Hat\G^j\z^I \)\Bigg]\\
&+\frac{i}{N}\pa_I\cw\[\lbar\z^I\G\(\J_r-N^i\J_i \)+\(\lbar\J_r-N^i\lbar\J_i
\)\G\z^I +N\(\lbar\J_i\Hat\G^i\z^I+\lbar\z^I \Hat\G^i\J_i \) \]\\
& -\frac{1}{N}\pa_K \cg_{IJ}\[\(\dot\vf^J-N^i\pa_I\vf^J
\)\(\lbar\z^I\G\z^K-\lbar\z^K\G\z^I \)+N\pa_i\vf^J
\(\lbar\z^I\Hat\G^i\z^K-\lbar\z^K\Hat \G^i \z^I \) \]\\
& -2\cm_{IJ}\lbar\z^I\z^J\Bigg\}.\numberthis\label{lagrangian}
\eala

Given the radial Lagrangian $L$ we can derive the canonical momenta
\bsub\label{canonical-momenta}
\bal
\p_a^{\;i}=\; & \frac{\d L}{\d\dot e^a_{\;i}}=  \(\d^i_je_{ak}+\d^i_ke_{aj}
\)\frac{\sqrt{-\g}}{2\k^2}\Bigg[\(\g^{jk}\g^{lm} -\g^{jl}\g^{km}
\)K_{lm}+\frac12\g^{jk}\(\cg_{IJ}\lbar\z^I\z^J+\lbar\J_p\Hat\G^{pq}\J_q\)\NO\\
&\hskip0.5in-\frac{1}{4N}\(\lbar\J_p[\Hat\G^{jp},\Hat\G^k]\(\J_r-N^l\J_l\)-\(\lbar\J_r-N^l\lbar\J_l\)[\Hat\G^{jp},\Hat\G^k]\J_p\)\Bigg] \NO \\
&\hskip0.5in -\frac{\sqrt{-\g}}{2\k^2}\Bigg[e^{bi}\(\frac14\lbar\J_j\G\{\Hat\G^{jk},\G_{ab}\}\J_k-\frac12\cg_{IJ}\lbar\z^I\G_{ab}\G\z^J \)+e_{aj}\(\lbar\J^j\Hat\G^{ik}\J_k+\lbar\J_k\Hat\G^{ki}\J^j\)\Bigg],\label{pi-e}\\
\p^{\vf}_I=\;&\frac{\d
L}{\d\dot{\vf}^I}=\frac{\sqrt{-\g}}{2N\k^2}\Big[-2\cg_{IJ}\(\dot
\vf^J-N^i\pa_i\vf^J \)+N\pa_I\cg_{JK}\lbar\z^J\z^K-N\pa_K\cg_{IJ}\(\lbar\z^J\G\z^K-\lbar\z^K\G\z^J\)\NO\\
&\hskip0.5in-i\cg_{IJ}\(\lbar\z^J\(\J_r-N^i\J_i+N\Hat\G^i\G\J_i\)-\(\lbar\J_r-N^i\lbar\J_i+N\lbar\J_i\G\Hat\G^i\)\z^J\)\Big] ,\label{pi-vf}\\
\p^{\z}_I=\;&L\frac{\overleftarrow\d }{\d\dot{\z_-^I}}=\frac{\sqrt{-\g}}{\k^2}
\cg_{IJ}\lbar\z^J_+ ,\label{pi-bar-z}\\
\p^{\lbar \z}_I=\;&\frac{\overrightarrow\d }{\d\dot{\lbar\z}_-^I}L=\frac{\sqrt{-\g}}{\k^2}
\cg_{IJ}\z^J_+ ,\label{pi-z}\\
\p^{i}_\J=\;&L\frac{ \overleftarrow\d }{\d\dot{\J}_{+i}}=\frac{\sqrt{-\g}}{\k^2}
\lbar\J_{-j}\Hat\G^{ji},\label{pi-bar-psi}\\
\p^{i}_{\lbar \J}=\;&\frac{\overrightarrow\d }{\d\dot{\lbar\J}_{+i}}L=\frac{\sqrt{-\g}}{\k^2}
\Hat\G^{ij}\J_{-j}.\label{pi-psi}
\eal
\esub
One should keep in mind that $\p_{\lbar\J}^i$ and $\p_\J^i$ have minus radiality, and $\p_I^\z$ and $\p_I^\z$ have plus radiality.

From $K_{ij}=K_{ji}$, we obtain the constraint
\bala 
& 0=\cj_{ab}\equiv \frac{\k^2}{\sqrt{-\g}}(e_a^i\p_{bi}-e_b^i\p_{ai})-\frac14\lbar\J_j\G\{\Hat\G^{jk},\G_{ab}\}\J_k+\frac12\cg_{IJ}\lbar\z^I\G_{ab}\G\z^J\\
&\hskip1in -\frac12 e_a^i e_b^j(\lbar\J_i\Hat\G_{jk}\J^k+\lbar\J^k\Hat\G_{kj}\J_i-\lbar\J_j\Hat\G_{ik}\J^k-\lbar\J^k\Hat\G_{ki}\J_j) ,\numberthis\label{lorentz-constraint}  
\eala
which, as we will see, corresponds to the local Lorentz generator of the frame bundle on the slice $\S_r$ \cite{Kalkkinen:2000uk}.

Taking inverse of the canonical momenta\footnote{For instance, the inverse of the canonical momentum $\p_{\lbar\J}^i$ is $\J_{-i}=\frac{\k^2}{\sqrt{-\g}}\frac{1}{d-1}[\Hat\G_{ij}-(d-2)\g_{ij}]\p_{\lbar\J}^j$.} and implementing Legendre transformation we obtain the radial Hamiltonian
\bal
H=\;&\int d^d x\(\dot e^a_{\;i}\p_a^{\;i}+\dot \vf^I\p_I^\vf+\p_I^{\z}\dot \z_-^I+\dot{\lbar\z}_-^I\p_I^{\z}+\p_\J^{i}\dot\J_{+i}+\dot{\lbar\J}_{+i}\p_{\lbar \J}^{i} \)-L\NO\\
=\;&\int d^dx\[N\ch+N_i\ch^i+\(\lbar\J_r-N^i\lbar\J_i\)\cf+\lbar\cf\(\J_r-N^i\J_i\)\],
\eal
where
\bala
\ch=\;&\frac{\k^2}{2\sqrt{-\g}}\Bigg[\(\frac{1}{d-1}e_i^ae_j^b-e^a_je^b_i\)\p_a^i\p_b^j-\cg^{IJ}\p_I^\vf\p_J^\vf+\cg^{IJ}\(\p_I^\z\slashed{\bb D}\p_J^{\lbar\z}-\p_I^\z\overleftarrow{\slashed{\bb D}}\p_J^{\lbar\z}\) \\
&-\frac{1}{2(d-1)}\(e^{aj}\p_a^i+e^{ai}\p_a^j\)\Big[(d-1)(\lbar\J_{+i}\p_{\lbar \J j}+\p_{\J j}\J_{+i})+\p_\J^p\(\Hat\G_{pi}-(d-2)\g_{pi}\)\Hat\G_j{}^k\J_{+k}\\
&+\lbar\J_{+k}\Hat\G^k{}_j\(\Hat\G_{ip}-(d-2)\g_{ip}\)\p_{\lbar\J}^p\Big]
+\frac{1}{d-1}e^a_i\p_a^i\(-\lbar\z_-^I\p_I^{\lbar\z}-\p_I^\z\z_-^I+\lbar\J_{+j}\p_{\lbar\J}^j+\p_\J^j\J_{+j}\)\\
&+2\cg^{IJ}\G^L_{\;JK}[\cg]\p_I^\vf\(\lbar\z_-^K\p_L^{\lbar\z}+\p_L^\z\z_-^K\) +i\p_I^\vf\Big[\frac{1}{d-1}\(\lbar\z_-^I\Hat\G_i\p_{\lbar\J}^i+\p_\J^i\Hat\G_i\z_-^I\)\\
&-\cg^{IJ}\(\p_I^\z\Hat\G^i\J_{+i}+\lbar\J_{+i}\Hat\G^i\p_J^{\lbar\z}\)\Big]
-\p_{\J}^k\[\(\frac{1}{d-1}\Hat\G_k\Hat\G_j-\g_{kj}\)\slashed{\bb D}-\overleftarrow{\slashed{\bb D}}\(\frac{1}{d-1}\Hat\G_k\Hat\G_j-\g_{kj}\) \]\p_{\lbar\J}^j\\
&+\frac{i}{d-1}\(\p_I^\z\slashed\pa\vf^I\Hat\G_i\p_{\lbar\J}^i-\p_\J^i\Hat\G_i\slashed\pa\vf^I\p_I^{\lbar\z}\)-2i\pa_i\vf^I\(\p_I^\z\p_{\lbar\J}^i-\p_\J^i\p_I^{\lbar\z}\)\\
&+\cg^{IM}\cg^{KN}\pa_i\vf^J\(\pa_K\cg_{IJ}-\pa_I\cg_{KJ}\)\p_M^\z\Hat\G^i\p_N^{\lbar\z}\Bigg]\\
&-\frac12\cw\(\lbar\J_{+i}\p_{\lbar\J}^i+\p_\J^i\J_{+i}\)+\cm_{IJ}\(\cg^{IK}\p_K^\z\z_-^J+\cg^{JK}\lbar\z_-^I\p_K^{\lbar\z}\)\\
& -\frac i2\pa_I\cw\[\cg^{IJ}\(\lbar\J_{+i}\Hat\G^i\p_J^{\lbar\z}+\p_J^{\z}\Hat\G^i\J_{+i}\)+\frac{1}{d-1}\(\p_\J^i\Hat\G_i\z_-^I+\lbar\z_-^I\Hat\G_i\p_{\lbar\J}^i\) \]\\
&+\frac{\sqrt{-\g}}{2\k^2}\Bigg[-R[\g]+\cg_{IJ}\pa_i\vf^I\pa^i\vf^J+\cv(\vf)+\cg_{IJ}\lbar\z_-^I\(\slashed{\bb D}-\overleftarrow{\slashed{\bb D}}\)\z_-^J+\lbar\J_{+i}\Hat\G^{ijk}\(\bb D_j-\overleftarrow{\bb D}_j\)\J_{+k}\\
& +D_k\(\lbar\J_{+i}\(\g^{jk}\Hat\G^i-\g^{ik}\Hat\G^j\)\J_{+j} \)+i\cg_{IJ}\pa_i\vf^J\(\lbar\z_-^I\Hat\G^j\Hat\G^i\J_{+j}-\lbar\J_{+j}\Hat\G^i\Hat\G^j\z_-^I\)\\
&+\pa_K\cg_{IJ}\pa_i\vf^J\(\lbar\z_-^I\Hat\G^i\z_-^K-\lbar\z_-^K\Hat\G^i\z_-^I\)\Bigg],\numberthis\label{hamiltonian}
\eala
\bala
\ch^i=\;& -e^{ai}D_j\p_a^{\;j}+(\pa^i\vf^I)\p_I^\vf+ (\lbar\z_-^I\overleftarrow{\bb D}^i)\p_I^{\lbar \z}+\p_I^{\z}(\bb D^i\z_-^I)+\p_\J^{j}\(\bb D^i\J_{+j}\)+\(\lbar\J_{+j}\overleftarrow{\bb D}^i\)\p_{\lbar \J}^{j}\\
& -D_j(\p_\J^j\J_+^i+\lbar\J_+^i\p_{\lbar\J}^j),\numberthis\label{hi}
\eala
\bala
\cf=\;&\frac{2\k^2}{\sqrt{-\g}}\Bigg\{\frac{1}{4(d-1)}\Hat\G_i\p_{\lbar \J}^{i}e^a_j\p_a^j-\frac 18\G^a\g_{ik}\p_{\lbar \J}^{k}\p_a^i-\frac18 e^a_l\Hat\G_i\p_{\lbar \J}^{l}\p_a^i+\frac i4\cg^{IJ}\p_I^\vf\p_J^{\lbar \z}\Bigg\}\\
&+\frac14\G^a\J_{+i}\p_a^i+\frac14\Hat\G_i\J_{+j}e^{aj}\p_a^i+\frac i2\p_I^\vf\z_-^I-\bb D_i\p_{\lbar \J}^{i}-\frac{1}{2(d-1)}\cw\Hat\G_i\p_{\lbar \J}^{i}-\frac i2\pa_i\vf^I\Hat\G^i\p_I^{\lbar \z}\\
&-\frac i2\cg^{IJ}\pa_I\cw\p_J^{\lbar \z}+\frac{\sqrt{-\g}}{2\k^2}\( 2\Hat\G^{ij}\bb D_i\J_{+j}+\cw\Hat\G^i \J_{+i} +i\cg_{IJ}\pa_i\vf^J\Hat\G^i\z_-^I+i(\pa_I\cw)\z_-^I\).\numberthis\label{cf}
\eala
We note that in the above computations we used the local Lorentz constraint \eqref{lorentz-constraint}.

By radiality we split $\cf$ into two parts
\bal
\cf_+\equiv\G_+\cf=\;&\frac{\k^2}{2\sqrt{-\g}}\Bigg[\p_a^j e^{ak}\(\frac{1}{d-1}\g_{jk}\Hat\G_i-\frac12\g_{ij}\Hat\G_k-\frac12\g_{ik}\Hat\G_j\)\p_{\lbar\J}^i +i\cg^{IJ}\p_I^\vf\p_J^{\lbar \z}\Bigg]\NO\\
&-\frac {1}{2(d-1)}\cw\Hat\G_i\p_{\lbar \J}^{i} -\frac i2\cg^{IJ}\pa_I\cw \p_J^{\lbar \z}+ \frac {\sqrt{-\g}}{2\k^2}\(2\Hat\G^{ij}\bb D_i\J_{+j}+ i\cg_{IJ}\pa_i\vf^J \Hat\G^i\z_-^I \),\label{cf-pos}
\eal
and
\bal
\cf_-\equiv\G_-\cf=\;&\frac14\(\Hat\G_i\J_{+j}+\Hat\G_j\J_{+i}\)e^{aj}\p_a^i+\frac i2\p_I^\vf\z_-^I\NO\\
& -\bb D_i\p_{\lbar \J}^{i}-\frac i2\pa_i\vf^I\Hat\G^i\p_I^{\lbar \z}+\frac{\sqrt{-\g}} {2\k^2}\(\cw\Hat\G^i\J_{+i}+i\pa_I\cw\z_-^I\).\label{cf-neg}
\eal

The canonical momenta for $N$, $N_i$ and $\J_r$ vanish identically, and it then follows from the Hamilton's equation that
\be\label{first-class-constraint}
\ch=\ch^i=\cf_-=\cf_+=0.
\ee
These first class constraints reflect respectively radial reparameterization and diffeomorphism and supersymmetry and super-Weyl invariance along the radial cut-off $\S_r$, which can be seen by comparing with \eqref{superdiffeo-induced}.

Inserting \eqref{canonical-momenta} in \eqref{lorentz-constraint}, we obtain
\be\label{local-Lorentz} 
0= e_a^i\p_{bi}-e_b^i\p_{ai}+\frac12\lbar\z_-^I\G_{ab}\p_I^{\lbar\z}-\frac12\p_I^\z\G_{ab}\z_-^I-\frac12\p_\J^i\G_{ab}\J_{+i}+\frac12\lbar\J_{+i}\G_{ab}\p_{\lbar\J}^i, 
\ee
which reflects the local frame rotation symmetry of the theory according to \eqref{superdiffeo-induced}. We emphasize that on the bosonic level this local Lorentz constraint is reduced into
\be 
e^i_a\p_{bi}=e^i_b\p_{ai}, 
\ee 
which implies that we can define symmetric canonical momenta for the metric such as
\be \label{pij-pai}
\frac{\d}{\d\dot \g_{ij}}L^B\equiv \p^{ij}=\frac12 e^{aj}\frac{\d}{\d\dot  e^a_i}L^B.
\ee 
Here $L^B$ denotes bosonic part of the radial Lagrangian \eqref{lagrangian}.

We emphasize that linearity of the constraints $\ch^i=\cf_-=0$ and local Lorentz constraint reflect the fact that their corresponding symmetries are not broken by the cut-off. Meanwhile, $\ch=0$ and $\cf_+=0$ constraints look much different from their corresponding symmetries, due to the quadratic terms, implying that in fact the cut-off breaks these symmetries, though they are non-linearly realized in the bulk.

\subsection{Hamilton-Jacobi equations and the holographic renormalization}\label{counterterms-HJ-one-point}

The HJ equations are obtained by plugging
\be\label{pi-cs}
\p_a^{\;i}=\frac{\d }{\d e^a_{\;i}}\bb S ,\quad \p_I^\vf=\frac{\d}{\d \vf^I}\bb S ,\quad \p_I^{\z}=\bb S \frac{\overleftarrow\d}{\d \z_-^I},\quad \p_I^{\lbar \z}=\frac{\overrightarrow\d}{\d \lbar\z_-^I}\bb S,\quad \p_\J^{i}=\bb S\frac{\overleftarrow\d}{\d \J_{+i}},\quad \p_{\lbar \J}^{i}=\frac{\overrightarrow\d}{\d \lbar\J_{+i}}\bb S
\ee
into the first class constraints \eqref{first-class-constraint}. Here $\bb S[e,\vf,\z_-,\J_+]$ is the Hamilton's principal functional. 

The Hamilton's principal functional $\bb S$ is particularly important since it can be identified as the on-shell action evaluated on the radial slice $\S_r$. For the sake of renormalization of the on-shell action we have only to solve these HJ equations for $\bb S$ up to the finite terms without relying on the specific solution of the equations of motion. Because this asymptotic solution of the HJ equations are obtained for the generic sources we can identify them as the counterterms for cancellation of the divergence of the on-shell action as well as all the correlation functions.

As pointed out in \cite{Lindgren:2015lia}, the constraint $\ch^i=0$ and the local Lorentz constraint \eqref{local-Lorentz} which reflects the bulk diffeomorphism invariance along the transverse direction is automatically satisfied as long as we look for a local and covariant solution. Hence the equations which we have to solve are the constraints $\ch=\cf_-=\cf_+=0$.

Let us briefly review the algorithm of solving the HJ equation in AlAdS geometry. In general, the Hamiltonian constraint is solved asymptotically by using the formal expansion of $\bb S$ with respect to the dilatation operator $\d_D$ \cite{Papadimitriou:2004ap} (see section 5.2 of \cite{Papadimitriou:2016yit} for a recent review)
\be \label{delta-D}
\d_D=\int d^dx\;\sum_\F (\D_\F-d)\frac{\d}{\d\F} ,
\ee 
where $\F$ refers to every field in the theory and $\D_\F$ denotes to the scaling dimension of the operator dual to $\F$. The solution takes form of
\be 
\bb S=\int_{\S_r}d^dx\sqrt{-\g}\;\cl=\int_{\S_r}d^dx\sqrt{-\g}\;\Big(\cl_{[0]}+\cl_{[1]}+\cdots +\wt \cl_{[d]}\log e^{-2r}+\cl_{[d]}+\cdots\Big),
\ee 
where
\be 
\d_D \cl_{[n]}=-n\cl_{[n]},\;0\leq n<d,\;\d_D\wt \cl_{[d]}=-d\wt\cl_{[d]}.
\ee 
Since the dilatation operator $\d_D$ is asymptotically identical to the radial derivative
\be 
\d_r=\int_{\S_r}d^dx\;\sum_\F \dot\F\frac{\d}{\d\F} 
\ee 
in AlAdS, one can see that $\cl_{[n]} $ for $n<d$ and $\wt\cl_{[d]}$ are asymptotically divergent, which we can identify as counterterms, namely
\be 
\bb S_{\rm ct}=-\int_{\S_r}d^dx\sqrt{-\g}\;\Big(\cl_{[0]}+\cl_{[1]}+\cdots +\wt \cl_{[d]}\log e^{-2r}\Big).
\ee 
By construction, this is collection of all possible divergent terms.

This general argument of finding $\bb S_{\rm ct}$ is not suitable in our case, since the operator $\d_D$ requires knowledge of scaling dimensions in the theory a priori. To avoid this disadvantage, here we employ a universal operator
\be \label{delta-e}
\d_{e}=\int d^dx\;\(e^a_i\frac{\d}{\d e^a_i}+\frac12 \lbar\J_{+i}\frac{\d}{\lbar\J_{+i}}+\frac12 \frac{\overleftarrow\d}{\d\J_{+i}}\J_{+i}\),
\ee 
rather than $\d_D$ \cite{Papadimitriou:2011qb,Lindgren:2015lia}, since we know that the scaling dimension of the operators dual to $e^a_i$ and $\J_{+i}$ in AlAdS are $d+1$ and $d+1/2$ respectively, see appendix \ref{asymptotics-fermion}. Note that $\d_e$ basically counts the number of vielbein and gravitino. The formal expansion of the Hamilton's principal function $\bb S[e,\vf,\z_-,\J_+]$ with respect to $\d_e$ is thus
\be\label{expand-e}
\bb S= \bb S_{(0)}+\bb S_{(1)}+\bb S_{(2)}+\cdots,\quad \bb S_{(k)}\equiv \int d^dx\;\bb L_{(k)},
\ee
where $\d_{e}\bb S_{(k)}=(d-k)\bb S_{(k)}$. This implies that
\be 
\p^i_{(k)a} e^a_i+\frac12 \p^i_{(k)\J}\J_{+i}+\frac12 \lbar\J_{+i}\p_{(k)\lbar\J}^i =(d-k)\bb L_{(k)}+\pa_iv_{(k)}^i,
\ee 
for certain $v_{(k)}^i$. However, the Lagrangian $\bb L_{(k)}$ is defined up to a total derivative, and thus we can put \cite{Papadimitriou:2011qb}
\be \label{L-k}
\p^i_{(k)a} e^a_i+\frac12 \p^i_{(k)\J}\J_{+i}+\frac12 \lbar\J_{+i}\p_{(k)\lbar\J}^i :=(d-k)\bb L_{(k)}.
\ee 
As we see later, this identification of $\bb L_{(k)}$ greatly simplifies the HJ equation and makes it almost algebraic.

By using \eqref{L-k} we can solve the HJ equation recursively, but this procedure stops at $\bb S_{(d)}$ due to poles. The reason why higher-order terms, which are finite in $r\rightarrow\infty$ limit, cannot be determined in this recursive procedure is that they are related to the arbitrary integration constants which form a complete integral together with the integration constants from the flow equations.

Since the scaling dimension of the other operators is less than $d$ under the general assumption that there is neither exactly marginal nor irrelevant deformation, all the divergent terms appear up to $\bb S_{(d)}$ so that we can identify the counterterms as
\be 
S_{\rm ct}=-\sum_{k=0}^{2d}\bb S_{(k/2)},
\ee 
if we consider the theory only up to quadratic terms in fermions. Note that the logarithmically divergent terms are distributed in almost all of $\bb S_{(k)}$s. Since our radial slice is 4 dimensional, these terms appear with the pole $1/(d-4)$. Converting this pole by (dimensional regularization) \cite{Papadimitriou:2004ap,Papadimitriou:2011qb}
\be 
\frac{1}{d-4}\rightarrow -\frac12\log e^{-2r}
\ee 
and summing up all of them, we obtain the logarithmically divergent terms $\wt \cl_{[d]}$. We emphasize that the two algorithms we described in fact give the same result for $S_{\rm ct}$.

Once the local counterterms $S_{\rm ct}$ are obtained, we renormalize the on-shell action by

\be \label{ren-onshell-action}
\Hat S_{ren}=\lim_{r\rightarrow+\infty}(S_{\rm full}+S_{\rm ct})=\lim_{r\rightarrow+\infty}\int_{\S_r}d^dx\;\cl_{[d]}.
\ee
The canonical momenta are automatically renormalized by $\bb S_{\rm ct}$, namely
\be 
\Hat \p^\F\equiv \p^\F+\frac{\d}{\d\F}S_{\rm ct}, \text{ for every field }\F,
\ee 
and variation of the renormalized on-shell action under any symmetry transformation is given by the chain rule
\be 
\d \Hat S_{ren}=\lim_{r\rightarrow+\infty}\int d^dx\(\Hat \p_a^i\d e^a_i+\Hat \p^\vf_I\d\vf^I+\d\lbar \z_-^I\Hat \p_I^\z+\Hat{\p}_I^\z\d\z_-^I+\d\lbar \J_{+i}\Hat \p_\J^i+\Hat{\p}_\J^i\d\J_{+i} \).
\ee

\subsection{Flow equations and leading asymptotics}

After Solving the HJ equations we insert \eqref{pi-cs} into \eqref{canonical-momenta}s to get the flow equations for the sources. Whereas for the bosonic sector it works, for the fermionic sector this procedure does not since \eqref{pi-bar-z}, \eqref{pi-z}, \eqref{pi-bar-psi}, \eqref{pi-psi} just play a role of field-redefinition and are just reminiscence of the second-class constraints in the Lagrangian \eqref{lagrangian}, Notice that the second-class constraints are completely eliminated from the radial Hamiltonian $H$. Therefore from now on we regard that the theory is originally defined in the Hamiltonian formalism and the radial Hamiltonian $H$ is more fundamental object than the radial Lagrangian $L$, even though both the Hamiltonian and Lagrangian formalism basically give the same equations of motion.

The flow equations are then obtained by substituting \eqref{pi-cs} into the Hamilton's equations of motion
\bsub \label{Hamilton-eom}
\bal
& \dot e^a_i=\frac{\d H}{\d \p_a^i},\quad \dot\p_a^i=-\frac{\d H}{\d e^a_i},\\
& \dot\vf^I=\frac{\d H}{\d\p_I^\vf},\quad \dot\p_I^\vf=-\frac{\d H}{\d \vf^I},\\
&\dot\z_-^I=\frac{\d }{\d\p^\z_I}H,\quad \dot{\p}_I^\z=-H\frac{ \d }{\d\z_-} ,\quad\dot{\lbar \z}_-^I=H\frac{\d }{\d \p^{\lbar\z}_I},\quad \dot{\p}_I^{\lbar\z}=-\frac{\d }{\d\lbar\z_-^I}H ,  \\
& \dot\J_{+i}=\frac{\d }{\d\p_\J^i}H,\quad \dot{\p}_\J^i=-H\frac{\d}{\d\J_{+i}} ,\quad \dot{\lbar\J}_{+i}=H\frac{\d }{\d \p_{\lbar\J}^i},\quad \dot{\p}_{\lbar\J}^i=-\frac{\d}{\d\lbar \J_{+i}}H,
\eal
\esub 
namely
\bala 
\dot e^a_i
=\;&\frac{\k^2}{2\sqrt{-\g}}\Bigg\{2\(\frac{1}{d-1}e^a_i e^b_j-e^a_j e^b_i\)\p^j_b -\frac{1}{2(d-1)}e^{aj}\Big[(d-1)(\lbar\J_{+i}\p_{\lbar\J j}+\p_{\J j}\J_{+i})\\
&-\p_\J^p[\Hat\G_{pi}-(d-2)\g_{pi}]\Hat\G_j{}^{k}\J_{+k}+\lbar\J_{+k}\Hat\G^{k}{}_j[\Hat\G_{ip}-(d-2)\g_{ip}]\p_{\lbar\J}^p+(i\leftrightarrow j)\Big]\\
&+\frac{1}{d-1}e^a_i\(-\lbar\z_-^I\p_I^{\lbar\z}-\p_I^\z\z_-^I+\lbar\J_{+j}\p_{\lbar\J}^j+\p_\J^j\J_{+j}\)\Bigg\},\numberthis \label{flow-e}
\eala 

\bala
\dot\vf^I
=\;&\frac{\k^2}{\sqrt{-\g}}\cg^{IJ}\Bigg[-\p_J^\vf +\G^K_{JL}[\cg]\(\p_K^\z\z_-^L+\lbar\z_-^L\p_K^{\lbar\z}\)-\frac i2\(\p_J^\z \Hat\G^i\J_{+i}+\lbar\J_{+i}\Hat\G^i\p_J^{\lbar\z}\)\Bigg]\\
&+\frac{\k^2}{\sqrt{-\g}}\frac{i}{2(d-1)}\(\lbar\z_-^I\Hat\G_i\p_{\lbar\J}^i+\p_\J^i\Hat\G_i\z_-^I\),\numberthis \label{flow-vf}
\eala 

\bala 
\dot\J_{+i}=\;&\frac{\k^2}{2\sqrt{-\g}}\Bigg[-\frac12\(\d_i^ke^{aj}+\g^{jk}e^a_{i}\)\p_a^k\J_{+j}+\frac{1}{d-1}e^a_j\p_a^j\J_{+i}+i\p_I^\vf \Hat\G_i\z_-^I\\
&-\frac{1}{2(d-1)}\(e^{aj}\p_a^l+e^{al}\p_a^j\)\(\Hat\G_{il}-(d-2)\g_{il}\)\Hat\G_j{}^k\J_{+k}-\frac{i}{d-1}\Hat\G_i\slashed\pa\vf^I\p_J^{\lbar\z}\\
&-\frac{2}{d-1}\(\Hat\G_{ijk}-(d-2)\g_{ij}\Hat\G_k\)\bb D^k\p_{\lbar\J}^j+2i\pa_i\vf^I\p_I^{\lbar\z}\Big]-\frac12\cw\J_{+i}\\
&-\frac{i}{2(d-1)}\pa_I\cw\Hat\G_i\z_-^I.\numberthis \label{flow-J}
\eala 
and
\bala
\dot\z_-^I=\;& \frac{\k^2}{2\sqrt{-\g}}\Bigg[2\cg^{IJ}\slashed{\bb D}\p_J^{\lbar\z}+\pa_i\cg^{IJ}\Hat\G^i\p_J^{\lbar\z}-\frac{1}{d-1}e^a_i\p_a^i\z_-^I+2\cg^{LJ}\G^I_{JK}[\cg]\p_L^\vf\z_-^K\\
&-i\cg^{IJ}\p_\vf^J\Hat\G^i\J_{+i}-2i\pa_i\vf^I\p_{\lbar\J}^i+\cg^{IM}\cg^{KN}\pa_i\vf^J(\pa_K\cg_{JM}-\pa_M\cg_{KJ})\Hat\G^i\p_N^{\lbar\z}\Bigg]\\
&+\cm_{JK}\cg^{IK}\z_-^J-\frac i2\pa^I\cw\Hat\G^i\J_{+i}.\numberthis \label{flow-z}
\eala
Here for simplicity we choose the gauge \eqref{eFG-gauge}, which makes the radial Hamiltonian $H$ reduced into $H=\int d^dx\;\ch$. We emphasize that these flow \eqref{flow-e}, \eqref{flow-vf}, \eqref{flow-J} and \eqref{flow-z} together with the HJ equations form a complete set of equations of motion of the theory.\footnote{One can use the flow equations \eqref{flow-J} and \eqref{flow-z} to determine the asymptotic behavior of $\J_{+i}$ and $\z_-^I$ in appendix \ref{asymptotics-fermion} instead of using the Euler-Lagrange equations of motion \eqref{psi-eom} and \eqref{zeta-eom}.} 

\section{Solution of the Hamilton-Jacobi equation}
\label{generic-HJ}
\setcounter{equation}{0}

To solve the HJ equation efficiently we divide the Hamilton's principal function into several parts according to the structure of terms. Namely, we first split $\bb S$ into two sectors: $\bb S^B$, purely bosonic part and $\bb S^F$, quadratic in fermions. The terms in $\bb S^F$ are further split into 3 parts: $\bb S^{\z\z}$ quadratic terms in $\z_-^I$s,  $\bb S^{\J\J}$ quadratic terms in $\J_{+i}$ and $\bb S^{\z\J}$ bilinear in $\z_-^I$ and $\J_{+i}$. In total,
\be 
\bb S=\bb S^B+\bb S^{\z\z}+\bb S^{\J\J}+\bb S^{\z\J}.
\ee 
Due to radiality and Lorentz structure of the fermionic sources, the asymptotic expansion of $\bb S^B$, $\bb S^{\z\J}$, $\bb S^{\z\z}$ and $\bb S^{\J\J}$ should be
\bsub
\bal 
& \bb S^B=\bb S_{(0)}^B+\bb S_{(2)}^B+\bb S_{(4)}^B+\cdots ,\\
& \bb S^{\z\J}=\bb S_{(3/2)}^{\z\J}+\bb S_{(7/2)}^{\z\J}+\cdots,\\
& \bb S^{\z\z}=\bb S_{(1)}^{\z\z}+\bb S_{(3)}^{\z\z}+\bb S_{(5)}^{\z\z}+\cdots,\\
& \bb S^{\J\J}=\bb S_{(2)}^{\J\J}+\bb S_{(4)}^{\J\J}+\cdots.
\eal 
\esub

How to solve the HJ equation for the bosonic sector has been discussed in many literature regarding some special models \cite{Papadimitriou:2004ap,Papadimitriou:2011qb,Papadimitriou:2004rz}, though it is difficult to solve the HJ equation for the general model.\footnote{One might try to solve the HJ equation for the general scalar-gravity model by using the argument in \cite{Papadimitriou:2011qb}.} The key feature is that after finding solution of the HJ equation to leading order, we only need to solve (almost algebraic) first-order differential equation from the next order, thanks to the relation \eqref{L-k}. Nevertheless, these first-order differential equations are not easy to solve at the first attempt.

Here we have another set of first-order differential equations, namely $\cf_-=\cf_+=0$. These are relatively simpler than the Hamiltonian constraint $\ch=0$, so one can try to solve these constraints first. Not surprisingly, it works well, in particular for the fermionic sector, and the solution is totally consistent with the other constraints, as we will see soon.

\subsection{Bosonic sector}
\label{bosonic sector}

Let us first consider the bosonic sector. The corresponding Hamiltonian constraint $\ch=0$ is
\bal\label{HJ-boson}
& \frac{\k^2}{2\sqrt{-\g}}\[4\(\frac{1}{d-1}\g_{ij}\g_{kl}-\g_{ik}\g_{jl}\)\frac{\d \bb S^B}{\d \g_{ij}}\frac{\d \bb S^B}{\d \g_{kl}}- \cg^{IJ}\frac{\d \bb S^B}{\d \vf^I}\frac{\d \bb S^B}{\d \vf^J}\]\NO\\
& \hskip2in +\frac{\sqrt{-\g}}{2\k^2}\(-R[\g]+\cg_{IJ}\pa_i\vf^I\pa^i\vf^J+\cv(\vf)\)=0.
\eal
One can readily see that the HJ equation for $\bb S_{(0)}$ is
\be\label{HJ-0}
\frac{\k^2}{2\sqrt{-\g}}\[4\(\frac{1}{d-1}\g_{ij}\g_{kl}-\g_{ik}\g_{jl}\)\frac{\d \bb S_{(0)}}{\d \g_{ij}}\frac{\d \bb S_{(0)}}{\d \g_{kl}}- \cg^{IJ}\frac{\d \bb S_{(0)}}{\d \vf^I}\frac{\d \bb S_{(0)}}{\d \vf^J}\]+\frac{\sqrt{-\g}}{2\k^2}\cv(\vf)=0.
\ee

The leading term of $\bb S$, $\bb S_{(0)}$ should not contain any derivatives and must be purely bosonic so that its ansatz becomes
\be \label{S0-ansatz}
\bb S_{(0)}=-\frac{1}{\k^2}\int d^dx\sqrt{-\g}\;U(\vf).
\ee
Substituting this ansatz into the constraint $\cf_-=0$, we obtain
\be 
\frac14\(\Hat\G_i\J_{+j}+\Hat\G_j\J_{+i}\)e^{aj}\frac{\d\bb S_{(0)}}{\d e^a_i}+\frac{\sqrt{-\g}}{2\k^2}\cw\Hat\G_i\J_{+i}=0,
\ee 
and find the unique solution for $U(\vf)$ given by $U=\cw(\vf)$, or
\be \label{cs-0}
\bb S_{(0)}=-\frac{1}{\k^2}\int d^dx\sqrt{-\g}\;\cw.
\ee 
As promised, we obtain \eqref{cs-0} regardless of the sign of \eqref{GH-Z}. It follows that leading asymptotics of the scalar field $\vf^I$ is also determined whatever the sign of \eqref{GH-Z} was chosen, as we see in \eqref{leading-flow-vf}. From \eqref{cs-0} we can now determine the leading asymptotics of the fields by using the above flow equations, namely
\bsub 
\bal
& e^a_i(r,x)\sim e^r e^a_{(0)i}(x),\\
& \J_{+i}(r,x)\sim e^{r/2}\J_{(0)+i}(x) ,\\
& \dot \vf^I\sim \cg^{IJ}\pa_J\cw,\quad \text{or }\vf^I\sim e^{-\m^Ir}\vf^I_{(0)} \label{leading-flow-vf} ,\\
& \dot \z_-^I\sim-\frac12\z_-^I+(\cg^{IK}\pa_J\pa_K\cw)\z_-^J,\quad \text{or }\z_-^I\sim e^{-(\m^I+\frac12)r}\z^I_{-(0)} \label{leading-flow-z},
\eal
\esub
where $\m^I$ stands for radial weight of $\vf^I$ when the scalars are properly diagonalized.

Now let us go to the next level of the bosonic sector. The HJ equation for $\bb S^B_{(2)}$ is then
\be \label{HJ-2-B}
-\frac{2}{d-1}\cw\g_{ij}\frac{\d}{\d\g_{ij}}\bb S_{(2)}^B+\cg^{IJ}\pa_I\cw\frac{\d}{\d \vf^J}\bb S_{(2)}^B+\frac{\sqrt{-\g}}{2\k^2}\(-R[\g]+\cg_{IJ}\pa_i\vf^I\pa^i\vf^J\)=0.
\ee
The most general ansatz for $\bb S_{(2)}^B$ is as follows:
\be
\bb S_{(2)}^B=\frac{1}{\k^2}\int d^dx\sqrt{-\g}\(\X(\vf) R+A_{IJ}(\vf)\pa_i\vf^I\pa^i\vf^J\).
\ee
Then,
\bal 
\g_{ij}\frac{\d}{\d\g_{ij}}\bb S_{(2)}^B=\; & \frac{\sqrt{-\g}}{\k^2}\frac{d-2}{2}\(\X R+A_{IJ}\pa_i\vf^I\pa^i\vf^J\)-\frac{\sqrt{-\g}}{\k^2}(d-1)\Box\X,\label{trace-pij(2)}\\
\frac{\d}{\d\vf^J}\bb S_{(2)}^B=\; & \frac{\sqrt{-\g}}{\k^2}\(R\pa_J\X+\pa_J A_{IK}\pa_i\vf^I\pa^i\vf^K-2D_i\(A_{JK}\pa^i\vf^K\)\),
\eal
where we used the relation
\be 
\g^{ij}\d R_{ij}=D^i D^j\d\g_{ij}-\g^{ij}\Box\(\d\g_{ij}\).
\ee
One can notice from \eqref{trace-pij(2)} that
\be 
\bb L^B_{(2)}=\frac{\sqrt{-\g}}{\k^2}\(\X R+A_{IJ}\pa_i\vf^I\pa^i\vf^J-\frac{2(d-1)}{d-2}\Box\X \).
\ee 
Therefore, \eqref{HJ-2-B} becomes
\bal
0=\;& R\(-\frac{d-2}{d-1}\cw\X^{[1]}+\cg^{IJ}\pa_I\cw\pa_J\X-\frac12\)+\pa_i\vf^I\pa^i\vf^J\Big(-\frac{d-2}{d-1}\cw A_{IJ}+2\cw\pa_I\pa_J\X\NO\\
& +\cg^{KL}\pa_L\cw\pa_K A_{IJ}-2\cg^{KL}\pa_K\cw\pa_I A_{LJ}+\frac12\cg_{IJ}\Big)+2\Box\vf^I\(\cw\pa_I\X-\cg^{JK}\pa_J\cw A_{IK}\),
\eal
and we obtain the equations for $\X$ and $A_{IJ}$
\bsub\label{eqn-cs-2B}
\bal 
0=\;& -\frac{d-2}{d-1}\X+V^I\pa_I\X-\frac{1}{2\cw},\label{eqn-X}\\
0=\;&-\frac{d-2}{d-1}A_{IJ}+V^K \pa_K A_{IJ}+\pa_I V^K A_{JK}+\pa_J V^K A_{IK}+\frac{1}{2\cw}\cg_{IJ},\label{eqn-A}\\
0 =\;& \pa_I\X-V^J A_{IJ},\label{eqn-X-A}
\eal 
\esub 
where
\be 
V^I\equiv \frac{1}{\cw}\cg^{IJ}\pa_J\cw .
\ee 
Note that $A_{IJ}$ should satisfy the condition
\be 
\pa_I(V^K A_{JK})=\pa_J(V^K A_{IK}).
\ee 

We emphasize that we do not discuss existence of the solution for $A_{IJ}$ and $\X$ here. Nevertheless, the equations \eqref{eqn-cs-2B} are useful for determination of $\bb S^{\z\z}_{(1)}$, $\bb S^{\J\J}_{(2)}$ and $\bb S^{\z\J}_{(3/2)}$.

$\bb S^B_{(2n)}$ ($n\geq 2$) is obtained by the following recursive equation
\bal 
0=\;&-\frac{2}{d-1}\cw\g^{ij}\p_{(2n)ij}^{B}+\cw V^I\p_{(2n)I}^{B}\NO\\
&+\frac{\k^2}{2\sqrt{-\g}}\sum_{m=1}^{n-1}\[ 4\(\frac{1}{d-1}\g_{ij}\g_{kl}-\g_{ik}\g_{jl}\)\p^{ij}_{B(2m)}\p^{kl}_{B(2n-2m)} -\cg^{IJ}\p_I^{B(2m)}\p_J^{B(2n-2m)}\].\label{H-bos-recursive}
\eal 
In particular, when $d=4$ the inhomogeneous terms on the RHS become
\be
2\frac{\k^2}{\sqrt{-\g}}\(\frac{1}{d-1}\g_{ij}\g_{kl}-\g_{ik}\g_{jl}\)\p^{ij}_{(2)}\p^{kl}_{(2)}=\frac{\sqrt{-\g}}{\k^2}\X^2\(\frac{d}{2(d-1)}R^2-2R_{kl}R^{kl}\), \label{H-bos-inhomo-4}
\ee
where 
\be 
\X= \frac{1}{2(d-2)}+\co(\vf^2)
\ee 
is the solution of \eqref{eqn-X}, while other inhomogeneous terms are asymptotically suppressed.

\subsection{Fermionic sector}
\label{gen-fermion}

After substituting the leading order solution \eqref{cs-0} into the Hamiltonian constraint \eqref{hamiltonian}, we get the following first-order differential equation for $\wt{\bb S}\equiv \bb S-\bb S_{(0)}$
\bala 
0=\;&\cw\(-\frac{1}{d-1}e^a_i\wt\p_a^i+V^I\wt\p_I^\vf\)-\frac{1}{2(d-1)}\cw\(\lbar\J_{+i}\p_\J^i+\p_\J^i\J_{+i}\)+\cw\(\frac{1}{2(d-1)}\d_I^J+\pa_I V^J\)\times\\
&\times \(\lbar\z_-^I\p_J^\z+\p_J^\z\z_-^I\)+\frac{\k^2}{2\sqrt{-\g}}\Bigg\{\(\frac{1}{d-1}e_i^ae_j^b-e^a_je^b_i\)\wt \p_a^i\wt \p_b^j-\cg^{IJ}\wt \p_I^\vf\wt \p_J^\vf+\cg^{IJ}\(\p_I^\z\slashed{\bb D}\p_J^\z-\p_I^\z\overleftarrow{\slashed{\bb D}}\p_J^\z\) \\
&-2\wt\p^{ij}\Big[(\lbar\J_{+i}\p_{\J j}+\p_{\J j}\J_{+i})+\frac{1}{d-1}\p_\J^p\(\Hat\G_{pi}-(d-2)\g_{pi}\)\Hat\G_j{}^k\J_{+k}\\
&+\frac{1}{d-1}\lbar\J_{+k}\Hat\G^k{}_j\(\Hat\G_{ip}-(d-2)\g_{ip}\)\p_\J^p-i\frac{\pa_I\cw}{\cw}\g_{jk}\(\lbar\z_-^I\Hat\G_i\p_\J^k+\p_\J^k\Hat\G_i\z_-^I\)\Big]\\
& 
+\frac{2}{d-1}\g_{ij}\wt\p^{ij}\(-\lbar\z_-^I\p_I^\z-\p_I^\z\z_-^I+\lbar\J_{+k}\p_\J^k+\p_\J^k\J_{+k}-i\frac{\pa_I\cw}{\cw}\lbar\z_-^I\Hat\G_k\p_\J^k-i\frac{\pa_I\cw}{\cw}\p_\J^k\Hat\G_k\z_-^I\)\\
&+\[\cg^{IJ}\cg^{LM}(\pa_J\cg_{MK}-\pa_M\cg_{JK})-\cw\pa_K\(\frac{\cg^{IL}}{\cw}\) \]\wt \p_I^\vf\(\lbar\z_-^K\p_L^\z+\p_L^\z\z_-^K\) +i\wt \p_I^\vf\Big[\frac{1}{d-1}\(\lbar\z_-^I\Hat\G_i\p_\J^i+\p_\J^i\Hat\G_i\z_-^I\)\\
&-\cg^{IJ}\(\p_J^\z\Hat\G^i\J_{+i}+\lbar\J_{+i}\Hat\G^i\p_J^\z\)\Big]
-\p_\J^k\[\(\frac{1}{d-1}\Hat\G_k\Hat\G_j-\g_{kj}\)\slashed{\bb D}-\overleftarrow{\slashed{\bb D}}\(\frac{1}{d-1}\Hat\G_k\Hat\G_j-\g_{kj}\) \]\p_\J^j\\
&+\frac{i}{d-1}\(\p_I^\z\slashed\pa\vf^I\Hat\G_i\p_\J^i-\p_\J^i\Hat\G_i\slashed\pa\vf^I\p_I^\z\)-2i\pa_i\vf^I\(\p_I^\z\p_\J^i-\p_\J^i\p_I^\z\)\\
&+\cg^{IM}\cg^{KN}\pa_i\vf^J\(\pa_K\cg_{IJ}-\pa_I\cg_{KJ}\)\p_M^\z\Hat\G^i\p_N^\z\Bigg\}\\
&+\frac{\sqrt{-\g}}{2\k^2}\Bigg[-R[\g]+\cg_{IJ}\pa_i\vf^I\pa^i\vf^J+\cg_{IJ}\lbar\z_-^I\(\slashed{\bb D}-\overleftarrow{\slashed{\bb D}}\)\z_-^J+\lbar\J_{+i}\Hat\G^{ijk}\(\bb D_j-\overleftarrow{\bb D}_j\)\J_{+k}\\
& +D_k\(\lbar\J_{+i}\(\g^{jk}\Hat\G^i-\g^{ik}\Hat\G^j\)\J_{+j} \)+i\cg_{IJ}\pa_i\vf^J\(\lbar\z_-^I\Hat\G^j\Hat\G^i\J_{+j}-\lbar\J_{+j}\Hat\G^i\Hat\G^j\z_-^I\)\\
&+\pa_K\cg_{IJ}\pa_i\vf^J\(\lbar\z_-^I\Hat\G^i\z_-^K-\lbar\z_-^K\Hat\G^i\z_-^I\)-2i V_I\(\lbar\z_-^I\Hat\G^{ij}\bb D_i\J_{+j}+\lbar\J_{+j}\overleftarrow{\bb D}_i\Hat\G^{ij}\z_-^I\)\\
&+V_I \cg_{JK}\pa_i\vf^J\(\lbar\z_-^I\Hat\G^i\z_-^K-\lbar\z_-^K\hat\G^i\z_-^I\)\Bigg],\numberthis\label{H-wt-cs}
\eala 
where
\be 
\wt\p_a^i\equiv \frac{\d\wt{\bb S}}{\d e^a_i},\quad \wt\p_I^\vf\equiv \frac{\d\wt{\bb S}}{\d \vf^I}.
\ee 
From this one could write the recursive equation for every $\bb S_{(k)}$. It, however, looks too complicated, and thus we first write down equations for $\bb S_{(1)}^{\z\z}$, $\bb S_{(3/2)}^{\z\J}$ and $\bb S_{(2)}^{\J\J}$, namely
\bsub \label{H-123}
\bala
0=\; & -\cw\bb L_{(1)}^{\z\z}+\cw\Bigg\{V^I\pa_I+\frac{1}{2(d-1)}\(\lbar\z_-^I\frac{\d}{\d\lbar\z_-^I}+\frac{\overleftarrow \d}{\d\z_-^I}\z_-^I\)+\pa_I V^J\(\lbar\z_-^I\frac{\d}{\d\lbar\z_-^J}+\frac{\overleftarrow \d}{\d\z_-^J}\z_-^I\)\Bigg\}\bb S_{(1)}^{\z\z} \\
&+\frac{\sqrt{-\g}}{2\k^2}\Big[\cg_{IJ}\(\lbar\z_-^I\slashed{\bb D}\z_-^J-\lbar\z_-^I\overleftarrow{\slashed{\bb D}}\z_-^J\)+\(V_I\cg_{JK}+\pa_K\cg_{IJ}\) \pa_i\vf^J\(\lbar\z_-^I\Hat\G^i\z_-^K-\lbar\z_-^K\hat\G^i\z_-^I\) \Big].\numberthis\label{H-zz-1} \\
0=\;&-\frac{d-\frac 32}{d-1}\cw \bb L_{(3/2)}^{\z\J}+\cw\Bigg[V^I\pa_I+\frac{1}{2(d-1)}\(\lbar\z_-^I\frac{\d}{\d\lbar\z_-^I}+\frac{\overleftarrow \d}{\d\z_-^I}\z_-^I\)+\pa_L V^K\(\lbar\z_-^L\frac{\d}{\d\lbar\z_-^K}+\frac{\overleftarrow \d}{\d\z_-^K}\z_-^L\)\Bigg]\bb S_{(3/2)}^{\z\J}\\
&+\frac{\sqrt{-\g}}{2\k^2}i\Big[-2V_I\(\lbar\z_-^I\Hat\G^{ij}\bb D_i\J_{+j}+\lbar\J_{+j}\overleftarrow{\bb D}_i\Hat\G^{ij}\z_-^I\)+\cg_{IJ}\(\lbar\z_-^I\Hat\G^j\slashed\pa\vf^J\J_{+j}-\lbar\J_{+j}\slashed\pa\vf^J\Hat\G^j\z_-^I\) \Big],\numberthis\label{H-zJ-32} \\
0=\;&-\frac{d-2}{d-1}\cw\bb L_{(2)}^{\J\J} +\cg^{IJ}\pa_I\cw\frac{\d}{\d \vf^J}\bb S_{(2 )}^{\J\J}\\
&\hskip1in +\frac{\sqrt{-\g}}{2\k^2}\Big[\lbar\J_{+i}\Hat\G^{ijk}(\bb D_j-\overleftarrow{\bb D}_j)\J_{+k}+D_k\(\lbar\J_{+i}(\g^{jk}\Hat\G^i-\g^{ik}\Hat\G^j)\J_{+j}\) \Big],\numberthis\label{H-JJ-2}
\eala
\esub
where we used \eqref{L-k}.

While \eqref{H-zz-1} and \eqref{H-zJ-32} are not so easy to treat at first sight, the solution of \eqref{H-JJ-2} is obvious, namely
\be
\bb L^{\J\J}_{(2)}=-\frac{\sqrt{-\g}}{\k^2}\X\[\lbar\J_{+i}\Hat\G^{ijk}(\bb D_j-\overleftarrow{\bb D}_j)\J_{+k}+D_k\(\lbar\J_{+i}(\g^{jk}\Hat\G^i-\g^{ik}\Hat\G^j)\J_{+j}\) \],\label{cs-2-JJ}
\ee 
once we take into account \eqref{eqn-X}. Instead of solving \eqref{H-zz-1} and \eqref{H-zJ-32} directly, we now try $\cf_+$ constraint \eqref{cf-neg}, which greatly reduces the amount of efforts. They are respectively at the 'level' 1 and 3/2
\bsub \label{cf-pos-1-32}
\bal
& i\cg^{IJ}\pa_I\cw\frac{\d}{\d\lbar\z_-^J}\bb S_{(1)}^{\z\z}+\frac{1}{d-1}\cw\Hat\G_i \frac{\d}{\d\lbar\J_{+i}}\bb S_{(3/2)}^{\z\J}=\frac{\sqrt{-\g}}{2\k^2}i\cg_{IJ}\pa_i\vf^J\hat\G^i\z_-^I ,\label{cf-pos-1}\\
& \frac{1}{d-1}\cw\Hat\G_i\frac{\d}{\d\lbar\J_{+i}}\bb S_{(2)}^{\J\J}+i\cg^{IJ}\pa_I\cw \frac{\d}{\d\lbar\z_-^J}\bb S_{(3/2)}^{\z\J}=\frac{\sqrt{-\g}}{\k^2}\Hat\G^{ij}\bb D_i\J_{+j}.\label{cf-pos-32}
\eal
\esub 
The solution \eqref{cs-2-JJ} allows us to solve \eqref{cf-pos-32} immediately and we obtain
\be 
\frac{\d}{\d\lbar\z_-^I}\bb S_{(3/2)}^{\z\J}=i\frac{\sqrt{-\g}}{\k^2}\(-2\pa_I\X\lbar\z_-^I\Hat\G^{ij}\bb D_i\J_{+j}+A_{IJ}\lbar\z_-^I\Hat\G^i\slashed\pa\vf^J\J_{+i}\). 
\ee 
One can readily see that
\bal 
\bb S_{(3/2)}^{\z\J}=\;& \frac{i}{\k^2}\int d^dx\sqrt{-\g}\Big[2\pa_I\X\( \lbar\J_{+i}\overleftarrow{\bb D}_j\Hat\G^{ij}\z_-^I-\lbar\z_-^I\Hat\G^{ij}\bb D_i\J_{+j}\)+\NO\\
&\hskip1.5in +A_{IJ}\(\lbar\z_-^I\Hat\G^i\slashed\pa\vf^J\J_{+i}-\lbar\J_{+i}\slashed\pa\vf^I\Hat\G^i\z_-^J\)\Big].\label{cs-32-zJ}
\eal
In the same way, we find from \eqref{cf-pos-1} that
\be 
\bb S_{(1)}^{\z\z}=\frac{1}{\k^2}\int d^dx\sqrt{-\g}\;\(A_{IJ}\lbar\z_-^I(\slashed{\bb D}-\overleftarrow{\slashed{\bb D}})\z_-^J+(\pa_J A_{Ik}-\pa_I A_{JK})\lbar\z_-^I\slashed\pa\vf^K\z_-^J \).\label{cs-1-zz}
\ee 

Moreover, we can confirm that the solutions \eqref{cs-1-zz} and \eqref{cs-32-zJ} satisfy the Hamiltonian constraints \eqref{H-zz-1} and \eqref{H-zJ-32} respectively. That is not the whole story, and one has to convince himself that $\cf_-=0$ constraint also holds for these solutions. From \eqref{cf-neg}, we obtain
\bsub \label{cf-neg-gen}
\bal
& 0=\bb D_i\frac{\d}{\d\lbar\J_{+i}}\bb S_{(2k)}^{\J\J}+\frac i2\pa_i\vf^I\Hat\G^i\frac{\d}{\d\lbar\z_-^I}\bb S_{(2k-1/2)}^{\z\J}-\Hat\G_i\J_{+j}\frac{\d}{\d\g_{ij}}\bb S_{(2k)}^B,\label{cf-neg-2k+1}\\
& 0=\bb D_i\frac{\d}{\d \lbar\J_{+i}}\bb S_{(2k-1/2)}^{\z\J}+\frac i2\pa_i\vf^I\Hat\G^i\frac{\d}{\d\lbar\z_-^I}\bb S_{(2k-1)}^{\z\z}-\frac i2\z_-^I\frac{\d}{\d\vf^I}\bb S_{(2k)}^B,\label{cf-neg-2k}
\eal
\esub 
where $k$ is an arbitrary positive integer. It is not so difficult to check the solutions we obtained satisfy the constraints \eqref{cf-neg-2k} and \eqref{cf-neg-2k+1} for $k=1$, implying that the combination
\be 
\bb S^B_{(2)}+\bb S^{\J\J}_{(2)}+\bb S^{\z\z}_{(1)}+\bb S^{\z\J}_{(3/2)}
\ee 
is ($\e_+$) supersymmetric.

We have seen how to obtain the Hamilton's principal function in the fermionic sector from its bosonic supersymmetric partner, but at the lower 'level'. It was relatively easy because we could give the most general ansatz for $\bb S_{(2)}^B$ which has a small number of terms. To go further we should first find out $\bb S_{(4)}^B$, $\bb S_{(6)}^B$, $\cdots$ and obtain their SUSY partners by using the above trick. The ansatz for $\bb S_{(2n)}^B$ ($n\geq2$), however, has lots of terms and is complicated, hence finding its SUSY partner must be a boring subject.

Although we stop finding the general solution of the HJ equations in the fermionic sector here, we remark that the solution we have found is almost sufficient for providing the divergent counterterms in the low dimensions, say, $d=4$. This is because in the generic case that there are no scalar fields dual to marginal operators, $\bb S^{\z\z}_{(3)}$ and $\bb S^{\z\J}_{(7/2)}$ are asymptotically suppressed in 4 dimensions. As a result, what is remained in the case of $d=4$ is only to find out $\bb S^{\J\J}_{(4)}$ that are the logarithmically divergent terms, which are directly related to the holographic Weyl anomaly \cite{Henningson:1998gx}.

We should emphasize that from the general analysis here the divergent counterterms (except for $\bb S_{(0)}$) always satisfy the constraint $\cf_-=0$ and so is the renormalized on-shell action $\Hat S_{ren}$.

We finish this subsection by presenting the recursive relation obtainted from \eqref{cf-pos}, namely
\bala 
0=\;& -\frac{1}{d-1}\cw\Hat\G_i\p_{\lbar \J(n-1/2)}^i-i\cg^{IJ}\pa_I\cw\p_{J(n-1)}^{\lbar\z} +\frac{\k^2}{\sqrt{-\g}}\sum_{m=1}^{\left \lfloor{\frac{n}{2}}\right \rfloor-1}\Big[ \frac i2\cg^{IJ}\p_{I(2m)}^\vf\p_{J(n-2m-1)}^{\lbar\z}\\
&\hskip0.5in+\p_{(2m)}^{jk}\(\frac{1}{d-1}\g_{jk}\Hat\G_i-\g_{ij}\Hat\G_k\)\p_{\lbar \J(n-2m-1/2)}^i\Big],\numberthis\label{cf-pos-recursive}
\eala
where (integer or half-integer) $n\geq 4$. This will be useful for determination of the super-Weyl anomaly in section \ref{sWeyl-anomaly}.

\subsection{Logarithmically divergent terms in 4D}

As mentioned in subsection \ref{counterterms-HJ-one-point}, every $\bb S_{(k)}$ in the asymptotic expansion \eqref{expand-e} with respect to the operator $\d_e$ contains the poles related to the logarithmically divergent terms. Let us denote such terms by $\wt{\bb S}_{(k)}$. Whereas $\wt{\bb S}^B_{(4)}$ and $\wt{\bb S}^{\J\J}_{(4)}$ are purely gravitational (meaning that they are related only to the metric and the gravitino field) and universal, $\wt{\bb S}^{\z\z}_{(1)}$, $\wt{\bb S}^{\z\J}_{(3/2)}$, $\wt{\bb S}^B_{(2)}$ and $\wt{\bb S}^{\J\J}_{(2)}$ are model-dependent. We first discuss the former and then study the latter for a simple model.

$\wt{\bb S}^B_{(4)}$ is easily obtained from \eqref{H-bos-recursive} and \eqref{H-bos-inhomo-4}, namely
\bala
\wt{\bb S}^B_{(4)}\equiv\;&\int d^dx\sqrt{-\g}\;\wt\cl^B_{(4)}\log e^{-2r}\\
=\;&\frac{1}{4\k^2(d-2)^2}\int d^dx\sqrt{-\g}\(\frac{d}{4(d-1)}R^2-R_{ij}R^{ij} \)\log e^{-2r},\numberthis \label{log-metric}
\eala 
which is already well-known. Meanwhile, $\wt{\bb S}^{\J\J}_{(4)}$ is determined by the inhomogeneous terms of the Hamiltonian constraint \eqref{H-wt-cs} at the 'level' 4, namely\footnote{When the boundary metric is flat, \eqref{log-gravitino} matches with the result in \cite{Argurio:2014uca}.}
\bala 
\wt{\bb S}^{\J\J}_{(4)}\equiv \;& \int d^dx\sqrt{-\g}\;\wt\cl^{\J\J}_{(4)}\log e^{-2r}\\
=\;&\int d^dx\frac{\k^2}{4\sqrt{-\g}}\Bigg\{2\(\frac{1}{d-1}\g_{ij}\g_{kl}-\g_{ik}\g_{jl}\)\wt\p^{ij}_{(2)}e^{ak}\p_{a(2)}^{l\J}-\wt\p^{ij}_{(2)}(\lbar\J_{+k}\Hat\G_j\Hat\G^k\p_{\lbar \J i}^{(2)}+\p_{\J i}^{(2)}\Hat\G^k\Hat\G_j\J_{+k})\\
&-\frac{1}{d-1}\g_{ij}\wt\p^{ij}_{(2)}\(\lbar \J_+^{k}\Hat\G_{kl}\p_{(2)\lbar \J}^l+\p_\J^{(2)l}\Hat\G_{lk}\J_+^k\)+\frac{1}{2(d-1)}(\p_\J^{(2)k}\Hat\G_k\slashed{\bb D}\Hat\G_j\p_{\lbar \J}^{(2)j}-\p_\J^{(2)k}\Hat\G_k\slabD\Hat\G_j\p_{\lbar\J}^{(2)j})\\
&+\frac12(\p_\J^{(2)i}\slashed{\bb D}\p^{(2)}_{\lbar \J i}-\p^{(2)i}_\J\slabD\p^{(2)}_{\lbar \J i})\Bigg\}\log e^{-2r}\\
=\;&\frac{1}{8(d-2)^2\k^2}\int d^dx\sqrt{-\g}\Bigg\{(d-3)R(\lbar\J_{+i}\Hat\G^{ijk}\bb D_j\J_{+k}-\lbar\J_{+i}\labD_j\Hat\G^{ijk}\J_{+k})\\
&+\frac{d}{d-1}RD_j\[\lbar\J_{+i}(\g^{ij}\Hat\G^k-\g^{jk}\Hat\G^i)\J_{+k}\]-(d-4)R(\lbar\J_{+i}\Hat\G^i\Hat\G^{jk}\bb D_j\J_{+k}-\lbar\J_{+i}\labD_j\Hat\G^{ij}\Hat\G^k\J_{+k})\\
&+\frac{(d-2)^2}{d-1}R\Big[ \lbar\J_+^k\Hat\G^j\bb D_k\J_{+j}-\lbar\J_{+}^i\slashed{\bb D}\J_{+i}-\lbar\J_{+i}\labD^k\Hat\G^i\J_{+k}+\lbar\J_+^k\slabD\J_{+k}\Big]\\
&+2R_{kl}\Big[\lbar\J_{+i}[(\g^{ip}\Hat\G^k-\g^{ik}\Hat\G^p)\bb D^l-\labD^l(\g^{ip}\Hat\G^k-\g^{pk}\Hat\G^i)]\J_{+p}-\lbar\J_{+i}\Hat\G^i\Hat\G^{jl}\bb D_j\J_+^k+\lbar\J_+^k\labD_j\Hat\G^{lj}\Hat\G^i\J_{+i} \\
&-D_j[\lbar\J_+^l\Hat\G^{kji}\J_{+i}-\lbar\J_{+i}\Hat\G^{ijk}\J_+^l -\lbar\J_{+i}(\g^{jk}\g^{pl}\Hat\G^i-\g^{jk}\g^{il}\Hat\G^p+\g^{jp}\g^{il}\Hat\G^k-\g^{pl}\g^{ij}\Hat\G^k)\J_{+p}]\Big]\\
&-\frac{2(d-2)^2}{d-1}(\lbar\J_{+i}\labD_j\Hat\G^{ij}\slashed{\bb D}\Hat\G^{kl}\bb D_k\J_{+l}-\lbar\J_{+i}\labD_j\Hat\G^{ij}\slabD\Hat\G^{kl}\bb D_k\J_{+l})\\
&-2(\lbar\J_{+p}\labD_q\Hat\G^{pqi}\slashed{\bb D}\Hat\G_i{}^{jk}\bb D_j\J_{+k}-\lbar\J_{+p}\labD_q\Hat\G^{pqi}\slabD\Hat\G_i{}^{jk}\bb D_j\J_{+k})\Bigg\}\log e^{-2r}.\numberthis \label{log-gravitino}
\eala 
Although nontrivial, one can show that $\wt{\bb S}^B_{(4)}+\wt{\bb S}^{\J\J}_{(4)}$ satisfies the constraints $\ch=\cf_-=\cf_+=0$ (i.e. conformal, supersymmetry and super-Weyl invariance), namely
\bsub\label{weyl-consistency}
\bal
& 0=\(e^a_i\frac{\d}{\d e^a_i}+\frac12\lbar\J_{+i}\frac{\d}{\d \lbar\J_{+i}}+\frac{\overleftarrow \d}{\d\J_{+i}}\J_{+i}  \)\wt{\bb S}^{\J\J}_{(4)},\label{anomaly-inv-conformal}\\
& 0=\Hat\G_i\J_{+j}\frac{\d}{\d\g_{ij}}\wt{\bb S}^B_{(4)}+\bb D_i\frac{\d}{\d\lbar\J_{+i}}\wt{\bb S}^{\J\J}_{(4)},\label{anomaly-inv-susy}\\
& 0=\Hat\G_i\frac{\d}{\d\lbar\J_{+i}}\bb S^{\J\J}_{(4)}.\label{anomaly-inv-superweyl}
\eal 
\esub

\subsection{Generic finite counterterms in 4D and summary}\label{generic-finite-summary}

Up to now we obtained generic part of the divergent counterterms. $\cs_{\rm ct}$ can involve additional finite terms which satisfy the first class constraints \eqref{first-class-constraint}, though.\footnote{Otherwise, these finite terms would generate trivial cocycle terms, which do not have any physical implication.} The possible bosonic finite counterterms are Euler density and Weyl invariant in 4D, namely,
\be 
E_{(4)}=\frac{1}{64}\(R^{ijkl}R_{ijkl}-4R^{ij}R_{ij}+R^2\),\quad I_{(4)}=-\frac{1}{64}\(R^{ijkl}R_{ijkl}-2R^{ij}R_{ij}+\frac13 R^2\)
\ee 
Integral of the Euler density $E_{(4)}$ by itself satisfies all the first class constraints, since it is topological quantity, any local variation of which vanishes. Therefore, we find that the possible supersymmetric finite counterterms are linear combination of
\bala 
X_I=\;&64I_{(4)}+(d-3)R(\lbar\J_{+i}\Hat\G^{ijk}\bb D_j\J_{+k}-\lbar\J_{+i}\labD_j\Hat\G^{ijk}\J_{+k})+\frac{d}{d-1}RD_j\[\lbar\J_{+i}(\g^{ij}\Hat\G^k-\g^{jk}\Hat\G^i)\J_{+k}\]\\
&-(d-4)R(\lbar\J_{+i}\Hat\G^i\Hat\G^{jk}\bb D_j\J_{+k}-\lbar\J_{+i}\labD_j\Hat\G^{ij}\Hat\G^k\J_{+k})\\
&+\frac{(d-2)^2}{d-1}R\Big[ \lbar\J_+^k\Hat\G^j\bb D_k\J_{+j}-\lbar\J_{+}^i\slashed{\bb D}\J_{+i}-\lbar\J_{+i}\labD^k\Hat\G^i\J_{+k}+\lbar\J_+^k\slabD\J_{+k}\Big]\\
&+2R_{kl}\Big[\lbar\J_{+i}[(\g^{ip}\Hat\G^k-\g^{ik}\Hat\G^p)\bb D^l-\labD^l(\g^{ip}\Hat\G^k-\g^{pk}\Hat\G^i)]\J_{+p}-\lbar\J_{+i}\Hat\G^i\Hat\G^{jl}\bb D_j\J_+^k+\lbar\J_+^k\labD_j\Hat\G^{lj}\Hat\G^i\J_{+i} \\
&-D_j[\lbar\J_+^l\Hat\G^{kji}\J_{+i}-\lbar\J_{+i}\Hat\G^{ijk}\J_+^l -\lbar\J_{+i}(\g^{jk}\g^{pl}\Hat\G^i-\g^{jk}\g^{il}\Hat\G^p+\g^{jp}\g^{il}\Hat\G^k-\g^{pl}\g^{ij}\Hat\G^k)\J_{+p}]\Big]\\
&-\frac{2(d-2)^2}{d-1}(\lbar\J_{+i}\labD_j\Hat\G^{ij}\slashed{\bb D}\Hat\G^{kl}\bb D_k\J_{+l}-\lbar\J_{+i}\labD_j\Hat\G^{ij}\slabD\Hat\G^{kl}\bb D_k\J_{+l})\\
&-2(\lbar\J_{+p}\labD_q\Hat\G^{pqi}\slashed{\bb D}\Hat\G_i{}^{jk}\bb D_j\J_{+k}-\lbar\J_{+p}\labD_q\Hat\G^{pqi}\slabD\Hat\G_i{}^{jk}\bb D_j\J_{+k}),\numberthis\label{X1}
\eala 
and
\be 
X_E=E_{(4)},\quad X_P=\cp= \frac{1}{64}\e^{ijkl}R_{ijpq}R_{kl}{}^{pq},
\ee 
where $\cp$ is the Pontryagin density. Notice that integral of $\cp$ is the topological quantity and thus it can be a finite counterterm as in the case of the Euler density, as long as there is no other symmetry which prevents its appearance.

In summary, collecting all of these finite counterterms and the previous divergent ones we obtain
\bal\label{generic-counterterms}
\boxed{
\begin{aligned}
S_{\rm ct}=\;&-\(\bb S_{(0)}+\bb S_{(1)}+\bb S_{(2)}\)-\(\wt{\bb S}^B_{(4)}+\wt{\bb S}^{\J\J}_{(4)} \)+\cdots \\
=\;&\frac{1}{\k^2}\int d^dx\sqrt{-\g}\Big\{\cw-\X R-A_{IJ}\pa_i\vf^I\pa^i\vf^J-A_{IJ}\lbar\z_-^I(\slashed{\bb D}-\slabD)\z_-^J\\
&\hskip0.5in-(\pa_J A_{IK}-\pa_I A_{JK})\lbar\z_-^I\slashed\pa\vf^K\z_-^J-2i\pa_I\X(\lbar\J_{+i}\labD_j\Hat\G^{ij}\z_-^I-\lbar\z_-\Hat\G^{ij}\bb D_i\J_{+j})\\
&\hskip0.5in -i A_{IJ}(\lbar\z_-^I\Hat\G^i\slashed\pa\vf^J\J_{+i}-\lbar\J_{+i}\slashed\pa\vf^I\Hat\G^i\z_-^J)+\X\lbar\J_{+i}\Hat\G^{ijk}(\bb D_j-\labD_j)\J_{+k}\\
&\hskip0.5in +\lbar\J_{+i}(\pa^i\X\Hat\G^j-\pa^j\X \Hat\G^i)\J_{+j}\Big\}\\
&-\wt{\bb S}_{(4)}^B-\wt{\bb S}_{(4)}^{\J\J}+\frac{1}{\k^2}\int d^dx\sqrt{-\g}\(\a_I X_I+\a_E X_E+\a_P X_P\)+\cdots,
\end{aligned}
}
\eal
where $\wt{\bb S}^B_{(4)}$ and $\wt{\bb S}^{\J\J}_{(4)}$ are given in \eqref{log-metric} and \eqref{log-gravitino} and $\a_I$, $\a_E$ and $\a_P$ are arbitrary constants. Here the ellipsis stand for the model-dependent terms, which we discuss in section \ref{toy-app} for a simple toy model.

\subsection{Application to a toy model}
\label{toy-app}

For completeness, we present an application of our general procedure to a simple toy model.   

In the toy model there is only one scalar field, which corresponds to the operator with the scaling dimension $\D=d-1$ with $d=4$. It then implies that
\be 
\cw=-(d-1)-\frac12\vf^2+k_3\vf^3+k_4\vf^4+\co(\vf^5),
\ee 
where $k_3$ and $k_4$ are arbitrary constants, and therefore the solution of \eqref{eqn-X}, \eqref{eqn-A} and \eqref{eqn-X-A} becomes
\be 
\X=\frac{1}{2(d-2)}-\frac{1}{d-4}\cdot \frac{1}{4(d-1)}\vf^2+\cdots ,\quad A_{IJ}=-\frac{1}{d-4}\cdot \frac12+\cdots .
\ee 
The divergent counterterms that we need other than those in \eqref{generic-counterterms} are only the logarithmically divergent terms. Following the argument in section \ref{counterterms-HJ-one-point} again we can see them from the poles (when $d=4$) in $\X$ and $A_{IJ}$ and are responsible for additional logarithmically divergent terms.

We thus obtain
\bsub 
\bal 
& \wt{\bb S}^B_{(2)}=\frac{1}{4\k^2}\int d^dx\sqrt{-\g}\(\frac{1}{2(d-1)}\vf^2 R+\pa_i\vf\pa^i\vf  \)\log e^{-2r},\\
& \wt{\bb S}^{\z\z}_{(1)}=\frac{1}{4\k^2}\int d^dx\sqrt{-\g}\;(\lbar\z_-\slashed{\bb D}\z_-+\mathrm{h.c.})\log e^{-2r} ,\\
&\wt{\bb S}^{\z\J}_{(3/2)}=\frac{i}{4\k^2}\int d^dx\sqrt{-\g}\(\lbar\z_-\Hat\G^i\slashed \pa\vf \J_{+i}-\frac{2}{(d-1)}\vf\lbar\z_-\Hat\G^{ij}\bb D_i\J_{+j}+\mathrm{h.c.} \)\log e^{-2r},\\
&\wt{\bb S}^{\J\J}_{(2)}=-\frac{1}{4\k^2}\int d^dx\sqrt{-\g}\;\frac{1}{d-1}\(\frac{1}{2}\vf^2\lbar\J_{+i}\Hat\G^{ijk}\bb D_j\J_{+k}+\vf \lbar\J_{+i}\pa^i\vf\Hat\G^j\J_{+j}+\mathrm{h.c.} \)\log e^{-2r}.
\eal  
\esub 
One can easily check that $\wt{\bb S}^B_{(2)}+\wt{\bb S}^{\z\z}_{(1)}+\wt{\bb S}^{\z\J}_{(3/2)}+\wt{\bb S}^{\J\J}_{(2)}$ again satisfies the constraints $\ch=\cf_-=\cf_+=0$.

Besides $X_I$ and $X_E$, the possible finite counterterms (conformal and $\e_+$ supersymmetric) are
\bala
X_0=\;&\frac{1}{2(d-1)}\vf^2 R+\pa_i\vf\pa^i\vf+\lbar\z_-\slashed{\bb D}\z_-+i\lbar\z_-\Hat\G^i\slashed \pa\vf \J_{+i}\\
&-\frac{2i}{d-1}\vf\lbar\z_-\Hat\G^{ij}\bb D_i\J_{+j}-\frac{1}{2(d-1)}\vf^2\lbar\J_{+i}\Hat\G^{ijk}\bb D_j\J_{+k}-\frac{1}{d-1}\vf\lbar\J_{+i}\pa^i\vf\Hat\G^j\J_{+j}+\mathrm{h.c.},\numberthis\label{X3}
\eala
and the finite term $k_4\vf^4$ in $\cw$ should be in the counterterms without any ambiguity, due to the $\cf_-$ constraint.

In total, the divergent counterterms for the toy model are
\be 
\bb S^{div}_{\rm ct}=-\(\bb S_{(0)}+\bb S_{(1)}+\bb S_{(3/2)}+\bb S_{(2)}\)-\int d^dx\sqrt{-\g}\;\wt\cl_{[4]}\log e^{-2r},
\ee 
where the logarithmically divergent counterterms are
\be 
\int d^dx\sqrt{-\g}\;\wt\cl_{[4]}\log e^{-2r}=\wt{\bb S}^{\z\z}_{(1)}+\wt{\bb S}^{\z\J}_{(3/2)}+\wt{\bb S}^B_{(2)}+\wt{\bb S}^{\J\J}_{(2)}+\wt{\bb S}^B_{(4)}+\wt{\bb S}^{\J\J}_{(4)}.
\ee 
Adding possible finite ones, the whole counterterms are
\bal \label{whole-counterterm}
\boxed{ 
\begin{aligned}
\bb S_{\rm ct}=\;&\frac{1}{\k^2}\int d^dx\sqrt{-\g}\Big[-(d-1)-\frac12\vf^2+k_3\vf^3+k_4\vf^4-\frac{1}{2(d-2)}R\\
&+\frac{1}{2(d-2)}\lbar\J_{+i}\Hat\G^{ijk}(\bb D_j-\labD_j)\J_{+k}\Big]-\int  d^dx\sqrt{-\g}\;\wt\cl_{[4]}\log e^{-2r}\\
&+\frac{1}{\k^2}\int d^dx\sqrt{-\g}\;(\a_I X_I+\a_E X_E+\a_P X_P+ \a_0 X_0),
\end{aligned}
}
\eal 
where $\a_E$, $\a_I$, $\a_P$ and $\a_0$ are arbitrary constants and determine the renormalization scheme. 

\section{Holographic dictionary and Ward identities}\label{holographic-dic-ward}
\setcounter{equation}{0}

Now that all the counterterms are determined, we can relate by the holographic dictionary \cite{Witten:1998qj} the renormalized canonical momenta to the renormalized local operators of the boundary field theory, namely
\bsub \label{one-point-function}
\bal
& \ct _a^i=-\lim_{r\rightarrow\infty}e^{(d+1)r}\frac{1}{\sqrt{-\g}}\(\p_a^i+\frac{\d S_{\rm ct}}{\d e^a_i} \):=-\frac{1}{|\bm e_{(0)}|}\P_a^i,\\
& \co_I^\vf=\lim_{r\rightarrow\infty}e^{(d+\m^I)r}\frac{1}{\sqrt{-\g}}\(\p_I^\vf +\frac{\d S_{\rm ct}}{\d \vf^I} \):=\frac{1}{|\bm e_{(0)}|}\P_I^\vf ,\\
& \co_I^{\lbar\z} =\lim_{r\rightarrow\infty}e^{(d+\m^I+\frac12)r}\frac{1}{\sqrt{-\g}}\(\p_I^{\lbar\z}+\frac{\d S_{\rm ct}}{\d \lbar\z^I} \):=\frac{1}{|\bm e_{(0)}|}\P_I^{\lbar\z} ,\\
& \cs^i =\lim_{r\rightarrow\infty}e^{(d+\frac12)r}\frac{1}{\sqrt{-\g}}\(\p_{\lbar\J}^i+\frac{\d S_{\rm ct}}{\d \lbar \J_{+i}} \):=\frac{1}{|\bm e_{(0)}|}\P_{\lbar\J}^i,
\eal 
\esub 
where $\ct_a^i$ is the energy-momentum tensor,\footnote{Definition of the energy-momentum tensor is modified when the vielbein is used instead of the metric, see e.g. (2.198) in \cite{DiFrancesco:1997nk}.} $S^i$ is the supercurrent\footnote{Spinor index of the supercurrent $\cs^i$ is implicit.} and $\bm e_{(0)}=\det(e^a_{(0)i})$. We note that since these local renormalized operators are obtained in the presence of arbitrary sources we can obtain higher-point functions simply by taking functional derivative of them with respect to the sources.

\subsection{Ward identities and anomalies}

One can find from the computation of section \ref{generic-HJ} and \ref{toy-app} that $\bb S_{\rm ct}$ satisfies the first class constraints $\ch^i=\cf_-=0$ and local Lorentz constraint \eqref{lorentz-constraint}, and so does the renormalized on-shell action $\Hat S_{ren}$. This is also related to the fact that these constraints are linear functional derivative equations.

Since $\ch$ and $\cf_+$ are not linear constraints, one should expect that the counterterms do not satisfy the constraints $\ch=0$ and $\cf_+=0$ in general and thus generate the non-trivial cocycle terms, which appear in the constraints for the renormalized on-shell action. Also, the poles appearing in solving the constraints contribute to the corresponding anomaly. In total, after removing all divergent counterterms, the first class constraints \eqref{cf-pos}, \eqref{cf-neg}, \eqref{hamiltonian}, \eqref{hi} and \eqref{local-Lorentz} are reduced into
\bsub \label{ward-identities}
\bal
0=\;&-\frac12\G^a\J_{(0)+i}\ct_a^i+\frac i2\z_{(0)-}^I\co_I^\vf-\frac i2\slashed\pa\vf_{(0)}^I\co_I^{\lbar\z}-\bb D_i\cs^i,\label{ward-e+}\\
\ca_{\rm sW}=\;&-i\cg^{IJ}\pa_I\cw \co_J^{\lbar\z}+\Hat\G_i\cs^i,\label{ward-e-}\\
\ca_{\rm W}=\;&e^a_{(0)i}\ct_a^i-\cg^{IJ}\pa_I\cw \co_J^\vf-\frac12\(\lbar\J_{(0)+i}S^i+\mathrm{h.c.}\)+\NO \\
&\hskip1in+\(\frac12\d^J_I-\pa_I\pa^J\cw  \)\(\lbar\z_{(0)-}^I\co_J^{\lbar\z}+\mathrm{h.c.}\) ,\label{ward-weyl} \\
0=\;&e_{(0)}^{ai}D_j\ct_a^j+\pa^i\vf_{(0)}^I\co _I^\vf+\(\lbar\z_{(0)-}^I\labD^i\co_I^{\lbar\z}+\rm{h.c.}\)+\(\lbar\J_{(0)+j}\labD^i \cs ^j+\rm{h.c.} \)\NO \\ 
&-D_j\(\lbar\J_{(0)+}^i\cs ^j+{\rm h.c.}\),\label{ward-diffeo}\\
0=\;& e_{(0)ai}\ct_b^i-e_{(0)bi}\ct _a^i+\frac12\(\lbar\z_{(0)-}^I\G_{ab}\co_I^{\lbar\z}+\lbar\J_{(0)+i}\G_{ab}\cs^i+{\rm h.c.} \)\label{ward-lorentz},
\eal
\esub 
where $\ca_{\rm sW}$ and $\ca_{\rm W}$ are super-Weyl and Weyl anomaly densities respectively. In \eqref{ward-identities} we keep only up to the quadratic order and zero order in $\vf^I$ in the Taylor expansion of $\cw$ and $\cg_{IJ}$ respectively.

We call \eqref{ward-identities}s as Ward identities which relate the local sources and their dual operators of the field theory. These Ward identities that play a key role in the following discussions reflect the remained local symmetries of the bulk SUGRA after fixing the strong FG gauge \eqref{sFG}, on which we did HR for the bulk theory in section \ref{generic-HJ}. The remaining local symmetry transformations of SUGRA are called generalized Penrose-Brown-Henneaux (gPBH) transformations, whose action on the sources are carefully treated in appendix \ref{gPBH}. The resulting expressions are \eqref{superdiffeo-induced}. Before discussing about the gPBH action on the renormalized canonical momenta, let us first find the anomalies explicitly in the case of $d=4$.

\subsubsection{Weyl anomaly}

Although there are many ways to find the Weyl anomaly, a direct way is to read it from the HJ equation. One can see that in \eqref{H-wt-cs} substituted by $\cl_{[4]}$, the first linear terms are indeed the RHS of the trace Ward identity \eqref{ward-weyl} and the rest of the terms give us part of the trace anomaly. The terms with pole $1/(d-4)$ which appeared in the HJ equations for $\bb S_{(1)}$, $\cdots$, $\bb S_{(4)}$ are also inherited into \eqref{H-wt-cs} for $\bb S_{[4]}$. These non-homogeneous terms are already identified to the logarithmically divergent terms and thus we only need to multiply them by 2 to obtain the trace anomaly \cite{Papadimitriou:2016yit}. For graviton and gravitino parts, the trace anomaly density is then\footnote{The SUSY completion of the Weyl anomaly in the 4 dimensional supersymmetric theory was obtained in \cite{Anselmi:1997am,Bonora:2013rta} by using the superspace formalism. To get the fermionic sector explicitly, however, one has yet to expand it further around the bosonic coordinates.}
\bala
\ca^{(G)}_{\rm W}[e,\J_+]=\;&\frac{1}{4(d-2)^2\k^2}\Bigg\{\frac{d}{2(d-1)}R^2-2R_{ij}R^{ij}\\
&+(d-3)R(\lbar\J_{+i}\Hat\G^{ijk}\bb D_j\J_{+k}-\lbar\J_{+i}\labD_j\Hat\G^{ijk}\J_{+k})\\
&+\frac{d}{d-1}RD_j\[\lbar\J_{+i}(\g^{ij}\Hat\G^k-\g^{jk}\Hat\G^i)\J_{+k}\]\\
&-(d-4)R(\lbar\J_{+i}\Hat\G^i\Hat\G^{jk}\bb D_j\J_{+k}-\lbar\J_{+i}\labD_j\Hat\G^{ij}\Hat\G^k\J_{+k})\\
&+\frac{(d-2)^2}{d-1}R\Big[ \lbar\J_+^k\Hat\G^j\bb D_k\J_{+j}-\lbar\J_{+}^i\slashed{\bb D}\J_{+i}-\lbar\J_{+i}\labD^k\Hat\G^i\J_{+k}+\lbar\J_+^k\slabD\J_{+k}\Big]\\
&+2R_{kl}\Big[\lbar\J_{+i}[(\g^{ip}\Hat\G^k-\g^{ik}\Hat\G^p)\bb D^l-\labD^l(\g^{ip}\Hat\G^k-\g^{pk}\Hat\G^i)]\J_{+p}\\
&-\lbar\J_{+i}\Hat\G^i\Hat\G^{jl}\bb D_j\J_+^k+\lbar\J_+^k\labD_j\Hat\G^{lj}\Hat\G^i\J_{+i} -D_j[\lbar\J_+^l\Hat\G^{kji}\J_{+i}-\lbar\J_{+i}\Hat\G^{ijk}\J_+^l \\
&-\lbar\J_{+i}(\g^{jk}\g^{pl}\Hat\G^i-\g^{jk}\g^{il}\Hat\G^p+\g^{jp}\g^{il}\Hat\G^k-\g^{pl}\g^{ij}\Hat\G^k)\J_{+p}]\Big]\\
&-\frac{2(d-2)^2}{d-1}(\lbar\J_{+i}\labD_j\Hat\G^{ij}\slashed{\bb D}\Hat\G^{kl}\bb D_k\J_{+l}-\lbar\J_{+i}\labD_j\Hat\G^{ij}\slabD\Hat\G^{kl}\bb D_k\J_{+l})\\
&-2(\lbar\J_{+p}\labD_q\Hat\G^{pqi}\slashed{\bb D}\Hat\G_i{}^{jk}\bb D_j\J_{+k}-\lbar\J_{+p}\labD_q\Hat\G^{pqi}\slabD\Hat\G_i{}^{jk}\bb D_j\J_{+k})\Bigg\}.
\numberthis \label{weyl-anomaly-G}
\eala

The holographic computation of the supersymmetric Weyl anomaly in 4D is quite remarkable; even though its bosonic part has already been known for a long time, it seems really tough to obtain its SUSY partner terms by means of giving an ansatz and finding out the coefficients, whereas the holography enables us to compute them directly.

We comment that although the bosonic sector of $\ca^G_{\rm W}$ is the sum of the $a$ anomaly density $E_{(4)}$ and $c$ anomaly one $I_{(4)}$, the fermionic sector is in fact SUSY partner of $c$ anomaly density up to a total derivative. This is because integral of $E_{(4)}$ is supersymmetric by itself, as mentioned before.

For the toy model of section \ref{toy-app}, we have additional contribution to the Weyl anomaly density, which is
\bala 
\ca_W^{(\rm model)}[\F]=\;&\frac{1}{2\k^2}\Bigg(\frac{1}{2(d-1)}\vf^2 R+\pa_i\vf\pa^i\vf+\lbar\z_-\slashed{\bb D}\z_-+i\lbar\z_-\Hat\G^i\slashed \pa\vf \J_{+i}-\frac{2i}{d-1}\vf\lbar\z_-\Hat\G^{ij}\bb D_i\J_{+j}\\
&-\frac{1}{2(d-1)}\vf^2\lbar\J_{+i}\Hat\G^{ijk}\bb D_j\J_{+k}-\frac{1}{d-1}\vf\lbar\J_{+i}\pa^i\vf\Hat\G^j\J_{+j}+\mathrm{h.c.}\Bigg).\numberthis 
\eala 
The total Weyl anomaly density is thus given by\footnote{It seems like that the bosonic sector of the conformal anomaly density $\ca$ here is different from the one given in \cite{Papadimitriou:2016yit} (see between (5.61) and (5.62) there), because of the $\vf^4$ term in $\wt\cl_{(4)}$. One can, however, easily check that it actually vanishes, taking into account \eqref{sup-eq}. This is because in our model the superpotential $\cw$ is local by construction.}  
\bal\label{weyl-anomaly-total}
\boxed{ 
	\begin{aligned} 
		\ca_W[\F]=\;& \ca_W^{(G)}[\F]+\ca^{(\rm model)}_W[\F]\\
		=\;& \ca_W^{(G)}+\frac{1}{2\k^2}\Bigg(\frac{1}{2(d-1)}\vf^2 R+\pa_i\vf\pa^i\vf+\lbar\z_-\slashed{\bb D}\z_-+i\lbar\z_-\Hat\G^i\slashed \pa\vf \J_{+i}\\
		&-\frac{2i}{d-1}\vf\lbar\z_-\Hat\G^{ij}\bb D_i\J_{+j}-\frac{1}{2(d-1)}\vf^2\lbar\J_{+i}\Hat\G^{ijk}\bb D_j\J_{+k}-\frac{1}{d-1}\vf\lbar\J_{+i}\pa^i\vf\Hat\G^j\J_{+j}+\mathrm{h.c.}\Bigg).
	\end{aligned}
}
\eal

\subsubsection{Super-Weyl anomaly}
\label{sWeyl-anomaly}

Here we compute the super-Weyl anomaly for the toy model. As pointed out in section \ref{gen-fermion}, \eqref{cf-pos-32} hold up to the finite order. For the toy model, it means that the RHS of \eqref{cf-pos-32} is not canceled out and an additional finite term
\be 
+\frac{\sqrt{-\g}}{\k^2}\frac{\vf^2}{2(d-1)}\Hat\G^{ij}\bb D_i\J_{+j}
\ee
comes out from the LHS of \eqref{cf-pos-32}. As in the case of Weyl anomaly, we thus get from \eqref{cf-pos-recursive}
\bala
& -i\cg^{IJ}\pa_J\cw\p_{(7/2)I}^{\lbar \z}-\frac{1}{d-1}\cw\Hat\G_i \p_{(4)\lbar \J}^i\\
=\;& -\frac{\k^2}{\sqrt{-\g}}\p^{jk}_{(2)}\(\frac{1}{d-1}\g_{jk}\Hat\G_i-\g_{ij}\Hat\G_k\)\p_{(2)\lbar \J}^i  -\frac{\sqrt{-\g}}{2\k^2}i\pa_i\vf \Hat\G^i\z_-+\frac{\sqrt{-\g}}{\k^2}\frac{\vf^2}{2(d-1)}\Hat\G^{ij}\bb D_i\J_{+j}\\
=\;&\frac{\sqrt{-\g}}{\k^2}\[\frac{1}{4(d-2)^2}\(\frac{d}{d-1}R\Hat\G^{kl}-2R_i{}^k\Hat\G^{il}+2R_i{}^l\Hat\G^{ik}\)\bb D_k\J_{+l}-\frac i2\pa_i\vf\Hat\G^i\z_-+\frac{1}{2(d-1)}\vf^2\Hat\G^{ij}\bb D_i\J_{+j} \],\numberthis 
\eala
or
\bal 
\boxed{
	\begin{aligned} 
		\ca_{\rm sW}[\F]=\;&	\frac{1}{\k^2}\Bigg[\frac{1}{4(d-2)^2}\(\frac{d}{d-1}R\Hat\G^{kl}-2R_i{}^k\Hat\G^{il}+2R_i{}^l\Hat\G^{ik}\)\bb D_k\J_{+l}-\\
		&\hskip2in -\frac i2\pa_i\vf\Hat\G^i\z_-+\frac{1}{2(d-1)}\vf^2\Hat\G^{ij}\bb D_i\J_{+j} \Bigg].
	\end{aligned}
}\label{superweyl-anomaly}
\eal 
Notice that terms in the first bracket
\be \label{sW-G}
\ca^{(G)}_{\rm sW}[e,\J_+]=\frac{1}{\k^2}\frac{1}{4(d-2)^2}\(\frac{d}{d-1}R\Hat\G^{kl}-2R_i{}^k\Hat\G^{il}+2R_i{}^l\Hat\G^{ik}\)\bb D_k\J_{+l}
\ee 
are universal, in the sense that they do not depend on the model. Notice that \eqref{sW-G} is different from the result obtained by using Feynman diagram \cite{Abbott:1977xk} (see also \cite{Chaichian:2003kr}). The reason seems to be that here we computed the sum of $a$-anomaly and $c$-anomaly, while the super-trace anomaly in \cite{Abbott:1977xk} is a different linear combination of them. In any case, the result of \cite{Abbott:1977xk} does not satisfy the WZ consistency condition, as commented in footnote \ref{fn:sW-anomaly}.

\subsubsection{Wess-Zumino consistency condition}
\label{WZ-condition}

From the relation \eqref{anomaly-inv-conformal} and corresponding equation for the toy model we find that the Weyl anomaly \eqref{weyl-anomaly-G} and \eqref{weyl-anomaly-total} satisfy Wess-Zumino (WZ) consistency condition, which can be seen as follows. Defining the Weyl transformation operator $\d_\s$ by
\be 
\d_\s\equiv \int_{\pa\cm}d^dx\;\sum_{\F_{(0)}} \d_\s\F_{(0)}\frac{\d}{\d\F_{(0)}}, 
\ee 
where $\F_{(0)}$ refers to the source for every field $\F$, the WZ consistency condition becomes that $[\d_{\s_1},\d_{\s_2}]S_{ren}=0$. This is equivalent to $\d_{\s_1}\int d^dx\;\ca_{\rm W}\s_2$ is symmetric in $\s_1$ and $\s_2$, which can be seen from \eqref{anomaly-inv-conformal} since
\be 
\sum_{\F_{(0)}} \d_{\s_1}\F_{(0)}\frac{\d}{\d\F_{(0)}}\int d^dy\;\ca_{\rm W}\s_2=\s_1\pa^i(T\pa_i\s_2),
\ee  
for a certain scalar function $T$. We note that the SUSY and super-Weyl invariance of Weyl anomaly followed by \eqref{anomaly-inv-susy} and \eqref{anomaly-inv-superweyl} can be thought as the WZ consistency checks.

In order to see the super-Weyl anomaly \eqref{superweyl-anomaly} satisfies WZ consistency condition, first we need to find the algebra of relevant symmetries. From \eqref{superdiffeo-induced}, one can readily see that\footnote{Here the subscript $o$ is omitted again, which was used to denote the leading asymptotics of the variation parameters in appendix \ref{gPBH}.}
\be \label{commutator-e+e--boson}
[\d_{\e_+},\d_{\lbar\e'_-}]e^a_i=(\d_\s+\d_\l)e^a_i,\quad [\d_{\e_+},\d_{\lbar\e'_-}]\vf^I=(\d_\s+\d_\l)\vf^I,
\ee 
with the parameters $\s=\frac12\lbar\e_-'\e_+$,\quad $\l=\frac12\lbar\e_-'\G^{ab}\e_+$. Notice that in our stage it is impossible to see the above commutator for the fermionic sources, since our consideration is limited to quadratic order in fermions. However, \eqref{commutator-e+e--boson} provides us the WZ consistency condition for the super-Weyl anomaly, namely
\be \label{WZ-sW-1}
\(\d_{\e_+}\int d^dx|\bm e_{(0)}|\lbar\e'_-\ca_{\rm sW}[\F_{(0)}]\)\Big|_{\rm bosonic}=\([\d_{\e_+},\d_{\lbar\e'_-}]S_{ren}\)\Big|_{\rm bosonic}=-\int d^dx|\bm e_{(0)}|\s\ca_{W}^{(B)}[\F_{(0)}],
\ee 
since $\d_\l S_{ren}=0$. Here $\ca_W^{(B)}$ refers to the bosonic sector of the Weyl anomaly. In the following we show \eqref{WZ-sW-1} in detail, namely
\bala 
& \d_{\e_+}\int d^dx\sqrt{-\g}\;\lbar\e'_-\ca_{\rm sW}=\\
=\;&\frac{1}{\k^2}\int d^dx\sqrt{-\g}\lbar\e'_-\Big[\frac{1}{4(d-2)^2}\(\frac{d}{d-1}R\Hat\G^{kl}-2R_i{}^k\Hat\G^{il}+2R_i{}^l\Hat\G^{ik}\)\bb D_k\bb D_l\e_+\\
&\hskip0.5in-\frac 14\pa_i\vf\Hat\G^i\Hat\G^j\pa_j\vf \e_++\frac{1}{2(d-1)}\vf^2\Hat\G^{ij}\bb D_i\bb D_j\e_+ \Big]\\
=\;&\frac{1}{\k^2}\int d^dx\sqrt{-\g}\;\lbar\e'_-\Big[\frac{1}{32(d-2)^2}\(\frac{d}{d-1}R\Hat\G^{kl}-2R_i{}^k\Hat\G^{il}+2R_i{}^l\Hat\G^{ik}\)R_{mnkl}\Hat\G^{mn}\\
&\hskip0.5in-\frac 14\pa_i\vf\pa^i\vf +\frac{1}{16(d-1)}\vf^2\Hat\G^{ij}\Hat\G^{kl}R_{ijkl} \Big]\e_+\\
=\;&\frac{1}{\k^2}\int d^dx\sqrt{-\g}\;\lbar\e'_-\Big[\frac{1}{32(d-2)^2}\(-\frac{2d}{d-1}R^2+8R_{ij}R^{ij}\)-\frac14\pa_i\vf\pa^i\vf-\frac{1}{8(d-1)}\vf^2 R\Big]\e_+\\
=\;&-\int d^dx\sqrt{-\g}\;\s \ca_W^{(B)},\numberthis 
\eala
where we have again $\s=\frac12\lbar\e'_-\e_+$. In the above computation, we omitted the subscript ${}_{(0)}$ for simplicity. In the same spirit, one can find another WZ consistency condition for the super-Weyl anomaly from
\be 
[\d_{\e_-},\d_{\e'_-}]e^a_i=[\d_{\e_-},\d_{\e'_-}]\vf^I=0.
\ee 
We therefore have
\be 
\([\d_{\e_-},\d_{\e'_-}]S_{ren}\)\Big|_{\rm bosonic}=0,
\ee 
which can be shown in the same way. 

\subsection{SUSY transformation of operators}

Now that the Ward identities are completely given, we can use \eqref{ward-identities} to derive gPBH transformation of the renormalized canonical momenta, without using FG expansion of the induced fields and themselves \cite{Henneaux:1992ig,Papadimitriou:2010as,Cvetic:2016eiv}. In order to describe the gPBH transformation of the induced fields and their renormalized canonical momenta in an integrated way, we introduce concept of the generalized Poisson bracket, which is defined by (see e.g. (6.30) in \cite{Henneaux:1992ig})
\bala
&\{A[\F_{(0)},\P^\F],B[\F_{(0)},\P^\F]\}\equiv \int_{\pa\cm}d^dx\sum_{\F_{(0)}} \(\frac{\d A}{\d \F_{(0)}}\frac{\d B}{\d \P^\F}- \frac{\d B}{\d \F_{(0)}}\frac{\d A}{\d \P^\F}\)\\
&\hskip0.5in =\int_{\pa\cm}d^dx\Bigg(\frac{\d A}{\d e^a_{(0)i}}\frac{\d B}{\d \P_a^i}-\frac{\d B}{\d e^a_{(0)i}}\frac{\d A}{\d \P_a^i}
+\frac{\d A}{\d \vf_{(0)}^I}\frac{\d B}{\d \P_I^\vf }-\frac{\d B}{\d \vf_{(0)}^I}\frac{\d A}{\d \P^\vf_I} \\
&\hskip0.7in +A\frac{\overleftarrow \d }{\d \J_{(0)+i}}\frac{\overrightarrow\d }{\d \P_\J^i}B-B\frac{\overleftarrow\d }{\d \J_{(0)+i}}\frac{\overrightarrow\d }{\d \P_{\J}^i}A+A\frac{\overleftarrow\d }{\d \z_{(0)-}^I}\frac{\overrightarrow\d }{\d \P^\z_I}B-B\frac{\overleftarrow\d }{\d \z_{(0)-}^I}\frac{\overrightarrow\d }{\d \P^\z_I}A\\
&\hskip0.7in+B\frac{\overleftarrow\d }{\d \P_{\lbar\J}^i}\frac{\overrightarrow\d }{\d \lbar\J_{(0)+i}}A-A\frac{\overleftarrow\d }{\d \P_{\lbar\J}^i}\frac{\overrightarrow\d  }{\d \lbar \J_{(0)+i}}B +B\frac{\overleftarrow\d }{\d \P_I^{\lbar\z} }\frac{\overrightarrow\d }{\d \lbar\z_{(0)-}^I}A-A\frac{\overleftarrow\d }{\d \P_I^{\lbar\z} }\frac{\overrightarrow\d }{\d \lbar \z_{(0)-}^I}B\Bigg),\numberthis\label{poisson-def}
\eala
where $A[\F_{(0)},\P^\F]$ and $B[\F_{(0)},\P^\F]$ are arbitrary functions on the phase space ($\F_{(0)}$,$\P^\F$). The Ward identities \eqref{ward-identities} then allow us to define a constraint function on the phase space
\bala
\cc[\x,\s,\e_\pm,\l]\equiv\;& \int_{\pa\cm}d^dx\Bigg\{\x_i\Big(e_{(0)}^{ai}D_j\P_a^{\;j}-(\pa^i\vf_{(0)}^I)\P_I^\vf- (\lbar\z_{(0)-}^I\overleftarrow{\bb D}^i)\P_I^{\lbar \z}-\P_I^{\z}(\bb D^i\z_{(0)-}^I)\\
&-\P_\J^{j}\(\bb D^i\J_{(0)+j}\)-\(\lbar\J_{(0)+j}\overleftarrow{\bb D}^i\)\P_{\lbar \J}^{j}+D_j(\P_\J^j\J_{(0)+}^i+\lbar\J_{(0)+}^i\P_{\lbar\J}^j)\Big)\\
&+\s\Big[-e^a_{(0)i}\P^i_a-\cg^{IJ}\pa_I\cw\P^\vf_J-\frac12(\lbar\J_{(0)+i}\P_{\lbar\J} ^i+{\rm h.c.})\\
&+\big(\frac12\d^J_I-\pa_I\pa^J\cw\big)\big(\lbar\z_{(0)-}^I\P_J^{\lbar\z}+{\rm h.c.}\big) -|\bm e_{(0)}|\ca_{\rm W}[\F_{(0)}] \Big]\\
&+\lbar \e_+\Big(-\frac12\G^a\J_{(0)+i}\P^i_a-\frac i2\z_{(0)-}^I\P_I^\vf+\frac i2\slashed\pa\vf_{(0)}^I\P_I^{\lbar\z}+\bb D_i\P_{\lbar\J}^i\Big)\\
&+\Big(\frac12\P_a^i\lbar\J_{(0)+i}\G^a+\frac i2\P_I^\vf\lbar\z_{(0)-}^I+\frac i2\P_I^\z\slashed\pa\vf_{(0)}^I+\P_\J^i\labD_i\Big)\e_+ \\
&+\lbar\e_-\Big(i\cg^{IJ}\pa_I\cw\P_J^{\lbar\z}-\Hat\G_i\P_{\lbar\J}^i+|\bm e_{(0)}|\ca_{\rm sW}[\F_{(0)}]\Big)\\
&+\Big(\P_\J^i\Hat\G_i-i\cg^{IJ}\pa_I\cw\P_J^\z+|\bm e_{(0)}|\lbar\ca_{\rm sW}[\F_{(0)}]\Big)\e_-\\
&- \l^{ab}\Big[e_{(0)[ai}\P^i_{b]}+\frac14\big(\lbar\z_{(0)-}^I\G_{ab}\P_I^{\lbar\z}+\lbar\J_{(0)+i}\G_{ab}\P_{\lbar\J}^i+{\rm h.c.} \big) \Big]\Bigg\}, \numberthis\label{CC-ward}
\eala 
which generates the gPBH transformation \eqref{gPBH} through the Poisson bracket\footnote{It is obvious that variation of the sources can be obtained through this Poisson bracket. In appendix \ref{momenta-var-derivation} we show that the same thing holds for the canonical momenta.}
\bsub \label{PB-PBH}
\bal
& \d_{\s,\e_\pm,\l}\F_{(0)}=\{\cc[\s,\e_\pm,\l],\F_{(0)}\},\quad \d_{\s,\e_\pm,\l}\P^\F=\{\cc[\s,\e_\pm,\l],\P^\F\},\\
&\d^{(\rm cgct)}_\x\F_{(0)}=\{\cc[\x],\F_{(0)}\},\quad \d^{(\rm cgct)}_\x\P^\F=\{\cc[\x],\P^\F\}.
\eal
\esub 
Here $\d_\x^{(\rm cgct)}$ refers to the covariant general coordinate transformation (see e.g. section 11.3 of \cite{Freedman:2012zz}), under which variation of the fields is given by
\bsub 
\bal 
& \d^{(\rm cgct)}_\x e^a_{(0)i}=D_i\x^a,\quad \d^{(\rm cgct)}_\x \vf_{(0)}^I=\x^a\pa_a\vf_{(0)}^I\equiv \x^i\pa_i\vf_{(0)}^I,\\
& \d^{(\rm cgct)}_\x \J_{(0)+i}=\x^j\bb D_j\J_{(0)+i}+(D_i\x^j)\J_{(0)+j},\quad \d^{(\rm cgct)}_\x \z_{(0)-}^I=\x^a\bb D_a\z_{(0)-}^I\equiv\x^i\bb D_i\z_{(0)-}^I,
\eal 
\esub 
where $\x^a\equiv \x^i e^a_{(0)i}$. Meanwhile, $\d_\x$ given in \eqref{gPBH} is the general coordinate transformation and it is related to $\d_\x^{(\rm cgct)}$ by
\be 
\d_\x^{(\rm cgct)}=\d_\x-\d_{\l_{ab}=\o_{jab}\x^j}.
\ee 
The reason why diffeomorphism and local Lorentz transformation appear in a mixed way is that the constraint function and Poisson bracket can only give the covariant quantity but $\d_\x$ in \eqref{superdiffeo-induced} is not covariant by itself. Moreover, SUSY transformation demands the sources to be covariant and thus we are forced to see covariant general coordinate transformation rather than general coordinate transformation.

The useful variations of renormalized canonical momenta extracted from \eqref{PB-PBH} are
\bsub \label{var-e-pm-remomenta}
\bal 
& \d_{\e_+}\P_{\lbar\J}^i=\frac{\d}{\d\lbar\J_{(0)+i}}\cc[\e_+]=\frac12\P_a^i\G^a\e_+\label{susy-var-PJ}\\
& \d_{\e_-}\P_{\lbar\J}^i=\frac{\d}{\d\lbar\J_{(0)+i}}\cc[\e_-]=\frac{\d}{\d\lbar\J_{(0)+i}}\int d^dx|\bm e_{(0)}|\;\lbar\ca_{\rm sW}[\F_{(0)}]\e_-\NO \\
&\hskip0.5in=-\frac{|\bm e_{(0)}|}{\k^2}\frac{1}{8}\bb D_k\( \Big[\frac23 R_{(0)}\Hat\G_{(0)}^{ik}-R_{(0)j}{}^k\Hat\G_{(0)}^{ij}+R_{(0)j}{}^i\Hat\G_{(0)}^{kj}\Big]\e_-\)-\NO\\
&\hskip1in \underline{-\frac{|\bm e_{(0)}|}{\k^2}\frac{1}{6}\Hat\G^{ij}\bb D_j(\vf_{(0)}^2\e_-)} ,\\
&\d_{\e_+}\P_I^{\lbar\z}=\frac{\d}{\d\lbar\z_{(0)-}^I}\cc[\e_+]=\frac i2\P_I^\vf\e_+,\\
&\d_{\e_-}\P_I^{\lbar\z}=\frac{\d}{\d\lbar\z_{(0)-}^I}\cc[\e_-]=\frac{\d}{\d\lbar\z_{(0)-}^I}\int d^dx|\bm e_{(0)}|\;\lbar\ca_{\rm sW}[\F_{(0)}]\e_-=\underline{ -\frac{|\bm e_{(0)}|}{\k^2}\frac i2\pa_i\vf_{(0)}\Hat\G_{(0)}^i\e_-},\\
&\d_{\e_+}\P_I^\vf=\frac{\d}{\d\vf_{(0)} ^I}\cc[\e_+]=-\frac i2\pa_i\(\P^\z_I\Hat\G^i\e_+ \),\\
&\d_{\e_-}\P_I^\vf=\frac{\d}{\d\vf_{(0)} ^I}\cc[\e_-]=-i\pa_I(\cg^{JK}\pa_K\cw)\P_J^\z\e_-+\frac{\d}{\d\vf_{(0)}^I}\int d^dx|\bm e_{(0)}|\;\lbar\ca_{\rm sW}[\F_{(0)}]\e_-\NO \\
&\hskip0.5in= \underline{i\P^\z\e_-+ \frac{|\bm e_{(0)}|}{\k^2}\frac13\vf_{(0)}\lbar\J_{(0)+j}\labD_i\Hat\G_{(0)}^{ji}\e_-},
\eal 
\esub 
where $R_{(0)}$, $R_{(0)i}{}^j$ and $\Hat\G_{(0)}^i$ denote the Ricci scalar, Ricci tensor, Gamma matrix and determinant of the metric for the vielbein $e_{(0)i}^a$. Here the underline indicates that the terms over it are computed for the toy model. Notice that due to the super-Weyl anomaly $\e_-$ variation of the renormalized canonical momenta contain bosonic anomalous terms, which have similar origin to the Schwarzian derivative appearing in conformal transformation of the energy-momentum tensor of 2D CFT.

\subsection{BPS relations}

A bulk (bosonic) BPS configuration, which is a bosonic solution of the classical SUGRA action as well as is invariant under bulk SUSY transformation with a certain parameter, corresponds to a supersymmetric vacuum state of the dual field theory. Since vacuum expectation value (vev) of many observables are computed in SUSY field theories, it is necessary to pay a special attention to the bulk BPS solution.   Presence of the bulk BPS configuration implies that there exists a boundary SUSY parameter, under the gPBH transformation with which the fermionic sources are invariant, namely\footnote{Here we do not discuss the integrability condition of \eqref{CKS-condition}. For a discussion about some geometry of \eqref{twistor-equation}, which is also known as the twistor equation, see e.g. section 3.1 in \cite{Klare:2012gn}.}
\bsub \label{CKS-condition}
\bal
& \d_{\h}\J_{(0)+i}\equiv \d_{\h_+}\J_{(0)+i}+\d_{\h_-}\J_{(0)+i}=\bb D_i\h_+-\Hat\G_{(0)i} \h_-=0,\label{twistor-equation}\\
& \d_{\h}\z_{(0)-}^I=-\frac i2\Hat\G_{(0)}^i\pa_i\vf_{(0)}^I\h_++i\cg^{IJ}\pa_J\cw\h_-=0,
\eal 
\esub 
where the first equation is usually called as conformal Killing spinor (CKS) condition. Actually, the rigid supersymmetry of the boundary field theory is found by solving \eqref{CKS-condition} \cite{Festuccia:2011ws,Klare:2012gn,Cassani:2013dba}.\footnote{More precisely speaking, most of the rigid $\cn=1$ SUSY field theories on curved background are obtained when $U(1)$ $R$-symmetry gauge field is turned on. In this case, which is discussed in \cite{Papadimitriou:2017kzw}, the covariant derivative $\bb D_i$ in \eqref{twistor-equation} becomes $\bb D_i+igA_i$, where $g$ is the $R$-charge.}

Now we show that $\h$-variation of any renormalized canonical momenta vanishes on the BPS solution, i.e.
\be\boxed{  \label{BPS-statement}
\d_\h\P^\F\Big|_{\rm BPS}\equiv \d_{\h_+}\P^\F\Big|_{\rm BPS}+\d_{\h_-}\P^\F\Big|_{\rm BPS}=0,\quad \text{for any source }\F_{(0)}, }
\ee 
where for the fermionic operators we have from \eqref{var-e-pm-remomenta}
\bsub\label{eta-var-ops}
\bal 
& \d_\h\P_{\lbar\J}^i=\frac12\P^{i}_a\G^a\h_++\frac{\d}{\lbar\J_{(0)+i}}\int_{\S_r}d^dx|\bm e_{(0)}|\;\lbar\ca_{\rm sW}[\F_{(0)}]\h_-,\\
&\d_\h\P_I^{\lbar\z}=\frac i2\P_I^\vf\h_++\frac{\d}{\d\lbar\z_{(0)-}^I}\int_{\S_r}d^dx|\bm e_{(0)}|\;\lbar\ca_{\rm sW}[\F_{(0)}]\h_-.
\eal 
\esub 

This is in fact holographic version of that vev of any $Q$-exact operator vanishes on SUSY vacua. We only need to consider variation of the fermionic canonical momenta, since $\h$-variation of bosonic canonical momenta trivially vanishes on the bosonic solution. One can in principle see \eqref{BPS-statement} by expanding the bulk BPS equations. But since we have SUSY and super-Weyl Ward identities, the form of which are the same for all SCFTs, we take advantage of the Ward identities \eqref{ward-e+} for $\h_+$ and \eqref{ward-e-} for $\h_-$.

Taking into account the CKS condition \eqref{CKS-condition}, we obtain from the Ward identities that
\bala
0=\;&\int_{\pa\cm }d^dx\Bigg[\Big(-\frac12\lbar\J_{(0)+i}\G^a\P^i_a-\frac i2\lbar\z_{(0)-}^I\P_I^\vf-\frac i2\P_I^\z\slashed\pa\vf_{(0)}^I-\P_\J^i\labD_i\Big)\h_+\\
&\hskip1in +\Big(i\cg^{IJ}\pa_I\cw\P_J^\z-\P_\J^i\Hat\G_{(0)i}-|\bm e_{(0)}|\lbar\ca_{\rm sW}[\F_{(0)}]\Big)\h_-\Bigg]\\
=\;&\int_{\pa\cm}d^dx\Big(-\frac12\lbar\J_{(0)+i}\G^a\P_a^i\h_+-\frac i2\P_I^\vf \lbar\z_{(0)-}^I\h_+-|\bm e_{(0)}|\lbar\ca_{\rm sW}[\F_{(0)}]\h_-\Big).\numberthis\label{inter-susy-ward}
\eala
We emphasize that because the Ward identities are valid for any background, \eqref{inter-susy-ward} holds at least to linear order in fermions for any value of $\lbar \J_{(0)+i}$ and $\lbar\z_{(0)-}^I$ as long as the bosonic sources admit the CKS. There might be correction at order of $O\Big((\J_{(0)+})^2,(\z_{(0)-})^2\Big)$, though. Note that non-trivial dependence of bosonic momenta $\P_a^i$ and $\P_I^\vf$ on the fermionic sources occurs from the quadratic order in fermions, i.e.
\be 
\frac{\d}{\d\lbar\J_{(0)+i}}\P_a^i\Bigg|_{\J_{(0)+i}=\z_{(0)-}^I=\cdots=0}=0
\ee  
and so on. Therefore, by taking the functional derivative of \eqref{inter-susy-ward} with respect to the fermionic sources and evaluating on the (bosonic) supersymmetric background, we obtain the (bosonic) identities
\bsub\label{BPS-consequence}
\bal 
& \frac12 \P^{i}_a\G^a\h_++\frac{\d}{\lbar\J_{(0)+i}}\int_{\S_r}d^dx|\bm e_{(0)}|\;\lbar\ca_{\rm sW}[\F_{(0)}]\h_-=0,\\
&-\frac i2\P_I^\vf\h_+-\frac{\d}{\d\lbar\z_{(0)-}^I}\int_{\S_r}d^dx|\bm e_{(0)}|\;\lbar\ca_{\rm sW}[\F_{(0)}]\h_-=0,
\eal 
\esub 
where we used \eqref{pij-pai}. Therefore, we find that on the BPS backgrounds
\be 
\d_{\h}\P_{\lbar\J}^i=0,\quad \d_{\h}\P_I^{\lbar\z}=0,
\ee 
which confirms our claim.

From the field theory point of view, \eqref{BPS-statement} is quite natural, since supersymmetric vacua are annihilated by the preserved supercharge $Q$. It has, however, a deep implication. We should first emphasize that $\d_\h$ corresponds to the variation of quantum operators acting on Hilbert space and is different from the 'classical' SUSY variation which is considered in the context of SUSY localization. For instance $\P^{ij}\Hat\G_{(0)j}\h_+$, which is classically $Q$-exact here, has non-zero vev due to the anomalous contribution. This implies that the classical SUSY variation \emph{cannot} become as a total derivative in the path integral, so long as the anomalous terms in \eqref{BPS-consequence} do not vanish, see some comments on the assumption that SUSY should not be anomalous, in SUSY localization reviews such as \cite{Marino:2011nm,Cremonesi:2014dva}.

In order to convince ourselves, let us check \eqref{BPS-statement} for the toy model. First, let us remind that in the toy model, $d=4$ and scaling dimension of $\vf$ is 3. Then, \eqref{BPS-consequence}s become
\bal 
0=\;&-\G^a\h_+\P^{i}_a+\frac{1}{\k^2}\frac{1}{4(d-2)^2}\bb D_k\Big[\big(\frac {d}{d-1} R_{(0)}\Hat\G_{(0)}^{ki}-2R_{(0)j}{}^k\Hat\G_{(0)}^{ji}+2R_{(0)j}{}^i\Hat\G_{(0)}^{jk}\big)\h_-\Big]\NO \\
& +\frac{|\bm e_{(0)}|}{\k^2}\frac{1}{2(d-1)}\vf_{(0)}^2\Hat\G_{(0)}^{ij}\bb D_j\h_-,\label{bS-sC}\\
0=\;&-\frac i2\h_+\P^\vf+\frac{|\bm e_{(0)}|}{\k^2}\frac i2\Hat\G_{(0)}^i\h_-\pa_i\vf_{(0)}.\label{bS-Hyp}
\eal 
By combining \eqref{bS-Hyp} with the conformal Killing spinor equation for the toy model
\bsub 
\bal 
& \bb D_i\h_{+}=\Hat\G_{(0)i}\h_{-},\\
& \frac12\Hat\G_{(0)}^i\pa_i\vf_{(0)}\; \h_{+}+\vf_{(0)}\h_{-}=0,
\eal 
\esub 
we get
\be \label{vev-pvf}
-\vf_{(0)} \P^\vf+\frac{|\bm e_{(0)}|}{2\k^2}\pa_i\vf_{(0)}\pa^i\vf_{(0)}=0.
\ee 
This 'strange-looking' formula can be verified in the toy model, by using the bulk BPS equation.

From the bulk BPS equation for $\z$ with the bulk SUSY parameter $\Hat\e$
\be 
\d_{\Hat \e}\z=\(\slashed\pa\vf-\cw^\prime\)\Hat\e=0,\quad \cw'\equiv \frac{d}{d\vf}\cw(\vf),
\ee 
one can obtain
\be 
\dot\vf=-\sqrt{(\cw^\prime)^2+\pa_i\vf\pa^i\vf},
\ee 
where we fix the sign from leading asymptotics of $\vf$. It then follows from the definition of $\p^\vf$ that
\be 
\p^\vf=-\frac{\sqrt{-\g}}{\k^2}\dot\vf=\frac{\sqrt{-\g}}{\k^2}\sqrt{(-\cw^\prime)^2+\pa_i\vf\pa^i\vf}.
\ee 

On the other hand, the full bosonic counterterms are given by
\be 
S_{\rm ct}=\frac{1}{\k^2}\int d^dx\sqrt{-\g}\Big[\cw-\frac14 R-\frac12 \log e^{-2r}\Big(\frac{1}{6}\vf^2 R+\pa_i\vf\pa^i\vf+\cdots\Big)\Big], 
\ee 
where the ellipsis denote the terms which does not depend on $\vf$. The counterterms for the canonical momenta $\p_{\rm ct}^\vf$ is then given by
\be 
\p^\vf_{\rm ct}=\frac{\d}{\d\vf}S_{\rm ct}=\frac{\sqrt{-\g}}{\k^2}\Big[-(-\cw^\prime)-\frac12\log  e^{-2r}\Big(\frac13\vf R-2\Box\vf\Big)\Big].\label{toy-pvf-counter}
\ee
Furthermore, from the conformal Killing spinor condition \eqref{CKS-condition}, we obtain
\be 
0=\Big(\Box_{(0)}\vf_{(0)}-\frac{1}{6}\vf_{(0)}R_{(0)}\Big)\h_{+},
\ee 
which implies that the logarithmically divergent terms in \eqref{toy-pvf-counter} actually do not contribute to the counterterms. Eventually, the renormalized canonical momentum $\P^\vf$ becomes
\be 
\P^\vf=\frac{1}{\k^2}\lim_{r\rightarrow+\infty}e^{-3r}\sqrt{-\g}\frac{\pa_i\vf\pa^i\vf }{\sqrt{(-\cw^\prime)^2+\pa_i\vf\pa^i\vf}+(-\cw^\prime)}=\frac{|\bm e_{(0)}|}{2\k^2}\frac{\pa_i\vf_{(0)}\pa^i\vf_{(0)}}{\vf_{(0)}},
\ee 
which confirms the result \eqref{vev-pvf} as well as the anomalous SUSY variation of the renormalized canonical momenta \eqref{var-e-pm-remomenta}.

\subsection{Conserved charges and supersymmetry algebra with anomaly correction}

We recall that given a Killing vector $\x^i$ which satisfies the Killing condition\footnote{$g_{(0)ij}\equiv e^a_{(0)i}e_{(0)aj}$ is the induced metric on the boundary $\pa\cm$.}
\bsub 
\bal
& \cl_\x g_{(0)ij}=D_{(0)i}\x_j+D_{(0)j}\x_i=0,\\
& \cl_\x\vf^I_{(0)}=\x^i\pa_i\vf^I_{(0)}=0,\\
& \cl_\x\z_{(0)-}^I=\x^i\bb D_{(0)i}\z_{(0)-}^I+\frac14 D_{(0)i}\x_j\Hat\G^{ij}_{(0)}\z_{(0)-}^I=0,\\
& \cl_\x\J_{(0)+j}=\x^i\bb D_{(0)i}\J_{(0)+j}+(D_{(0)j}\x_i)\J_{(0)+}^i+\frac14D_{(0)k}\x_l\Hat\G^{kl}_{(0)}\J_{(0)+j}=0,
\eal
\esub 
we obtain a conservation law by combining \eqref{ward-diffeo} with \eqref{ward-lorentz}, namely
\be \label{conservation-diffeo}
D_i\[e^a_j\x^j\P_a^i+\x^j(\P_\J^i\J_{+j}+\lbar\J_{+j}\P_{\lbar\J}^i)\]=0.
\ee 
Note that we use the Kosmann's definition for the spinorial Lie derivative (see e.g. \cite{Figueroa-OFarrill:1999klq} and (A.11) of \cite{Dumitrescu:2012ha}\footnote{In many literatures including \cite{Figueroa-OFarrill:1999klq}, the spinoral Lie derivative is defined by $\cl_\x\z=\x^i\bb D_i\z-\frac14 D_i\x_j\Hat\G^{ij}\z$. The sign of the last term is minus, since the Gamma matrices there follow Grassman algebra in Euclidean signature, while here we use the Minkowskian signature.}) and the Lie derivative is related to gPBH transformations by
\be 
\cl_\x =\d^{\rm (cgct)}_\x+\d_{\l_{ab}=-e^i_a e^j_bD_{[i}\x_{j]}}.
\ee 
We emphasize that \eqref{conservation-diffeo} holds for any background. The conservation law \eqref{conservation-diffeo} allows us to define a conserved charge associated with $\x^i$, namely \cite{Papadimitriou:2005ii,An:2016fzu}
\be\boxed{  \label{conserved-charge-x}
\cq[\x]\equiv \int_{\pa\cm \cap \cc}d\s_i\; \(e^a_j\P^i_a+\P_\J^i\J_{+j}+\lbar\J_{+j}\P_{\lbar\J}^i\)\x^j,}
\ee 
which is independent on the choice of Cauchy surface $\cc$. Note that the conserved charge $Q_\x$ is related to the constraint function by
\be 
\cq[\x]=\cc[\x,\l_{ab}= -e^i_a e^j_bD_{[i}\x_{j]}].
\ee 

We have other conservation laws
\be \label{conservation-supercurrent}
D_i(\P_\J^i\h_+)=D_i(\lbar\h_+\P_{\lbar\J}^i)=0,
\ee
which follow from the SUSY and super-Weyl Ward identities \eqref{ward-e+} and \eqref{ward-e-} for the CKS parameters $\h_+$ and $\lbar\h_+$. Note that the conservation laws \eqref{conservation-supercurrent} hold only on the bosonic background. This allows us to define conserved charges
\be 
Q^s[\h_+]\equiv \int_{\pa\cm\cap\cc}d\s_i\;\P_\J^i\h_+,\quad Q^s[\lbar\h_+]\equiv \int_{\pa\cm\cap\cc}d\s_i\;\lbar\h_+\P_{\lbar\J}^i.
\ee 
On the bosonic background we can identify these conserved charges with the constraint functions, namely
\be 
Q^s[\h_+]=\cc[\h_+,\h_-],\quad Q^s[\lbar\h_+]=\cc[\lbar\h_+,\lbar\h_-].
\ee  
It then follows from \eqref{var-e-pm-remomenta} that on the bosonic background we have
\bal
\{Q^s[\h_+],Q^s[\lbar\h_+]\}\Big|_{\rm Bosonic}=\;&\int_{\pa\cm\cap\cc}d\s_i\;\lbar\h_+\{\cc[\h_+,\h_-],\P_\J^i\}\Big|_{\rm Bosonic}\NO\\
=\;&\int_{\pa\cm\cap\cc}d\s_i\;\[\frac12\P_a^i\lbar\h_+\G^a\h_++\lbar{\h}_+\Big(\frac{\d}{\d\lbar\J_{(0)+i}}\int_{\pa\cm}d^dx|\bm e_{(0)}|\lbar\ca_{\rm sW}\h_-\Big)  \]_{\rm Bosonic}.
\eal 

In the case where the conformal Killing vector\footnote{One can easily check $\ck^i$ satisfies the conformal Killing condition, by using \eqref{CKS-condition}.} 
\be\label{CKV} 
\ck^i\equiv i\lbar{\h}_+\Hat\G^i\h_+ 
\ee 
becomes a Killing vector, we can see that on the bosonic background the above commutator becomes
\be \label{Qeta-Qeta-commutator}
\{Q^s[\h_+],Q^s[\lbar\h_+]\}=-\frac i2\cq[\ck]+\int_{\pa\cm\cap\cc}d\s_i\;\lbar{\h}_+\Big(\frac{\d}{\d\lbar\J_{(0)+i}}\int_{\pa\cm}d^dx|\bm e_{(0)}|\lbar\ca_{\rm sW}\h_-\Big).
\ee 
Not surprisingly, the super-Weyl anomaly corrects the supersymmetry algebra, too.

We can obtain other commutators such as $\{\cq[\x],Q^s[\h_+]\}$. It is possible because $\cq[\x]$ for the Killing vector $\x^i$ is conserved for any background so that
\bala
\int_{\pa\cm\cap\cc}d\s_i\;\{\cq[\x],\P_\J^i\}\h_+=\;&\int_{\pa\cm\cap\cc}d\s_k\;\{\cc[\x,\l_{ab}=- e^i_a e^j_b D_{[i}\x_{j]}],\P_\J^k\}\h_+ \\
=\;&\int_{\pa\cm\cap\cc}d\s_i\Big[-\P_\J^i\cl_\x\h_++D_j[(\x^j\P_\J^i-\x^i\P_\J^j)\h_+]+\x^iD_j(\P_\J^j\h_+)\Big],
\eala
where the second term vanishes by using Stokes' theorem. The third term is also zero on the bosonic background, due to the conservation law. Therefore, we have
\be \label{commutator-xsi-eta}
\{\cq[\x],Q^s[\h_+]\}=-\int_{\pa\cm\cap\cc}d\s_i\;\P_\J^i\cl_\x\h_+=-Q^s[\cl_\x\h_+],
\ee 
and in the same way
\be \label{commutator-xsi-etabar}
\{\cq[\x],Q^s[\lbar\h_+]\}=-\int_{\pa\cm\cap\cc}d\s_i\;(\lbar\h_+\overleftarrow\cl_\x)\P_{\lbar\J}^i=- Q^s[\lbar\h_+\overleftarrow\cl_\x],
\ee 
since $\cl_\x\h_+$ and $\lbar\h_+\overleftarrow\cl_\x$ become conformal Killing spinors \cite{Figueroa-OFarrill:1999klq}, i.e.
\be 
\bb D_i(\cl_\x\h_+)=\frac 1d\Hat\G_i\Hat\G^j\bb D_j(\cl_\x\h_+),\quad (\lbar\h_+\overleftarrow\cl_\x)\labD_i=\frac 1d(\lbar\h_+\overleftarrow\cl_\x)\labD_j\Hat\G^j\Hat\G_i.
\ee
We note that \eqref{commutator-xsi-eta} and \eqref{commutator-xsi-etabar} can be obtained in the way around, namely by computing
\be
\{Q^s[\h_+],e^a_j\P^i_a+\P_\J^i\J_{+j}+\lbar\J_{+j}\P_{\lbar\J}^i\},\quad \{Q^s[\lbar\h_+],e^a_j\P^i_a+\P_\J^i\J_{+j}+\lbar\J_{+j}\P_{\lbar\J}^i\}.
\ee 

In summary, the supersymmetry algebra on the curved (bosonic) background is
\bal\label{curved-susy-algebra}
\boxed{ 
\begin{aligned}
& \{Q^s[\h_+],Q^s[\lbar\h_+]\}=-\frac i2\cq[\ck]+\int_{\pa\cm\cap\cc}d\s_i\;\lbar{\h}_+\Big(\frac{\d}{\d\lbar\J_{(0)+i}}\int_{\pa\cm}d^dx|\bm e_{(0)}|\lbar\ca_{\rm sW}\h_-\Big),\\
& \{\cq[\x],Q^s[\h_+]\}=-Q^s[\cl_\x\h_+],\\
&\{\cq[\x],Q^s[\lbar\h_+]\}=- Q^s[\lbar\h_+\overleftarrow\cl_\x].
\end{aligned}
}
\eal 
\eqref{curved-susy-algebra} closely resembles the SUSY algebra presented in the literatures (see e.g. \cite{Festuccia:2011ws,Martelli:2013aqa,Cassani:2014zwa}), except for the super-Weyl anomaly-effect term.

We comment that \eqref{curved-susy-algebra} can be obtained without using Poisson bracket, but in an equivalent and rather simple way. Recall that a symmetry of the field theory leads to a conservation of the corresponding (anomalous) Noether current $J^i$ (with the anomaly $\ca_J$)
\be 
D_i J^i=\ca_J,
\ee
from which we derive the variation of any operator $\co$ under the symmetry transformation (see e.g. (2.3.7) in \cite{Polchinski:1998rq}), namely
\be 
\d \co(x)+\int_{\pa\cm} d^dy\;[D_i J^i(y)-\ca_J(y)]\co(x)=0,
\ee 
where the second term can be computed by differentiating the relevant Ward identities with the source dual to operator $\co(x)$. Now one can readily see that the commutator of charges becomes
\be\label{algebra-prescription} 
\{Q_1,Q_2\}=\int_{\pa\cm\cap\cc}d\s_i\; (\d_1 J^i_2)=-\int_{\pa\cm\cap\cc}d\s_i\(\int_{\pa\cm} d^dy\;[D_j J^j_1(y)-\ca_J(y)]J^i_2\),
\ee 
and this prescription gives the same result with \eqref{curved-susy-algebra}. See e.g. appendix \ref{susyalgebra-without-Poisson} for derivation of $\{\cq[\x],Q^s[\h_+]\}$.

Now that we know from the last section that the LHS of \eqref{Qeta-Qeta-commutator} vanishes on BPS backgrounds, we can conclude that the conserved charge associate with $\ck^i$ on BPS backgrounds is totally fixed to be a functional derivative of the fermionic anomaly, namely
\be\boxed{  \label{susy-charge-identity}
\cq[\ck]\Big|_{\rm BPS}=-2i\int_{\pa\cm\cap \cc}d\s_i\;\lbar{\h}_+\Bigg\{\frac{\d}{\d\lbar\J_{(0)+i}}\int_{\pa\cm}d^dx|\bm e_{(0)}|\lbar\ca_{\rm sW}\h_-\Bigg\}.}
\ee 
Depending on the theory, $\ck^i$ can be combination of other Killing vectors such as $\pa_t$ and angular velocity. If this is the case, \eqref{susy-charge-identity} can be regarded as a relation of the conserved charges on the supersymmetric background, but accompanied with anomalous contribution. A similar relation is found in \cite{Papadimitriou:2017kzw}, which explains the discrepancy of the BPS condition (see e.g. (C.16) of \cite{Genolini:2016ecx})
\be 
\braket{H}+\braket{J}+\g\braket{Q}=0,
\ee  
for pure $AdS_5$ is precisely due to the anomalous contribution coming from the fermionic anomalies.

\section{Neumann boundary condition}\label{Neumann-BC}
\setcounter{equation}{0}

Most of computations so far are for plus sign choice of \eqref{GH-Z} at the beginning of section \ref{radial-Hamiltonian-Holography}. This plus sign is actually equivalent to imposing Dirichlet boundary condition on the spin $1/2$ field $\z$. Independently from this choice, we could determine the leading asymptotics of the scalar field, as emphasized before. This allows us to use the result of appendix \ref{asymptotics-fermion} and \ref{gPBH} to conclude that minus sign choice can be supersymmetric only when mass of its scalar SUSY-partner field belongs to the window \cite{Breitenlohner:1982bm,Balasubramanian:1998sn,Klebanov:1999tb}
\be 
-\(\frac d2\)^2\leq m^2\leq -\(\frac d2\)^2+1.
\ee 

In this window \eqref{GH-Z} is already finite, implying that the canonical momenta for $\z_-$ is not renormalized. Since $\z_+$ by itself becomes the renormalized canonical momentum, change of the sign from plus to minus is in fact Legendre transformation of the renormalized on-shell action $\Hat S_{ren}$, which is equivalent to impose Neumann boundary condition on $\z_-$ \cite{Papadimitriou:2007sj}. We have seen that $\Hat S_{ren}$ in the case of plus sign choice is ($\e_+$) supersymmetric (Dirichlet boundary condition for scalar field was implicitly imposed). Therefore, in order to preserve SUSY, one can expect that boundary condition for the scalar field should also be changed from Dirichlet to Neumann by Legendre transformation.

To see this, one has to prove that the total Legendre transformation action
\be 
S_L=-\int_{\S_r}(\Hat{\p}^\z \z_-+\lbar\z_-\Hat\p^{\lbar\z}+\vf \Hat\p^\vf) ,\quad \Hat\p^{\lbar\z}=\frac{\sqrt{-\g}}{\k^2}\z_+ 
\ee 
is invariant under $\e_+$ transformation. Note that variation of $\P^{\lbar\z}_I$ gives directly how gPBH transformations act on $\z_+$. We again consider only one scalar field, and it is straightforward to extend the result here to the case for several scalar fields. From \eqref{var-e-pm-remomenta}, one can find that the action of $\e_+$ on $S_L$ gives
\bala
\d_{\e_+}S_L\sim \;&-\int_{\S_r} \Big(\frac i2\Hat\p^\vf\lbar\z_-\e_+ -\frac i2\pa_i\vf\lbar\e_+\Hat\G^i\Hat\p^{\lbar\z}-\frac i2\Hat\p^\vf\lbar\z_-\e_+-\frac i2\vf\pa_i(\lbar\e_+\Hat\G^i\Hat\p^\vf)+{\rm h.c.}\Big)=0.
\eala
This confirms that the total action $S+S_L$ for the Neumann boundary condition is still invariant under $\e_+$ transformation.

When it comes to $\e_-$ variation of $S_L$, one can find that all the momenta-related terms are canceled, as before. The anomalous terms in $\e_-$ variation of the renormalized canonical momenta, however, are not canceled but contribute to $\e_-$ anomaly of $S+S_L$, together with $\ca_{\rm sW}$. Namely, we obtain for the toy model that
\be 
\d_{\lbar\e_-}(S+S_L)\sim \int_{\S_r} d^dx\sqrt{-\g}\;\lbar\e_-\(\ca^{(G)}_{\rm sW}-\frac{1}{6\k^2}\vf^2\Hat\G^{ij}\bb D_i\J_{+j}\)\equiv 
\int_{\S_r} d^dx\sqrt{-\g}\;\lbar\e_-\ca_{\rm sW}^N,
\ee 
where the super-Weyl anomaly for Neumann boundary condition is
\be 
\ca_{\rm sW}^N=\ca^{(G)}_{\rm sW}-\frac{1}{6\k^2}\vf^2\Hat\G^{ij}\bb D_i\J_{+j}.
\ee 

\section{Concluding remarks}
\label{conclusion}
\setcounter{equation}{0}

In this work we have considered a generic $\cn=2$ 5D supergravity with its fermionic sector in the context of holographic renormalization, through which we have obtained a complete set of supersymmetric counterterms. We have also found that scalars and their superpartners should satisfy the same boundary condition in order for the theory to be consistent with SUSY.

The Ward identities \eqref{ward-identities} and the anomalies lead to rather remarkable consequences. By means of them, we showed that SUSY transformation of operators and SUSY algebra of a theory which has $\cn=1$ 4D SCFT in curved space as a UV fixed point become anomalous at the quantum level, see \eqref{eta-var-ops} and \eqref{curved-susy-algebra}. We comment that once the $R$-symmetry gauge field is turned on, $R$-charge and the related terms appear on RHS of the first line \eqref{curved-susy-algebra}. Note that the anomalous terms are non-vanishing in general on curved backgrounds, even where all anomalies vanish.

\eqref{BPS-statement} stating that $Q$-exact operator has vanishing vev on the SUSY vacua, namely
\be 
\braket{\d_Q \co}=\braket{\d_Q^{\rm cl}\co}+(\rm Quantum\; correction)=0,
\ee
implies that the classical $Q$-variation cannot be a total derivative in the path integral unless the quantum correction vanishes, otherwise we have for instance
\be 
\braket{\d_Q^{\rm cl}\cs^i}=\int [D\f]\;\(\d_Q^{\rm cl}\cs^i\) e^{-S}=\int [D\f]\;\d_Q^{\rm cl}\(\cs^i e^{-S}\)=0.
\ee 
This is very important, since SUSY localization technique are justified only when the classical $Q$-variation is a total derivative in the path integral. See also footnote 10 in \cite{Witten:1988ze} stating a fundamental assumption used in the SUSY localization principle.

We should remark the importance of finding SUSY completion of the Weyl anomaly, because the supersymmetric Weyl anomaly automatically satisfies all the requirements for higher-derivative SUGRA in the superconformal way. For construction of higher-derivative SUGRA, see \cite{Kaku:1977rk} and some related references.

We emphasize that our whole analysis here crucially relies on existence of the superpotential $\cw$. If the theory does not possess any superpotential, one could introduce local and \emph{approximate} superpotential which is sufficient for reproducing all divergent terms of the scalar potential, as done in \cite{Bobev:2013cja}. Now one can see that the approximate superpotential should meet more restrictive criterion for the supersymmetric holographic renormalization. To make this point clear, let us discuss about the approximate superpotential suggested in \cite{Bobev:2013cja}, see (5.15) there. One can find from the BPS equations (3.20) and (3.25) and the algebraic equation (3.26) in \cite{Bobev:2013cja} that the BPS solution's flow to leading order is
\bsub 
\bal
& \frac{d\j}{dr}\sim-\j ,\\
& \frac{d\vf}{dr}\sim-\(2\vf+\sqrt{\frac 23}\j^2\),\\
& \frac{d\c}{dr}\sim-2\c\(1+\frac{\j^2}{\sqrt 6\vf}\),
\eal  
\esub 
where RHS of the last equation is a \emph{non-local} function of $\vf$ around $\vf=0$. Hence it is impossible to find the \emph{local and approximate} superpotential consistent with the BPS flow equations, which means that we need more generic $\cn=2$ gauged SUGRA model to study \cite{Bobev:2013cja}. Notice that this inconsistency of the approximate superpotential with the BPS flow equations imply that the supertential suggested in \cite{Bobev:2013cja} is \emph{not approximate} for the fermionic sector of SUGRA.

As long as there exists a superpotential (or at least approximate one for the whole sector of SUGRA), many of our results here can be extended straightforwardly to other dimensions. A direct application of the analysis of this paper to other dimensions is to obtain 2D super-Virasoro algebra with central extension. Let us explain this here schematically. The super-Weyl anomaly in 2D SCFT can be easily found by using the trick of section \ref{WZ-condition}, namely that the SUSY variation of the super-Weyl anomaly is equal to the Weyl anomaly. Since the Weyl anomaly is $e^a_i\ct^i_a=\frac{c}{24\p}R$, we see immediately that the super-Weyl anomaly in 2D is $\G_i\cs^i\sim i \frac{c}{24\p}\G^{ij}\bb D_i\J_j$ up to a constant coefficient, depending on the convention. It follows that the anomalous variation of the super-current operator is
\be \label{anomalous-var-2D-supercurrent}
\d_\h\cs^i=-\frac i4\G^a\h\ct_a^i+\frac{ic}{48\p}\Hat\G^{ij}\Hat\G^k\bb D_j\bb D_k\h,
\ee 
where $\h$ is the 2D CKS, satisfying the condition
\be \label{2D-CKS}
\Hat\G^i\bb D_i\h=\frac12\Hat\G_i\Hat\G^j\bb D_j\h,\quad  \text{or}\quad\Hat\G^j\Hat\G^i\bb D_j\h=0.
\ee 
Note that the anomalous term in \eqref{anomalous-var-2D-supercurrent} vanishes only when the 2D Ricci scalar $R=0$ and $\h$ is a spinor, all second derivatives of which vanish. Since \eqref{2D-CKS} gives infinite number of solutions, as 2D conformal Killing vector equation, one gets infinite number of conserved super-charges $G_r$, which are added to Virasoro algebra to form the super-Virasoro algebra. Now one can see that the central extension in (see e.g. (10.2.11b) in \cite{Polchinski:1998rr})
\be 
\{G_r,G_s\}=2L_{r+s}+\frac{c}{12}(4r^2-1)\d_{r,-s}
\ee 
of the super-Virasoro algebra in 2D flat background is derived from the anomalous term of \eqref{anomalous-var-2D-supercurrent}.

One should keep in mind, however, that since representation of the spinor fields strongly depends on the dimension of spacetime it might not be easy to put the SUGRA action into the form of \eqref{SUGRA-action} in other (especially odd) dimensions.

\section*{Acknowledgments}

I would like to thank Jin U Kang and Ui Ri Mun for interesting discussions. I am grateful to Jin U Kang for a careful reading and important comments on the draft.

\appendix

\renewcommand{\thesection}{\Alph{section}}
\renewcommand{\theequation}{\Alph{section}.\arabic{equation}}

\section*{Appendices}
\setcounter{section}{0}

\section{Notation, conventions for Gamma matrices and useful identities}\label{gamma-matrices}
\setcounter{equation}{0}

Throughout this paper Greek indexes $\m,\n$ and $\a,\b,\cdots$ refer to the coordinate and flat directions in the bulk respectively, and the Latin indexes $i,j,m,n,p,q,\cdots$ and $a,b,\cdots$ refer to the coordinate and flat directions on the slice respectively. The flat indices which correspond to radial like and time like directions are special, so we denote them by $\bar r$ and $\bar t$ respectively. The capital Latin letters $A,B,\cdots$ indicate the coordinate directions on the scalar and hyperino manifold. $\nabla_\m $, $D_i$ and $\bb D_i$ refer to the covariant derivative in the bulk and the covariant derivatives of the bosonic and fermionic fields on the slice sequentially.

We use the hermitian representation of the Lorentzian Gamma matrices, following the convention in \cite{Freedman:2012zz}. $\G^\a$ and $\G^a$ indicate the Gamma matrices along the flat directions in the bulk and the boundary, while $\G^\m$ and $\Hat\G^i$ refer to the Gamma matrices along the coordinate directions in the bulk and the boundary. The relations between these Gamma matrices are provided in appendix \ref{ADM-decomposition}. Both in the bulk and on the boundary the hermitian conjugation of the Gamma matrix is given by
\be 
\G^{\m\dagger}=\G^{\bar t}\G^\m\G^{\bar t},\quad \Hat\G^{i\dagger}=\G^{\lbar t}\Hat\G^i\G^{\bar t}.
\ee  

The following formulas, which hold in any $D$ dimensional spacetime (see e.g. section 3 in \cite{Freedman:2012zz}), are frequently used in this paper.
\bsub 
\bal
& \G^{\m\n\r}=\frac12\{\G^\m,\G^{\n\r}\},\\
& \G^{\m\n\r\s}=\frac12[\G^\m,\G^{\n\r\s}],\\
&\G^{\m\n\r}\G_{\s\t}=\G^{\m\n\r}{}_{\s\t}+6\G^{[\m\n}{}_{[\t}\d ^{\r]}{}_{\s]}+6\G^{[\m}\d^\n{}_{[\t}\d^{\r]}{}_{\s]},\\
& \G^{\m\n\r\s}\G_{\t\l}=\G^{\m\n\r\s}{}_{\t\l}+8\G^{[\m\n\r}{}_{[\l}\d^{\s]}{}_{\t]}+12\G^{[\m\n}\d^\r{}_{[\l}\d^{\s]}{}_{\t]},\\
& [\G_{\m\n},\G_{\r\s}]=2(g_{\n\r}\G_{\m\s}-g_{\m\r}\G_{\n\s}-g_{\n\s}\G_{\m\r}+g_{\m\s}\G_{\n\r} ) ,\\
& \G^{\m\n\r}\G_\r=(D-2)\G^{\m\n},\\
& \G^{\m\n\r}\G_{\r\s}=(D-3)\G^{\m\n}{}_\s+2(D-2)\G^{[\m}\d^{\n]}{}_\s,\\
& \G^{\m\n}\nabla_\m\nabla_\n\z =-\frac14 R\z,\\
&\G^{\m\n\r}\nabla_\n\nabla_\r\z=-\frac14(R\G^\m-2R_\n{}^\m \G^\n)\z,  
\eal 
\esub 
where $\d$ refers to the Kronecker delta.

There are left and right acting functional derivatives with respect to fermionic variable $\j$, namely
\be 
\frac{\overrightarrow \d}{\d\lbar \j},\quad \frac{\overleftarrow\d}{\d\j},
\ee  
and in most cases the rightarrow symbol $\rightarrow$ are omitted. Here $\lbar\j$ denotes the Dirac adjoint of the spinor $\j$, namely
\be 
\lbar\j\equiv \j^\dagger(i\G^{\bar t}).
\ee 

The affine connection $\G^\m_{\n\r}$ is related to spin connection by (see e.g. (7.100) in \cite{Freedman:2012zz})
\be 
\G^\r_{\m\n}=E^\r_\a(\pa_\m E^\a_\n+\o_\m{}^\a{}_\b E^\b_\n).
\ee 
In this work we consider the supergravity theory in the second order formalism. This means that our theory is torsionless and thus the spin connection is reduced into
\be
\o_{\m\a\b} =E_{\n\a}\pa_\m E_\b^\n+\G^{\r}_{\m\n}E_{\r\a}E_\b^\n.
\ee 
Variation of the torsionless spin connection
\be 
\d\o_{\m\a\b}=E^\n_{[\a}D_\m\d E_{\b]\n}-E^\n_{[\a}D_\n\d E_{\b]\m}+e^\r_\a E^\n_\b E_{\g\m}D_{[\n}\d E^\g_{\r]}.
\ee 
is useful for many of our computations. The covariant derivative of the fermionic fields are given by
\bal
\nabla_\m\J_\n=\;&\pa_\m\J_\n+\frac14\o_{\m\a\b}\G^{\a\b}\J_\n-\G^\r_{\m\n}\J_\r ,\\
\nabla_\m\z^I=\;&\pa_\m\z^I+\frac14\o_{\m\a\b}\G^{\a\b}\z^I.
\eal

\section{ADM decomposition and generalized PBH transformation}
\label{ADM-SBC}
\setcounter{equation}{0}

A preliminary step of the Hamiltonian analysis of the gravitational theory is to decompose the variables of theory including the metric (or the vielbeins) into a radial-like (or time-like) direction and  the other transverse directions (a.k.a. ADM decomposition \cite{Arnowitt:1960es}). Coupling gravity to spinor fields require vielbeins to appear in the action explicitly and thus the ADM decomposition of the vielbeins instead of the metric should be done.

The ADM decomposition brings us a natural choice of the gauge for variables of the theory, which is called as the Fefferman-Graham (FG) gauge. In the FG gauge, the Hamiltonian analysis becomes much simple.

\subsection{ADM decomposition of vielbein and the strong Fefferman-Graham gauge}

We begin with picking up a suitable radial coordinate $r$ and doing the ADM decomposition of the metric to run the Hamiltonian formalism. Since the vielbein explicitly appears in the action through the covariant derivative of the spinor fields we need to decompose the vielbein itself rather than the metric.

Choosing the radial coordinate $r$, we describe the bulk space as a foliation of the constant $r$-slices, which we denote by $\S_r$. Let $E^\a$ be vielbeins of the bulk and we decompose them as
\be
E^\a=\left(N n^\a+N^j e_j^\a\right)dr+e_j^\a dx^j,
\ee
such that
\be
g_{\m\n}=E_\m^\a E_\n^\b\h_{\a\b},\quad \g_{ij}=e_i^\a
e_j^\b\h_{\a\b},\quad n_\a e_i^\a=0,\quad \h_{\a\b}n^\a n^\b
=1,
\ee
where $\a,\b$ are bulk tangent space indices and $\h=\diag(1,-1,1,\ldots,1)$ (where $\h_{\bar t\bar t }=-1$). Note that $N$ and $N^\a$ are called as \emph{lapse} and \emph{shift} respectively. One can check that
\be 
ds^2\equiv g_{\m\n}dx^\m dx^\n=(N^2+N^i N_i)dr^2+2N_i dr dx^i+\g_{ij}dx^idx^j,
\ee 
which usually appears in the textbook. The inverse vielbeins are then given by 
\be
E_\a^r=\frac{1}{N}n_\a,\quad E_\a^i=e_\a^i-\frac{N^i}{N}n_\a.
\ee
It follows that 
\be
\G^r=\G^\a E_\a^r=\frac1N n_\a\G^\a\equiv \frac1N \G.
\ee
The extrinsic curvature on the radial slice $\S_r$ is defined as
\be 
K_{ij}\equiv \frac{1}{2N}(\dot\g_{ij}-D_i N_j-D_j N_i)
\ee 
and $K\equiv \g^{ij}K_{ij}$. Moreover, 
\be
\G^i=\G^\a E_\a{}^i=\Hat\G^i-\frac{N^i}{N}\G,
\ee
where $\Hat\G^i\equiv \G^\a e_\a^i$. These vielbeins satisfy the relation 
\be
e_\a^i e_i^\b+n_\a n^\b=\d_\a^\b.
\ee
One can also see that $\Hat\G^i$s satisfy the Clifford algebra on the slice and $\G$ anticommutes with all $\Hat\G^i$s, i.e.
\be 
\{\Hat\G^i,\Hat\G^j\}=2\g^{ij},\quad \{\Hat\G^i,\G\}=0.
\ee 
It follows that the matrix $\G$ can be used to define the 'radiality' (see e.g. \cite{Freedman:2016yue}) on the slice, so that, a generic spinor $\j$ on the slice can be split into two by radiality \footnote{When $d=D-1$ is even number, radiality can be regarded as chirality.},
\be\label{radiality}
\j_\pm\equiv \G_\pm \j,
\ee
where $\G_\pm\equiv \frac12\(1\pm\G\)$. 

We remind that splitting spinor fields by their radiality is inevitable because different radiality leads to different asymptotic behavior  \cite{Henneaux:1998ch,Henningson:1998cd} as well as the second-class constraints of the fermion action should be solved in a Lorentz invariant way \cite{Kalkkinen:2000uk}.

In order to simplify the calculations that follow it is convenient to pick a
particular vielbein frame so that 
\be\label{default-frame}
n_\a=(1,0),\quad e_{\bar r}^i=0,\quad e^{\lbar r}_i=0,
\ee
and $e_i^a$ becomes the vielbein on the slice $\S_r$. We will call the gauge \eqref{default-frame} combined with the traditional Fefferman-Graham (FG) gauge
\be \label{eFG-gauge}
N=1,\quad N^i=0,\quad \J_r=0
\ee
as the \emph{strong} FG gauge. Namely, the strong FG gauge refers to
\be \label{sFG}
E_r^{\bar r}=1,\quad E_r^a=0,\quad E^{\lbar r}_i=0,\quad \J_r=0.
\ee 

\subsection{Decomposition of the covariant derivatives}

We obtain (see also (88) and (89) in \cite{Kalkkinen:2000uk})
\bal
\o_{r\a\b}
=&\,n_{[\a}\dot n_{\b]}+e_{i[\a}\dot e_{\b]}{}^i+2n_{[\a}e_{\b]}{}^i\left(\pa_i
N-N^jK_{ji}\right)
-D_iN_je_{[\a}{}^ie_{\b]}{}^j,\\\NO\\
\o_{i\a\b} 
=&\,n_\a\pa_i n_\b+e_{j\a}\pa_i
e_\b{}^j+\G^k_{ij}[\g]e_{k\a}e_\b{}^j+2K^j_ie_{j[\a}n_{\b]},
\eal
where we have used the Christoffel symbols
\bea\label{Christoffel}
&&\G^r_{rr}=N^{-1}\left(\dot{N}+N^i\pa_i N-N^iN^jK_{ij}\right),\NO\\
&&\G^r_{ri}=N^{-1}\left(\pa_iN-N^jK_{ij}\right),\NO\\
&&\G^r_{ij}=-N^{-1}K_{ij},\NO\\
&&\G^i_{rr}=-N^{-1}N^i\dot{N}-ND^iN-N^{-1}N^iN^j\pa_jN+\dot{N}
^i+N^jD_jN^i+2NN^jK^i_j+N^{-1}N^iN^kN^lK_{kl},\NO\\
&&\G^i_{rj}=-N^{-1}N^i\pa_jN+D_jN^i+N^{-1}N^iN^kK_{kj}+NK^i_j,\NO\\
&&\G^k_{ij}=\G^k_{ij}[\g]+N^{-1}N^kK_{ij}.\NO 
\eea
Denoting the spin connection on the radial cut-off as $\Hat \o_{iab}$, we get
\bsub
\bal 
& \Hat\o_{iab}=e_{ja}\pa_i e_b^{\;j}+\G^k_{\;ij}[\g]e_{ka}e_b^{\;j}=\o_{iab},\\
& \o_{i\a\b}\G^{\a\b}=\Hat\o_{iab}\G^{ab}+2K_{ji}e_\a^{\;j}n_\b
\G^{\a\b}=\Hat\o_{iab}\G^{ab}+2 K_{ji}\Hat\G^j\G,\\
& \o_{r\a\b}\G^{\a\b} =e_{i\a}\dot e_\b^{\;i}\G^{\a\b}+2\G\Hat\G^i\(\pa_i N-N^j K_{ji}
\)-\Hat\G^{ij}D_i N_j,\\
& \nabla_i \J_j=\bb D_i \J_j+\frac 12K_{li}\Hat \G^l\G\J_j+\frac 1N K_{ij}(\J_r-N^k\J_k) ,\\
& \nabla_i \J_r=\bb D_i \J_r+\frac 12 K_{ji}\Hat \G^j\G\J_r-\G^j_{ir}\J_j-\G^r_{ir}\J_r,\\
& \nabla_r\J_i=\dot \J_i+\frac 14\[e_{ai}\dot e_b^{i}\G^{ab}+2\G\Hat\G^j \(\pa_j
N-N^l K_{lj} \)-\Hat\G^{jl}D_j N_l \]\J_i-\G^j_{ir}\J_j-\G^r_{ir}\J_r,\\
& \nabla_i\z=\bb D_i\z+\frac12 K_{ji}\Hat\G^j\G\z,\\
&\nabla_r\z=\dot\z+\frac 14\[e_{ai}\dot e_b^{i}\G^{ab}+2\G\Hat\G^j \(\pa_j
N-N^l K_{lj} \)-\Hat\G^{jl}D_j N_l \]\z,
\eal
\esub 
where
\bsub 
\bal 
& \bb D_i \J_j=\pa_i \J_j +\frac
14\Hat\o_{iab}\G^{ab}\J_j-\G^k_{ij}[\g]\J_k,\\
& \bb D_i \J_r=\pa_i \J_r +\frac 14\Hat\o_{iab}\G^{ab}\J_r,\\
& \bb D_i\z=\pa_i\z+\frac14\o_{iab}\G^{ab}\z 
\eal
\esub 
are the covariant derivatives of the spinors on the slice $\S_r$. Note that in the final computations we used the gauge \eqref{default-frame}.

\subsection{Equations of motion and leading asymptotics of fermionic fields}\label{asymptotics-fermion}

In order to discuss with the transformation law of the induced fields, we first study the leading asymptotic behavior of the fields, which can be understood from equations of motion. For $\J_\m$ and $\z^I$ they are respectively,
\be\label{psi-eom}
\G^{\m\n\r}\nabla_\n\J_\r-\cw\G^{\m\n}\J_\n -\frac
i2\cg_{IJ}\left(\slashed\pa\vf^I+\cg^{IK}\pa_K\cw\right)\G^\m\z^J=0,
\ee
and
\be\label{zeta-eom}
\cg_{IJ}\left(\d^J_K\slashed\nabla+\G^{J}_{KL}[\cg]
\slashed\pa\vf^L\right)\z^K+\cm_{IJ}(\vf)\z^J+\frac
i2\cg_{IJ}\G^\m\left(\slashed\pa\vf^J-\cg^{JK}\pa_K\cw\right)\J_\m=0.
\ee
Extracting the relevant terms, we obtain in the gauge \eqref{eFG-gauge}
\bal
0\sim\;& -\Hat\G^{ij}\(\dot \J_{+j}-\frac12\J_{+j}\)+\Hat\G^{ij}\(\dot\J_{-j}+\frac{2d-3}{2}\J_{-j}\)+\Hat\G^{ijk}\bb D_j(\J_{+k}+\J_{-k}),\label{asymptotics-gravitino}\\
0\sim \;& \dot\z_+ +\(\frac d2+M_\z  \)\z_+-\dot \z_--\(\frac d2-M_\z \)\z_-+\Hat\G^i\bb D_i\z_+-\Hat\G^i\bb D_i\z_- +\frac i2(\dot\vf+\m\vf)\Hat\G^i\J_{+i}\NO\\
&\hskip0.5in +\frac i2\Hat\G^i\Hat\G^j\pa_j\vf\J_{+i},\label{asymptotics-hyperino}
\eal  
where we assume that there is only one scalar $\vf$ and one spin-$1/2$ field $\z$ for simplicity, and $M_\z$ which is the mass of $\z$ and $\m$ are respectively
\be 
\m=-\pa_\vf\pa_\vf\cw\Big|_{\vf=0},\quad M_\z=\cm_{\vf\vf }\Big|_{\vf=0},
\ee 
under the assumption that the scalar manifold metric is canonically normalized. $\m$ and $M_\z$ are related by
\be 
M_\z=-\m+\frac{d-1}{2}.
\ee

When $d>2$, the leading asymptotics of $\J_{+i}$ and $\J_{-i}$ are 
\bal 
& \J_{+i}(r,x)\sim e^{\frac r2}\J_{(0)+i}(x),\label{Psi+asympt}\\
& \J_{-i}(r,x)\sim -\frac12 e^{-\frac12r} \(\frac{d-2}{d-1}\Hat\G^{(0)}{}_i\Hat\G^{(0)kl}-\Hat\G^{(0)}{}_i{}^{kl}\)\bb D^{(0)}_k\J_{(0)+l}(x),\label{Psi-asympt}
\eal
where we used $e^a_i(r,x)\sim e^r e^a_{(0)i}(x)$ in AlAdS geometry, and $\G^{(0){}_i}$ and $\bb D^{(0)}$ refer to the Gamma matrices and the covariant derivative with respect to $e^a_{(0)i}$.

We need to be more careful, regarding $\z$. First, we note that the leading asymptotics of $\vf$ should always be $\vf(r,x)\sim e^{-\m r}\vf_{(0)}(x)$ as can be seen from \eqref{leading-flow-vf}. Therefore, the final two terms in \eqref{asymptotics-hyperino} can be discarded from the argument. Now there are 3 cases to consider:
\begin{enumerate}
\item $M_\z> 1/2$ (or $\m< \frac d2-1$)

The leading asymptotics of $\z_-$ and $\z_+$ are respectively
\bal
& \z_-(r,x)\sim e^{-(\m+\frac12)r}\z_{(0)-}(x),\\
& \z_+(r,x)\sim -\frac{1}{\m+\frac32}\(e^{-(\m+\frac32)r}\Hat\G^{(0)i}\bb D_i^{(0)}\z_{(0)-} (x)-\frac i2\Hat\G^{(0)i}\Hat\G^{(0)j}\pa_j\vf_{(0)}(x)\J_{(0)+i}(x)\).
\eal	
\item $M_\z<-1/2$ (or $\m> \frac d2$)

Here the behavior of $\z_-$ and $\z_+$ is opposite to the first case, namely
\bal
& \z_+(r,x)\sim e^{-(d-\m-\frac12)r}\z_{(0)+}(x),\\
& \z_-(r,x)\sim \frac{1}{d-\m+\frac12}e^{-(d-\m+\frac12)r}\Hat\G^{(0)i}\bb D_i^{(0)}\z_{(0)+}(x).
\eal
	
\item $1/2\geq M_\z\geq -1/2$ (or $\frac d2\geq \m\geq  \frac d2-1$)

This case actually coincides with the double quantization window \cite{Breitenlohner:1982bm,Balasubramanian:1998sn,Klebanov:1999tb} of the scalar field. The leading asymptotics are
\bal
& \z_-(r,x)\sim e^{-(\m+\frac12)r}\z_{(0)-}(x),\\
&\z_+(r,x)\sim e^{-(d-\m-\frac12)r}\z_{(0)+}(x).
\eal 
\end{enumerate}

\subsection{Generalized PBH transformations}
\label{gPBH}

Let us find the most general bulk symmetry transformations that preserve the strong FG gauge \eqref{sFG}, which we call as generalized Penrose-Brown-Henneaux (gPBH) transformations. We can immediately see that local symmetries of the bulk SUGRA action \eqref{SUGRA-action} are diffeomorphism, local Lorentz and supersymmetry transformation. Their infinitesimal action on the bulk fields takes the form
\bsub \label{generic-var-fields}
\bal
& \d_{\x,\l,\e}E^\a_\m=\x^\n\pa_\n E^\a_\m+(\pa_\m\x^\n)E^\a_\n-\l^\a{}_\b E^\b_\m+\frac12(\lbar\e \G^\a\J_\m-\lbar\J_\m\G^\a\e),\\
&\d_{\x,\l,\e}\J_\m=\x^\n\pa_\n\J_\m+(\pa_\m\x^\n)\J_\n-\frac14\l^{\a\b}\G_{\a\b}\J_\m+(\nabla_\m+\frac{1}{2(d-1)}\cw\G_\m)\e,\\
& \d_{\x,\l,\e}\vf^I=\x^\m\pa_\m\vf^I+\frac i2(\lbar\e\z^I-\lbar\z^I\e),\\
&\d_{\x,\l,\e}\z^I=\x^\m\pa_\m\z^I-\frac14\l^{\a\b}\G_{\a\b}\z^I-\frac i2(\slashed\pa\vf^I-\cg^{IJ}\pa_J\cw)\e,
\eal
\esub 
with the parameters $\x^\m $, $\l^{\a\b}$ ($\l^{\a\b}=-\l^{\b\a}$) and $\e$ respectively. The condition which keeps the strong FG gauge invariant is then
\bsub
\bal 
& 0=\dot\x^r,\\
& 0=\dot \x^i e^a_i-\l^a{}_{\bar r},\\
& 0=\pa_i\x^r-\l^{\bar r}{}_a e^a_i+\frac12(\lbar\e_-\J_{+i}+\lbar\J_{+i}\e_--\lbar\e_+\J_{-i}-\lbar\J_{-i}\e_+),\\
& 0=\dot\e_++\dot\e_- +\dot\x^i(\J_{+i}+\J_{-i}) +\frac14 e_{ai}\dot e^i_b\G^{ab}(\e_++\e_-)+\frac{1}{2(d-1)}\cw(\e_+-\e_-),
\eal 
\esub 
and its solution is
\bsub\label{PBH-solution}
\bal 
& \x^r=\s(x),\\
& \x^i(r,x)=\x^i_o(x)-\int^r dr^\prime \;\g^{ij}(r^\prime,x)\Big[\pa_j\s+\frac12(\lbar\e_-\J_{+j}+\lbar\J_{+j}\e_--\lbar\e_+\J_{-j}-\lbar\J_{-j}\e_+) \Big],\\
& \l^{\bar r a}=e^{ai}\Big[\pa_i\s+\frac12(\lbar\e_-\J_{+i}+\lbar\J_{+i}\e_--\lbar\e_+\J_{-i}-\lbar\J_{-i}\e_+) \Big],\\
& \l^a{}_b=\l_o{}^a{}_b(x)+\cdots,\\
& \e_+(r,x)=\exp\Bigg[\frac r2+\int^r dr^\prime \Big(-\frac{\cw+(d-1)}{2(d-1)}+\g^{ij}(r^\prime,x)\pa_j\s-\frac14 e_{ai}\dot e_b^i\G^{ab} +O(\J^2)  \Big) \Bigg]\e_{o+}(x),\\
& \e_-(r,x)=\exp\Bigg[-\frac r2+\int^r dr^\prime \Big(\frac{\cw+(d-1)}{2(d-1)}+\g^{ij}(r^\prime,x)\pa_j\s-\frac14 e_{ai}\dot e_b^i\G^{ab} +O(\J^2)  \Big) \Bigg]\e_{o-}(x),
\eal 
\esub 
where $\s(x)$, $\x^i_o(x)$, $\l_o{}^a{}_b(x)$ and $\e_{o\pm}(x)$ are 'integration constants' which depend only on transverse coordinates. Taking into account leading behavior of the vielbeins and gravitino one can find the integral terms are subleading in \eqref{PBH-solution}. It follows that leading asymptotics of the generalized PBH transformations are parameterized by the arbitrary independent transverse functions
\be 
\s(x),\quad \x_o^i(x),\quad \l_o{}^a{}_b(x),\quad \e_{o\pm}(x),
\ee 
which in fact correspond to the local conformal, diffeomorphism, Lorentz, SUSY, and super-Weyl transformations of the induced fields on the radial slice $\S_r$ respectively, as we see soon.

Extracting the leading terms in \eqref{generic-var-fields} and taking into account asymptotic behavior of the induced fields, we obtain how the sources transform, namely (from now on and also in the main text we do not write the subscript $o$)
\bsub 
\bal
\d_{\x,\l,\e}e_i^a &\sim \x^j\pa_j e_i^a+\pa_i\x^j e^a_j+e^a_i\s-\l^a{}_b e^b_i+\frac12\left(\lbar\e_+\G^a\J_{+i}+\mathrm{h.c.}\right),\label{gamma-sdif-1}\\
\d_{\x,\l,\e}\J_{+i}&\sim\frac12 \J_{+i}\s+\x^j\pa_j\J_{+i}+(\pa_i\x^j)\J_{+j}+\bb D_i\e_+-\Hat\G_i\e_--\frac14\l^{ab}\G_{ab}\J_{+i},\label{J-sdif-1}\\
\d_{\x,\l,\e}\vf^I&\sim-\cg^{IJ}\pa_J\cw\s+\x^i\pa_i\vf^I+\frac i2\left(\lbar\e_+\z_-^{I}+\mathrm{h.c.}\right)+\frac i2\left(\lbar\e_-\z_+^{I}+\mathrm{h.c.}\right),\label{vf-sdif-1}
\eal
\esub 
where we do not write down the variation of $\J_{-i}$ since unlike $\J_{+i}$ its leading term \eqref{Psi-asympt} does not transform as a source so that it cannot be used as a generalized coordinate \cite{Arutyunov:1998ve}.

As for $\z^I$, we need a careful discussion, since its leading behavior changes according to its mass. In the first case where  $M_\z\geq 1/2$, $\z_+^I$ cannot be served as a source, like the case of gravitino $\J_{-i}$. We also find that in the second case where $M^I_\z\leq-1/2$ \eqref{vf-sdif-1} is not consistent with leading behavior of $\vf\sim e^{-\m r}$ due to the term $\frac i2(\lbar\e_-\z_+^I+{\rm h.c.})\sim e^{-(d-\m^I)r}$, which implies $\z_+^I$ can not be turned on as a source, in order to have the theory supersymmetric. In the final case where $1/2>M_\z>-1/2$, both $\z_+^I$ and $\z_-^I$ can be used as sources. The transformation law in this case is discussed in section \ref{Neumann-BC}. In summary, what we obtain is
\bsub\label{superdiffeo-induced}
\bal 
\d_{\x,\l,\e}e_i^a &\sim \x^j\pa_j e_i^a+\pa_i\x^j e^a_j+e^a_i\s-\l^a{}_b e^b_i+\frac12\left(\lbar\e_+\G^a\J_{+i}+\mathrm{h.c.}\right),\label{gamma-sdif}\\
\d_{\x,\l,\e}\J_{+i}&\sim\frac12 \J_{+i}\s+\x^j\pa_j\J_{+i}+(\pa_i\x^j)\J_{+j}+\bb D_i\e_+-\Hat\G_i\e_--\frac14\l^{ab}\G_{ab}\J_{+i},\label{J-sdif}\\
\d_{\x,\l,\e}\vf^I&\sim\cg^{IJ}\pa_J\cw\s+\x^i\pa_i\vf^I+\frac i2\left(\lbar\e_+\z_-^{I}+\mathrm{h.c.}\right),\label{vf-sdif}\\
\d_{\x,\l,\e}\z_-^{I}& \sim-\(\frac d2\d^I_K-\cg^{IJ}\cm_{JK}\)\z_-^{K}\s+\x^i\pa_i\z_-^{I}+i\cg^{IJ}\pa_J\cw\e_--\frac i2\Hat\G^i\pa_i\vf^I\e_+-\frac14\l^{ab}\G_{ab}\z_-^I,
\eal 
\esub 
where we inverted mass of $\z_{-}^I$ into the (scalar) $\s$-manifold language.

\section{Decomposition of the action and the fermion boundary terms}\label{ADM-decomposition}
\setcounter{equation}{0}

In this appendix we decompose the terms in the fermionic sector of the action \eqref{SUGRA-action}.

\subsection{Decomposition of the kinetic action of the hyperino field}
\label{hyperino-kinetic-decomp}
The kinetic term for $\z^I$ in the action \eqref{SUGRA-action} is decomposed as
\bal
&\cg_{IJ}\left(\lbar\z^I\G^\m\nabla_\m\z^J-(\nabla_\m\lbar\z^I) \G^\m\z^J\right)\NO\\ 
=\;&\cg_{IJ}\lbar\z^I\(\G^r\nabla_r\z^J+\G^i\nabla_i\z^J\)-\cg_{IJ}\lbar\z^I\overleftarrow \nabla_r\G^r\z^J\z^J-\cg_{IJ}\lbar\z^I\overleftarrow\nabla_i\G^i\z^J\NO\\
=\;&\cg_{IJ}\lbar\z^I\Bigg[\frac 1N\G\dot\z^J+\frac{1}{4N}\G\(e_{ai}\dot e_b^i\G^{ab}+ 2\G\Hat\G^i\(\pa_i N-N^j K_{ij}\)-\Hat\G^{ij}D_i N_j\)\z^J\NO\\
&\hskip0.5in+\(\Hat\G^i-\frac{N^i}{N}\G\)\(\bb D_i\z^J+\frac 12 K_{ij}\Hat\G^j\G\z^J\) \Bigg]\NO\\
&-\cg_{IJ}\Bigg[\dot{\lbar\z}^I-\frac 14\lbar\z^I\[e_{ai}\dot e_b^i\G^{ab}+2\G\Hat\G^i \(\pa_i N-N^j K_{ij}\)-\Hat\G^{ij}D_i N_j\]\Bigg]\frac 1N \G\z^J\NO\\
&-\cg_{IJ}\(\lbar\z^I\overleftarrow{\bb D}_i-\frac 12K_{ij}\lbar\z^I\Hat\G^j\G\)\(\Hat\G^i-\frac{N^i}{N}\G\)\z^J\NO\\
=\;&\frac 1N\cg_{IJ}\(\lbar\z_-^I\dot\z_+^J-\lbar\z_+^I\dot\z_-^J- \dot{\lbar\z}^I_-\z_+^J+\dot{\lbar\z}_+^I\z_-^J\)+\frac{1}{2N}\cg_{IJ}e_{ai}\dot e_b^i \lbar\z^I\G\G^{ab}\z^J\NO\\
&-\frac{1}{2N}D_i N_j\cg_{IJ}\lbar\z^I\G\Hat\G^{ij}\z^J+\cg_{IJ}\(\lbar\z^I\slashed{\bb D}\z^J-\lbar\z^I\overleftarrow{\slashed{\bb D}}\z^J\)\NO\\
&-\frac{N^i}{N}\cg_{IJ}\(\lbar\z^I \G\bb D_i\z^J-\lbar\z^I\overleftarrow{\bb D}_i\G\z^J\),
\eal
where the terms in the first bracket can be recast into
\bal
& \cg_{IJ}\(\lbar\z_-^I\dot\z_+^J-\lbar\z_+^I\dot\z_-^J- \dot{\lbar\z}^I_-\z_+^J+\dot{\lbar\z}_+^I\z_-^J\)=\cg_{IJ}\pa_r\(\lbar\z_-^I\z_+^J+\lbar\z_+^I\z_-^J \)-2\cg_{IJ}\lbar \z_+^I\dot\z_-^J-2\cg_{IJ}\dot{\lbar\z}^I_-\z_+^J \NO\\
=\;&\frac{1}{\sqrt{-\g}}\pa_r\(\cg_{IJ}\sqrt{-\g}\lbar\z^I\z^J\)-\(NK+D_k N^k\) \cg_{IJ}\lbar\z^I\z^J\NO\\
&-\(\dot\vf^K-N^i\pa_i\vf^K+N^i\pa_i\vf^K\)\pa_K\cg_{IJ}\lbar\z^I\z^J-2\cg_{IJ}\lbar \z_+^I\dot\z_-^J-2\cg_{IJ}\dot{\lbar\z}^I_-\z_+^J.
\eal

Finally, the hyperino kinetic terms are decomposed into
\bala
&\cg_{IJ}\left(\lbar\z^I\G^\m\nabla_\m\z^J-(\nabla_\m\lbar\z^I)\G^\m \z^J\right)\\ 
=\; & \frac{1}{N\sqrt{-\g}}\pa_r\left(\sqrt{-\g}\;\cg_{IJ}\lbar\z^I\z^J\right)-\frac{2}{N}\cg_{IJ}\left(\lbar\z^I_+\dot\z^J_-+\dot{\lbar\z}^I_-\z^J_+\right)-\left(K+ \frac{1}{N}D_k N^k\right)\cg_{IJ}\lbar\z^I\z^J\\
& +\frac{1}{2N}\cg_{IJ}e_{ai}\dot e_b^{i}\lbar\z^I\G^{ab}\G\z^J -\frac{1}{N}\( \dot\vf^K -N^i\pa_i \vf^K+N^i\pa_i\vf^K \)\pa_K\cg_{IJ}\lbar\z^I\z^J\\
& + \cg_{IJ}\(\lbar\z^I\Hat \G^i \bb D_i \z^J-\lbar\z^I \overleftarrow{\bb D_i}\Hat \G^i \z^J \)\\
& +\frac{1}{N}\cg_{IJ} \[-\frac12 D_i N_j \(\lbar \z^I\Hat\G^{ij}\G\z^J \)-N^i\lbar \z^I\G \bb D_i \z^J +N^i (\lbar\z^I\overleftarrow{\bb D}_i)\G\z^J \].\numberthis\label{hyp-kin}
\eala

\subsection{Gravitino part}
\label{gravitino-kinetic-decomp}

Repeating the same computation for the kinetic terms for gravitino as before, we obtain
\bala
& \(\lbar\J_\m\G^{\m\n\r}\nabla_\n\J_\r-\lbar\J_\m\overleftarrow{\nabla}_\n \G^{\m\n\r}\J_\r\)+\frac{1}{(D-2)} \lbar\J_\m\G^{\m\n\r}\(\cw\G_\n\)\J_\r\\
= \; & \frac{1}{N\sqrt{-\g}}\pa_r\(\sqrt{-\g}\lbar\J_i\Hat\G^{ij}\J_j \)-\frac 2N\(\dot{\lbar\J}_{+i}\Hat\G^{ij}  \J_{-j}+\lbar\J_{-i} \Hat \G^{ij} \dot\J_{+j} \)\\
& - \(K+\frac 1N D_k N^k \)\lbar \J_i\Hat\G^{ij}\J_j-\frac{1}{4N}e_{ak}\dot e_b^{\;k}\lbar \J_i \G\{ \Hat \G^{ij},\G^{ab}\}\J_j\\
& +\frac{1}{2N}K_{lk}\lbar \J_i\Big( N[\Hat \G^{ikj},\Hat \G^l ]\G +N^i [\Hat \G^{kj},\Hat \G^l]-N^j[\Hat \G^{ki},\Hat \G^l]\Big)\J_j\\
& +\frac{1}{2N}K_{ki}\(\lbar \J_j[\Hat \G^{ij},\Hat \G^k]\J_r-\lbar \J_r[\Hat \G^{ij},\Hat \G^k]\J_j\) \\
& +\frac {1}{N}\(\lbar \J_j\overleftarrow{\bb D}_i\G\Hat \G^{ij}\J_r+\lbar\J_r \G\Hat \G^{ij}\bb D_i\J_j-\lbar\J_j\G\Hat\G^{ij}\bb D_i \J_r-\lbar\J_r\overleftarrow{\bb D}_i\G\Hat\G^{ij}\J_j \)\\
&-\frac{1}{N}\cw\(\lbar\J_r \G\Hat\G^i\J_i +\lbar\J_i\Hat \G^i\G \J_r \) -\frac{1}{4N}\lbar\J_i \(2\pa_k N[\Hat \G^{ij},\Hat \G^k]-(D_k N_l)\G\{\Hat \G^{ij},\Hat\G^{kl} \} \)\J_j\\
& +\frac{1}{N}\lbar\J_j\(N\Hat\G^{jik}-N^j\G\Hat\G^{ik}-N^i\G\Hat\G^{kj}-N^k\G\Hat\G^{ji} \)\bb D_i\J_k\\
& +\frac{1}{N}\lbar\J_k\overleftarrow{\bb D}_i\(N\Hat\G^{jik}-N^j\G\Hat\G^{ik}-N^i\G\Hat\G^{kj}-N^k\G\Hat\G^{ji} \)\J_j\\
& -\frac{1}{N}\cw \lbar\J_i\(N\Hat\G^{ij}-N^i\G\Hat\G^j+N^j\G\Hat\G^i \)\J_j.\numberthis 
\eala

\subsection{Decomposition of the other terms}

For the other terms, we get
\bala
& i\cg_{IJ}\lbar\z^I\G^\m\left(\slashed\pa\vf^J-\cg^{JK}\pa_K\cw \right)\J_\m-i \cg_{IJ}\lbar \J_\m (\slashed \pa \vf^I+\cg^{IK}\pa_K \cw)\G^\m \z^J \\
= \; & \frac{i}{N}\cg_{IJ}\Bigg\{ \frac 1N \(\dot \vf^J-N^j\pa_j \vf^J\)\Big[ \lbar\z^I \(\J_r-N^i\J_i+ N\Hat\G^i\G\J_i \)-\(\lbar \J_r-N^i \lbar\J_i +N\lbar\J_i \G \Hat\G^i \)\z^I\Big]\\
& \hskip0.2in+\pa_i \vf^J\[\lbar\z^I\G\Hat\G^i\(\J_r-N^j \J_j \)-\(\lbar\J_r-N^j\lbar\J_j \)\Hat\G^i\G\z^I \]+N\pa_i\vf^I \(\lbar\z^I\Hat\G^j\Hat\G^i\J_j-\lbar\J_j\Hat\G^i\Hat\G^j\z^I \)\Bigg\}\\
&-\frac{i}{N}\pa_I\cw\[\lbar\z^I\G\(\J_r-N^i\J_i \)+\(\lbar\J_r-N^i\lbar\J_i \)\G\z^I +N\(\lbar\J_i\Hat\G^i\z^I+\z^I \Hat\G^i\J_i \) \],\numberthis 
\eala
and
\bala 
& \cg_{IJ}\left[\lbar \z^I \(\G^J_{\;KL}\slashed\pa\vf^L\)\z^K -\lbar \z^K \(\G^J_{\;KL}\slashed\pa\vf^L\)\z^I \right]\\
=\; & \frac{1}{N}\pa_K \cg_{IJ}\[\(\dot\vf^J-N^i\pa_I\vf^J \)\(\lbar\z^I\G\z^K-\lbar\z^K\G\z^I \)+N\pa_i\vf^J \(\lbar\z^I\Hat\G^i\z^K-\lbar\z^K\Hat \G^i \z^I \) \].\numberthis 
\eala

\section{Variation of the canonical momenta under the generalized PBH transformation}
\label{momenta-var-derivation}
\setcounter{equation}{0}

By chain rule,
\be 
\d \Hat S_{ren}=\int d^dx\;\sum_\F \P^\F \d\F,
\ee 
and let us define a symmetry transformation of $\Hat S_{ren}$ by
\be 
\d_\x=\int d^dx\;\sum_\F \d_\x\F(x)\frac{\d}{\d\F(x)}.
\ee 
Let us also assume that this symmetry has anomaly, i.e.
\be 
\d_\x \Hat S_{ren}=\int d^dx\;\sum_\F \P^\F \d_\x\F=\int d^dx|\bm e_{(0)}|\;\x\ca_\x.
\ee 
Then, definition of the constraint function $\cc[\x]$ \eqref{CC-ward} can be written as
\be 
\cc[\x]= -\int d^dx\(\sum_\F \P^\F\d_\x\F-|\bm e_{(0)}|\;\x\ca_\x\).
\ee 

Now we derive how $\x$-symmetry acts on $\P^\F$. It is
\bala
\d_\x\P^\F(x)=\;&\d_\x\frac{\d}{\d\F(x)}\Hat S_{ren}=\[\d_\x,\frac{\d}{\d\F(x)}\]\Hat S_{ren}+\frac{\d}{\d\F(x)}\d_\x \Hat S_{ren}\\
=\;&-\int d^dy\;\sum_{\F'}\(\frac{\d}{\d\F(y)}\d_\x\F'(x)\)\P^{\F\prime}(x)+\frac{\d}{\F(x)}\int d^dy|\bm e_{(0)}|\;\x\ca_\x \\
=\;&-\frac{\d}{\d\F(x)}\int d^dy\;\sum_{\F'} \(\P^{\F\prime}(y)\d_\x\F'(y)-|\bm e_{(0)}|\x\ca_\x\)\\
=\;&\{\cc[\x],\P^\F\},\numberthis\label{var-P-under-x} 
\eala
\eqref{var-P-under-x} confirms \eqref{PB-PBH}.

\section{Derivation of the SUSY algebra without using Poisson bracket}
\label{susyalgebra-without-Poisson}
\setcounter{equation}{0}

In this appendix we derive $\{\cq[\x],Q^s[\h_+]\}$. By differentiating the diffeomorphism Ward identity \eqref{ward-diffeo} in the integral form with respect to $\J_{+k}(y)$, we get
\bala
0=\;&\int_{\pa\cm}d^dx\;\x_i\Big[e_{(0)}^{ai}D_j\P_a^{\;j}-(\pa^i\vf_{(0)}^I)\P_I^\vf- (\lbar\z_{(0)-}^I\overleftarrow{\bb D}^i)\P_I^{\lbar \z}-\P_I^{\z}(\bb D^i\z_{(0)-}^I)\\
&-\P_\J^{j}\(\bb D^i\J_{(0)+j}\)-\(\lbar\J_{(0)+j}\overleftarrow{\bb D}^i\)\P_{\lbar \J}^{j}+D_j(\P_\J^j\J_{(0)+}^i+\lbar\J_{(0)+}^i\P_{\lbar\J}^j)\Big]_x \P_\J^k(y)\\
&+\(\x^i\P_\J^k\)\labD_i(y)-D_j\x^k\P_\J^j(y).\numberthis 
\eala 
From the local Lorentz Ward identity \eqref{ward-lorentz}, we obtain
\be 
0=\int_{\pa\cm}d^dx\;\l^{ab}\Big[e_{(0)[ai}\P^i_{b]}+\frac14\big(\lbar\z_{(0)-}^I\G_{ab}\P_I^{\lbar\z}+\lbar\J_{(0)+i}\G_{ab}\P_{\lbar\J}^i+{\rm h.c.} \big) \Big]_x \P_\J^k(y)-\frac14\l^{ab}\P_\J^k\G_{ab}(y). 
\ee 
Summing these two expressions for the parameter $\l_{ab}=e^i_ae^j_bD_{[i}\x_{j]}$, we obtain
\bal
0=\;&\int_{\pa\cm}d^dx\;D_j[\x^i(e^{ai}\P_a^j+\P_\J^j\J_{(0)+i}+\lbar\J_{(0)+i}\P_{\lbar\J}^j)]_x \P_\J^k(y)+\NO \\
&\hskip1in+ \(\x^i\P_\J^k\)\labD_i(y)-D_j\x^k\P_\J^j(y)-\frac14 D_i\x_j\P_\J^k\Hat\G^{ij}(y).
\eal
It follows from \eqref{algebra-prescription} that
\bala
& \{\cq[\x],Q^s[\h_+]\}=-\int_{\pa\cm\cap\cc}d\s_k(y)\int_{\pa\cm}d^dxD_j[\x^i(e^{ai}\P_a^j+\P_\J^j\J_{(0)+i}+\lbar\J_{(0)+i}\P_{\lbar\J}^j)]_x (\P_\J^k\h_+)_y\\
&=\int_{\pa\cm\cap\cc}d\s_k\[\(\x^i\P_\J^k\)\labD_i-D_j\x^k\P_\J^j-\frac14 D_i\x_j\P_\J^k\Hat\G^{ij} \]\h_+\\
&=\int_{\pa\cm\cap\cc}d\s_k\[D_i(\x^i\P_\J^k\h_+-\x^k\P_\J^i\h_+)+\x^kD_j(\P_\J^j\h_+)-\P_\J^k\cl_\x\h_+ \]\\
&=-Q^s[\cl_\x\h_+],\numberthis
\eala 
where the first term in the third line is zero by using the Stokes' theorem and the second term vanishes due to the conservation law. One can confirm that the other commutators in \eqref{curved-susy-algebra} can be obtained in the above way.

\addcontentsline{toc}{section}{References}


\bibliographystyle{jhepcap}
\bibliography{susy,ads-cft,holo-ren}

Department of Mathematics, SISSA, Italy

ICTP, Italy

\emph{E-mail address}: \href{mailto: oan@sissa.it}{\tt oan@sissa.it}

\end{document}

%% file: macros.tex
\def\){\right)}
\def\({\left( }
\def\]{\right] }
\def\[{\left[ }

\def\NO{\nonumber}

\newcommand{\be}{\begin{equation}}
\newcommand{\ee}{\end{equation}}

\def\bea{\begin{eqnarray}}
\def\eea{\end{eqnarray}}

\def\bal#1\eal{\begin{align}#1\end{align}}
\def\bala#1\eala{\begin{align*}#1\end{align*}}
\newcommand\numberthis{\addtocounter{equation}{1}\tag{\theequation}}

\def\bald{\begin{aligned}}
\def\eald{\end{aligned}}

\def\bsub{\begin{subequations}}
\def\esub{\end{subequations}}

\def\beqx{\begin{displaymath}}
\def\eeqx{\end{displaymath}}

\newcommand{\bmat}{\left(\begin{array}}
\newcommand{\emat}{\end{array}\right)}

\def\a{\alpha}
\def\b{\beta}
\def\c{\chi}
\def\d{\delta}
\def\e{\epsilon}
\def\f{\phi}
\def\g{\gamma}
\def\h{\eta}
\def\j{\psi}
\def\k{\kappa}
\def\l{\lambda}
\def\m{\mu}
\def\n{\nu}
\def\o{\omega}
    
\def\p{\pi}

\def\r{\rho}
\def\s{\sigma}
\def\t{\tau}
\def\x{\xi}
\def\z{\zeta}
\def\D{\Delta}
\def\F{\Phi}
\def\G{\Gamma}
\def\J{\Psi}

\def\P{\Pi}

\def\S{\Sigma}

\def\X{\Xi}

\def\vf{\varphi}

\def\ca{{\cal A}}

\def\cc{{\cal C}}

\def\cf{{\cal F}}
\def\cg{{\cal G}}
\def\ch{{\cal H}}

\def\cj{{\cal J}}
\def\ck{{\cal K}}
\def\cl{{\cal L}}
\def\cm{{\cal M}}
\def\cn{{\cal N}}
\def\co{{\cal O}}
\def\cp{{\cal P}}
\def\cq{{\cal Q}}

\def\cs{{\cal S}}
\def\ct{{\cal T}}

\def\cv{{\cal V}}
\def\cw{{\cal W}}

\def\bb#1{\ensuremath{\mathbb{#1}}} 

\def\bo{{\raise-.3ex\hbox{\large$\Box$}}}               
\def\pa{\partial}                                       
\def\face{{\raise.2ex\hbox{$\displaystyle \bigodot$}\mskip-2.2mu \llap {$\ddot
        \smile$}}}                                   
\def\>{\rangle}                                      
\def\<{\langle}                                      

\def\slash#1{\rlap{\hbox{$\mskip 1 mu /$}}#1}        
\def\wt#1{\widetilde{#1}}                            
\def\Hat#1{\widehat{#1}}                             
\def\lbar#1{\ensuremath{\overline{#1}}}              
\def\leftrightarrowfill{$\mathsurround=0pt \mathord\leftarrow \mkern-6mu
        \cleaders\hbox{$\mkern-2mu \mathord- \mkern-2mu$}\hfill
        \mkern-6mu \mathord\rightarrow$}        
\def\dvec#1{\vbox{\ialign{##\crcr
        \leftrightarrowfill\crcr\noalign{\kern-1pt\nointerlineskip}
        $\hfil\displaystyle{#1}\hfil$\crcr}}}           
\def\diag{{\rm diag \,}}                                

\def\-{\hphantom{-}}

\def\labD {\overleftarrow{\bb D}}
\def\slabD {\overleftarrow{\slashed{\bb D}}}